\PassOptionsToPackage{unicode}{hyperref}
\PassOptionsToPackage{hyphens}{url}
\PassOptionsToPackage{dvipsnames,svgnames,x11names}{xcolor}
\documentclass[
  english,
  ,man,floatsintext]{apa6}
\title{Data aggregation can lead to biased inferences in Bayesian linear mixed models and Bayesian ANOVA: A simulation study}
\author{Daniel J. Schad\textsuperscript{1}, Bruno Nicenboim\textsuperscript{2}, \& Shravan Vasishth\textsuperscript{3}}
\date{}

\usepackage{amsmath,amssymb}
\usepackage{lmodern}
\usepackage{iftex}
\ifPDFTeX
  \usepackage[T1]{fontenc}
  \usepackage[utf8]{inputenc}
  \usepackage{textcomp} 
\else 
  \usepackage{unicode-math}
  \defaultfontfeatures{Scale=MatchLowercase}
  \defaultfontfeatures[\rmfamily]{Ligatures=TeX,Scale=1}
\fi
\IfFileExists{upquote.sty}{\usepackage{upquote}}{}
\IfFileExists{microtype.sty}{
  \usepackage[]{microtype}
  \UseMicrotypeSet[protrusion]{basicmath} 
}{}
\makeatletter
\@ifundefined{KOMAClassName}{
  \IfFileExists{parskip.sty}{%
    \usepackage{parskip}
  }{
    \setlength{\parindent}{0pt}
    \setlength{\parskip}{6pt plus 2pt minus 1pt}}
}{
  \KOMAoptions{parskip=half}}
\makeatother
\usepackage{xcolor}
\IfFileExists{xurl.sty}{\usepackage{xurl}}{} 
\IfFileExists{bookmark.sty}{\usepackage{bookmark}}{\usepackage{hyperref}}
\hypersetup{
  pdftitle={Data aggregation can lead to biased inferences in Bayesian linear mixed models and Bayesian ANOVA: A simulation study},
  pdfauthor={Daniel J. Schad1, Bruno Nicenboim2, \& Shravan Vasishth3},
  pdflang={en-EN},
  pdfkeywords={data aggregation, Bayes factors, Bayesian model comparison, simulation-based calibration, null hypothesis testing, sphericity assumption, items},
  colorlinks=true,
  linkcolor={Maroon},
  filecolor={Maroon},
  citecolor={Blue},
  urlcolor={blue},
  pdfcreator={LaTeX via pandoc}}
\usepackage{graphicx}
\makeatletter
\def\maxwidth{\ifdim\Gin@nat@width>\linewidth\linewidth\else\Gin@nat@width\fi}
\def\maxheight{\ifdim\Gin@nat@height>\textheight\textheight\else\Gin@nat@height\fi}
\makeatother
\setkeys{Gin}{width=\maxwidth,height=\maxheight,keepaspectratio}
\makeatletter
\def\fps@figure{htbp}
\makeatother
\ifx\paragraph\undefined\else
  \let\oldparagraph\paragraph
  \renewcommand{\paragraph}[1]{\oldparagraph{#1}\mbox{}}
\fi
\ifx\subparagraph\undefined\else
  \let\oldsubparagraph\subparagraph
  \renewcommand{\subparagraph}[1]{\oldsubparagraph{#1}\mbox{}}
\fi
\newlength{\cslhangindent}
\newlength{\csllabelwidth}
\newlength{\cslentryspacingunit} 
\newenvironment{CSLReferences}[2] 
 {
  \setlength{\parindent}{0pt}
  \ifodd #1
  \let\oldpar\par
  \def\par{\hangindent=\cslhangindent\oldpar}
  \fi
  \setlength{\parskip}{#2\cslentryspacingunit}
 }%
 {}
\usepackage{calc}

\usepackage{upgreek}
\usepackage{longtable}
\usepackage{lscape}
\usepackage{multirow}		
\usepackage{tabularx}		
\usepackage[flushleft]{threeparttable}	
\usepackage{threeparttablex}            

\makeatletter
\newcommand\LastLTentrywidth{1em}
\newlength\longtablewidth
\newcommand{\getlongtablewidth}{\begingroup \ifcsname LT@\roman{LT@tables}\endcsname \global\longtablewidth=0pt \renewcommand{\LT@entry}[2]{\global\advance\longtablewidth by ##2\relax\gdef\LastLTentrywidth{##2}}\@nameuse{LT@\roman{LT@tables}} \fi \endgroup}

\makeatletter
\renewcommand{\paragraph}{\@startsection{paragraph}{4}{\parindent}%
  {0\baselineskip \@plus 0.2ex \@minus 0.2ex}%
  {-1em}%
  {\normalfont\normalsize\bfseries\itshape\typesectitle}}

\renewcommand{\subparagraph}[1]{\@startsection{subparagraph}{5}{1em}%
  {0\baselineskip \@plus 0.2ex \@minus 0.2ex}%
  {-\z@\relax}%
  {\normalfont\normalsize\itshape\hspace{\parindent}{#1}\textit{\addperi}}{\relax}}
\makeatother

\makeatletter
\patchcmd{\HyOrg@maketitle}
  {\section{\normalfont\normalsize\abstractname}}
  {\section*{\normalfont\normalsize\abstractname}}
  {}{\typeout{Failed to patch abstract.}}
\patchcmd{\HyOrg@maketitle}
  {\section{\protect\normalfont{\@title}}}
  {\section*{\protect\normalfont{\@title}}}
  {}{\typeout{Failed to patch title.}}
\makeatother

\usepackage{xpatch}
\makeatletter
\xapptocmd\appendix
  {\xapptocmd\section
    {\addcontentsline{toc}{section}{\appendixname\ifoneappendix\else~\theappendix\fi\\: #1}}
    {}{\InnerPatchFailed}%
  }
{}{\PatchFailed}
\usepackage{lineno}

\usepackage{csquotes}
\usepackage{amsmath}
\usepackage[utf8]{inputenc}
\usepackage[T1]{fontenc}
\usepackage[backend=biber,useprefix=true,style=apa,url=false,doi=false,sorting=nyt,eprint=false]{biblatex}
\DeclareLanguageMapping{american}{american-apa}
\usepackage{fancyvrb}
\usepackage{newfloat}
\usepackage{graphicx}
\usepackage{caption}
\usepackage{listings}
\usepackage{glossaries}
\makeglossaries
\DeclareFloatingEnvironment[fileext=los, listname=List of Schemes, name=Listing, placement=!htbp, within=section]{listing}
\usepackage{placeins}
\usepackage{todonotes}
\ifXeTeX
  \usepackage{polyglossia}
  \setmainlanguage[]{english}
\else
  \usepackage[main=english]{babel}

\def\languageshorthands#1{}
\fi
\ifLuaTeX
  \usepackage{selnolig}  
\fi

\shorttitle{Data aggregation}

\affiliation{\vspace{0.5cm}\textsuperscript{1} HMU Health and Medical University, Potsdam, Germany\\\textsuperscript{2} Tilburg University, Tilburg, Netherlands\\\textsuperscript{3} University of Potsdam, Potsdam, Germany}

\note{

Author Note \newline Correspondence concerning this article should be addressed to Daniel J. Schad, HMU Health and Medical University, Potsdam, Germany. E-mail: \href{mailto:danieljschad@gmail.com}{\nolinkurl{danieljschad@gmail.com}}. A preprint of this manuscript has been published on arXiv (\href{https://arxiv.org/abs/2203.02361}{https://arxiv.org/abs/2203.02361}). Reproducible code is available from \url{https://osf.io/mjf47/}. This study was not preregistered.
Acknowledgements: This work was partly funded by the Deutsche Forschungsgemeinschaft (DFG, German Research Foundation), Sonderforschungsbereich 1287, project number 317633480 (Limits of Variability in Language). We thank Paul-Christian Bürkner and Michael Betancourt for discussion on the formulation of SBC for Bayes factors.

}

\abstract{%
Bayesian linear mixed-effects models and Bayesian ANOVA are increasingly being used in the cognitive sciences to perform null hypothesis tests, where a null hypothesis that an effect is zero is compared with an alternative hypothesis that the effect exists and is different from zero. While software tools for Bayes factor null hypothesis tests are easily accessible, how to specify the data and the model correctly is often not clear. In Bayesian approaches, many authors use data aggregation at the by-subject level and estimate Bayes factors on aggregated data. Here, we use simulation-based calibration for model inference applied to several example experimental designs to demonstrate that, as with frequentist analysis, such null hypothesis tests on aggregated data can be problematic in Bayesian analysis. Specifically, when random slope variances differ (i.e., violated sphericity assumption), Bayes factors are too conservative for contrasts where the variance is small and they are too liberal for contrasts where the variance is large. Running Bayesian ANOVA on aggregated data can - if the sphericity assumption is violated - likewise lead to biased Bayes factor results. Moreover, Bayes factors for by-subject aggregated data are biased (too liberal) when random item slope variance is present but ignored in the analysis. These problems can be circumvented or reduced by running Bayesian linear mixed-effects models on non-aggregated data such as on individual trials, and by explicitly modeling the full random effects structure.
Reproducible code is available from \url{https://osf.io/mjf47/}.
}

\begin{document}
\maketitle

\hypertarget{introduction}{%
\section{Introduction}\label{introduction}}

Bayesian linear mixed-effects models (LMMs, also known as hierarchical or multi-level models) and Bayesian ANOVA are increasingly being used in the cognitive sciences to analyze repeated measures data (Heck et al., 2022). Bayes factors can be used to quantify evidence for an alternative hypothesis (with specific assumptions about the distribution of possible effect sizes detailed in a parameter prior) relative to a null hypothesis that a parameter is zero. Bayes factor null hypothesis tests arguably provide a better alternative to frequentist p-values (Jeffreys, 1939; Kass \& Raftery, 1995; Oberauer, 2022; Rouder, Haaf, \& Vandekerckhove, 2018; Schad, Nicenboim, Bürkner, Betancourt, \& Vasishth, 2022; Tendeiro \& Kiers, 2019, 2021; van Doorn, Aust, Haaf, Stefan, \& Wagenmakers, 2021; van Ravenzwaaij \& Wagenmakers, 2021; Wagenmakers, Lodewyckx, Kuriyal, \& Grasman, 2010). In recent years, software has been developed that allows easy access to Bayesian hypothesis testing for lay users, such as the Bayesian analysis software WinBUGS (Lunn, Thomas, Best, \& Spiegelhalter, 2000), JAGS (Plummer, 2003), PyMC3 (Salvatier, Wiecki, \& Fonnesbeck, 2016), Stan (Carpenter et al., 2017), Turing (Ge, Xu, \& Ghahramani, 2018), and others. Moreover, easy-to-use tools have been developed to provide access to complex Bayesian analyses for lay users. This includes the R-package \texttt{brms} (Bürkner, 2017, 2018), which provides an interface from R to Stan in combination with the R-package \texttt{bridgesampling} (Gronau, Singmann, \& Wagenmakers, 2020), the R-package \texttt{BayesFactor} (Morey \& Rouder, 2018), which implements several tools for (default) Bayes factor computations, \texttt{rstanarm} (Goodrich, Gabry, Ali, \& Brilleman, 2020), and the GUI-based software JASP (JASP Team, 2022), which is designed as a (frequentist and Bayesian) alternative to SPSS. In the present work, we use the R-packages \texttt{brms} (Bürkner, 2017, 2018) with \texttt{bridgesampling} (Gronau et al., 2020) as well as \texttt{BayesFactor} (Morey \& Rouder, 2018). Importantly, how to use these tools responsibly, e.g., how to represent the data and models correctly, is -- as it should be -- the responsibility of the user.

When data are clustered by subjects and repeated measures are available per subject (e.g., for different trials/items), one widely used approach in data analysis is to aggregate data to the by-subject level, such that one averaged data point is computed for each subject and each experimental condition. For example, this is standard practice in repeated measures ANOVA. For frequentist tools, it has been demonstrated (Box, 1954; Clark, 1973) that such aggregation can lead to biases in hypothesis tests in two situations that we explain below: (i) when the \emph{sphericity} assumption is violated; and (ii) when item variance is present in addition to subject variance.

First, the sphericity assumption is important in data aggregation, since sphericity is assumed by classical repeated measures ANOVA. To illustrate the concept using an example: let's assume a design with one repeated-measures factor with three levels F1, F2, and F3. In this setting, it's possible to compute for each subject the difference in the dependent variable between two factor levels, e.g., F1-F2, but also F1-F3 and F2-F3. That is, there are three difference variables that can be computed per subject. For each of these difference variables, we can compute the variance across subjects. The sphericity assumption means that these three variances are assumed to be all equal. If there are (true) differences in the variances, the sphericity assumption is violated. In classical repeated measures ANOVA, aggregating data to the by-subject level conflates residual noise with individual differences in an effect, since individual differences in an effect are used as the error term. A violation of the sphericity assumption is known to lead to biased p-values, such that p-values become anti-conservative (Box, 1954). According to standard frequentist procedures for repeated measures ANOVA, violation of sphericity is tested using the Mauchly test of sphericity (Mauchly, 1940), and if violated, corrections to (i.e., reductions of) the degrees of freedom are applied, such as Greenhouse Geisser correction (Greenhouse \& Geisser, 1959) or Huynh-Feldt correction (Huynh \& Feldt, 1976). Indeed, in our experience, empirically, violations of the sphericity assumption occur quite frequently in the cognitive sciences (e.g., Greenhouse \& Geisser, 1959, have been cited 5745 times in google-scholar in January 2023, and 3390 of these citations contain the word {``cognitive''}).

The result from an LMM on aggregated fully balanced data is the same as from an ANOVA. Thus, unsurprisingly (Magezi, 2015), the same problem as in repeated measures ANOVA also occurs when fitting frequentist LMMs to aggregated data, namely biased p-values when the sphericity assumption is violated (we show this using simulations in Appendix \ref{app:freqSphericity}). The big advantage with the LMM is that the remedy is easy: estimate all random effects terms by analyzing data in its unaggregated form.

A second factor that can lead to biased hypothesis tests when aggregating data is the presence of a second random factor, e.g., when item variance is present in addition to subject variance. Items play a key role in many psychological and cognitive studies. Some examples of items in cognitive research include stories, sentences, words, pictures, or actors. It is well known for frequentist repeated measures ANOVA that data aggregation (per subject and condition) in the presence of item variance leads to an inflation of type I (alpha) error (Clark, 1973; Forster \& Dickinson, 1976). Despite these concerns, some authors still recommend data aggregation (Raaijmakers, Schrijnemakers, \& Gremmen, 1999) even if it is not necessary in LMMs. (See Appendix~\ref{app:FreqItem} for simulations showing that this is indeed problematic.)

\hypertarget{implications-of-sphericity-and-item-variance-for-bayes-factors-analyses}{%
\subsection{Implications of sphericity and item variance for Bayes factors analyses}\label{implications-of-sphericity-and-item-variance-for-bayes-factors-analyses}}

Perhaps surprisingly, here we show that the insights gained from frequentist methods apply to Bayesian methods as well; specifically, Bayes factors end up being biased just as p-values are if sphericity is violated or if item variance is present. This point has gone unappreciated: many authors aggregate data at the by-subject level for Bayes factor analyses (Oberauer, 2022; van Doorn et al., 2021), or discuss the advantages of such an approach. Indeed, such data aggregation has the advantage that Bayesian models require less time for model fitting and for Bayes factor estimation.

One limitation of prior research on Bayes factors has been that it was often unclear whether a Bayes factor estimate from a complex LMM or from Bayesian ANOVA is accurate, i.e., whether it corresponds to the true Bayes factor, making investigation of potential biases associated with data aggregation difficult. To investigate this issue, we have used simulation-based calibration (SBC) for Bayes factors (Schad et al., 2022) (concerning SBC for parameter posteriors, see Cook, Gelman, \& Rubin, 2006; Schad, Betancourt, \& Vasishth, 2021; Talts, Betancourt, Simpson, Vehtari, \& Gelman, 2018). SBC allows us to test the accuracy of Bayes factor estimates, i.e., whether the estimates are correct and correspond to the true value. Here, we use these new developments and apply them to the question of data aggregation in Bayesian analyses.

Before explaining the details of the performed analyses, we summarize the performed simulation analyses with their results. We first study violations of the sphericity assumption (Issue 1). For this, we use a range of simulations to demonstrate that when the sphericity assumption is violated, Bayes factors estimated from aggregated data are biased, whereas this bias can be reduced by running analyses on non-aggregated data. We demonstrate this for the \texttt{brms} package with LMM analyses, where we compute Bayes factors for individual contrasts, for repeated-measures designs using (a) two designs with a factor with three-levels, (b) one design with a factor with four levels, and (c) one 2 \(\times\) 2 design. Next, we study omnibus Bayesian ANOVA Bayes factors. Using the \texttt{brms} package to perform ANOVA-type comparisons, we find that a violation of the sphericity assumption does not exhibit biased Bayes factor estimates even when estimated from aggregated data, at least in the two studied example designs (which are taken from a + b above). This may reflect some cancelling-off of different conservative versus liberal biases, which we think cannot be taken for granted to yield accurate results in general.

In an Appendix, we moreover look at the \texttt{BayesFactor} package, which assumes sphericity for different contrasts coding one factor. In our example analyses, we find that Bayes factors in the \texttt{BayesFactor} package exhibit a conservative bias. While this bias is stronger for aggregated than non-aggregated data, it does not disappear for non-aggregated data. In the simulations, the Bayes factor is only accurate for a simulated data set where the sphericity assumption is not violated. We also use the \texttt{BayesFactor} package to investigate a 2 \(\times\) 2 design, where we find that Bayes factors from aggregated data are biased when the sphericity assumption is violated. This bias is reduced for non-aggregated analyses. Again, Bayes factor estimates from non-aggregated data are accurate in simulated data sets where the sphericity assumption is not violated.

The second issue we study is data aggregation in the presence of item variability. We use a range of simulations to demonstrate that data aggregation in this context leads to liberal Bayes factor estimates, that this bias increases with larger item variability, and that this bias is avoided or reduced when fitting Bayesian LMMs to non-aggregated data and estimating the maximal random effects structure. We demonstrate that these results hold across a range of experimental designs and software packages (\texttt{brms}, and in the Appendix: \texttt{BayesFactor}). Specifically, these conclusions are supported by analyses of repeated-measures designs with crossed random effects for subjects and items, including (a) a 2-level factor design (for the \texttt{brms} and the \texttt{BayesFactor} packages), (b) a 4-level factor design (\texttt{BayesFactor} package), and (c) a 2 \(\times\) 2 design (\texttt{brms} package). However, analyses of (c) a 2 \(\times\) 2 design (\texttt{brms} package) also yielded biased Bayes factors for the non-aggregated analyses, suggesting possible limits of the bridge sampling approach to Bayes factor estimation.

Next, we provide a brief introduction to Bayesian data analysis and simulation-based calibration.

\hypertarget{brief-introduction-to-bayesian-analyses-and-simulation-based-calibration}{%
\section{Brief introduction to Bayesian analyses and simulation-based calibration}\label{brief-introduction-to-bayesian-analyses-and-simulation-based-calibration}}

Before performing the analyses, we provide a brief introduction to Bayesian inference and Bayes factor null hypothesis testing. This introduction is derived from Schad et al. (2022). Many other introductory treatments are available (Chow \& Hoijtink, 2017; Etz, Gronau, Dablander, Edelsbrunner, \& Baribault, 2018; Etz \& Vandekerckhove, 2018; Hoijtink \& Chow, 2017; Lee, 2011; Mulder \& Wagenmakers, 2016; Nicenboim \& Vasishth, 2016; van Doorn et al., 2021; Vandekerckhove, Rouder, \& Kruschke, 2018; Vasishth, Nicenboim, Beckman, Li, \& Kong, 2018).

\hypertarget{a-quick-review-of-bayesian-methodology}{%
\subsection{A quick review of Bayesian methodology}\label{a-quick-review-of-bayesian-methodology}}

Statistical analyses in the cognitive sciences often pursue two goals: to estimate parameters and their uncertainty, and to test hypotheses. The present work focusses on inference on hypotheses. Both of these goals can be achieved using Bayesian data analysis. Bayesian analyses focus on an ``observational'' model \(\mathcal{M}\), which specifies the probability density of the data \(y\) given the vector of model parameters \(\Theta\) and the model \(\mathcal{M}\), i.e., \(p(y \mid \Theta, \mathcal{M})\), or by dropping the model, \(p(y \mid \Theta)\). It is possible to use the observational model to simulate data, by selecting some model parameters \(\Theta\) and drawing random samples for the data \(\tilde{y}\). When the data is given (fixed, e.g., observed or simulated), then the observational model turns into a likelihood function: \(p(y \mid \Theta) = L_y(\Theta)\); this can be used to estimate model parameters or to compute the evidence for the model relative to other models.

\paragraph{Inference over hypotheses}

Bayes factors provide a way to compare any two model hypotheses (i.e., arbitrary hypotheses) against each other by comparing their marginal likelihoods (Betancourt, 2018; Kass \& Raftery, 1995; Ly, Verhagen, \& Wagenmakers, 2016; Schad et al., 2022). The Bayes factor tells us, given the data and the model priors, how much we need to update our relative belief between the two models.

To derive Bayes factors, we first compute the model posterior, i.e., the posterior probability for a model \(\mathcal{M}_i\) given the data:

\begin{equation}
p(\mathcal{M}_i \mid y) = \frac{p(y \mid \mathcal{M}_i) \times p(\mathcal{M}_i)}{p(y \mid \mathcal{M}_1) \times p(\mathcal{M}_1) + p(y \mid \mathcal{M}_2) \times p(\mathcal{M}_2)}= \frac{p(y \mid \mathcal{M}_i) \times p(\mathcal{M}_i)}{p(y)} .
\end{equation}

Here, \(p(\mathcal{M}_i)\) is the prior probability for each model \(i\). Based on the posterior model probability \(p(\mathcal{M}_i \mid y)\), we can compute the model odds for one model over another as:

\begin{equation}
\frac{p(\mathcal{M}_1 \mid y)}{p(\mathcal{M}_2 \mid y)} = \frac{[p(y \mid \mathcal{M}_1) \times p(\mathcal{M}_1)] / p(y)}{[p(y \mid \mathcal{M}_2) \times p(\mathcal{M}_2)] / p(y)} = \frac{p(y \mid \mathcal{M}_1)}{p(y \mid \mathcal{M}_2)} \times \frac{p(\mathcal{M}_1)}{p(\mathcal{M}_2)} \label{eq:PostRatio}
\end{equation}

In words:

\begin{equation}
Posterior\;ratio = Bayes\;factor \times prior\;odds
\end{equation}

The Bayes factor is thus a measure of relative evidence, the comparison of the predictive performance of one model (\(\mathcal{M}_1\)) against another one (\(\mathcal{M}_2\)). This comparison (\(BF_{12}\)) is a ratio of marginal likelihoods:

\begin{equation}
BF_{12} = \frac{p(y \mid \mathcal{M}_1)}{p(y \mid \mathcal{M}_2)}
\end{equation}

\(BF_{12}\) indicates the evidence that the data provide for \(\mathcal{M}_1\) over \(\mathcal{M}_2\); in other words, which of the two models is more likely to have generated the data, or the relative evidence that we have for \(\mathcal{M}_1\) over \(\mathcal{M}_2\). Bayes factor values larger than one indicate that \(\mathcal{M}_1\) is more compatible with the data, values smaller than one indicate \(\mathcal{M}_2\) is more compatible with the data, and values close to one indicate that both models are equally compatible with the data.

In the present work, we will consider the case of nested model comparison, where a null model hypothesizes that a model parameter is zero or absent (a point hypothesis), whereas an alternative model hypothesizes that the model parameter is present with some prior distribution and has some value different from exactly zero (a ``general'' hypothesis).

For most interesting problems and models in cognitive science, Bayes factors cannot be computed analytically; approximations are needed. One major approach is to estimate Bayes factors based on posterior MCMC draws via bridge sampling (Bennett, 1976; Meng \& Wong, 1996), implemented in the R package \texttt{bridgesampling} (Gronau et al., 2020), which we use in the present work. We also use Bayes factors computed by the R package \texttt{BayesFactor} (Morey \& Rouder, 2018). However, in general, we expect that similar results can be obtained using other software packages implementing Bayesian analyses.

\hypertarget{simulation-based-calibration-sbc}{%
\subsection{Simulation-based calibration (SBC)}\label{simulation-based-calibration-sbc}}

We have recently used an SBC-inspired method for Bayes factors (Schad et al., 2022), which is a statistical technique designed to test whether a Bayes factor estimated in a given analysis is accurate, or whether it is biased and deviates from the true Bayes factor. Here, we first provide a short description of SBC for Bayes factors (derived partly from Schad et al., 2022) and then perform SBC to test the accuracy of Bayes factors for analyses of non-/aggregated data in the presence of non-spherical random effects.

\paragraph{Introduction to SBC}

We can formulate an SBC-inspired method for Bayes factors (i.e., for model inference), where \(\mathcal{M}\) is a true model used to simulate artificial data \(y\), and \(\mathcal{M}'\) is a model inferred from the simulated data. SBC for model inference makes use of model priors p(\(\mathcal{M}\)), which are the prior probabilities for each model before observing any data. SBC can then be formulated as follows (Schad et al., 2022).

\begin{equation}
p(\mathcal{M}') = \sum_{\mathcal{M}} \int p(\mathcal{M}' \mid y) p(y \mid \mathcal{M}) p(\mathcal{M}) \, \mathrm{d} y \label{eq:SBC}
\end{equation}

We can read this equation sequentially (from right to left):
first, we sample a model from the model prior, \(p(\mathcal{M})\). Next, we simulate data based on this model, \(p(y \mid \mathcal{M})\). This in fact involves two steps: simulating parameters from the parameter prior, \(p(\Theta \mid \mathcal{M})\), and then simulating artificial data from the parameters and model, \(p(y \mid \Theta, \mathcal{M})\), i.e., \(p(y \mid \mathcal{M}) = \int p(y \mid \Theta, \mathcal{M}) \times p(\Theta \mid \mathcal{M}) \, \mathrm{d} \Theta\). Next, we estimate the posterior model probabilities from the simulated data, \(p(\mathcal{M}' \mid y)\). This again involves several steps: we estimate the model parameters based on the simulated data, \(p(\Theta' \mid y)\), for each of the two models, use this to compute marginal likelihoods, \(p(y \mid \mathcal{M}')\), for both models using bridge sampling, compute Bayes factors, and then compute the posterior probability for each model given the data, \(p(\mathcal{M}' \mid y)\), by adding the model prior.
That is, we obtain a posterior model probability for each simulated data set. We can now compare, whether the posterior model probability, \(p(\mathcal{M}' \mid y)\), -- averaged across all simulated data sets -- is the same as the prior model probability, \(p(\mathcal{M})\).
That is, the average posterior should be exactly the same as the prior.

The key idea is that if the computation of Bayes factors and posterior model probabilities is performed correctly (and of course the data simulation is implemented correctly), then the average posterior probability for a model should be the same as its prior probability. By contrast, if the average posterior probability for a model deviates from its prior probability, then this indicates that the Bayes factor estimate is biased, i.e., that the obtained Bayes factor does not correspond to the true Bayes factor.

For a given run of SBC, to test whether the Bayes factor estimates are biased, we here perform Bayesian t-tests on the posterior model probabilities, i.e., computing null hypothesis Bayes factors to test whether the posterior model probabilities differ from the prior model probability (here either \(0.5\) or \(0.2\)). For these tests, the t-test in the \texttt{BayesFactor} package is a simpler and faster alternative to the \texttt{brms} package, and we therefore use this here to test whether SBC indicates bias in the estimation of Bayes factors. Crucially, the results from null hypothesis Bayes factor analyses inherently depend on the prior for the effect size of the bias. Therefore, for each Bayesian t-test we perform a sensitivity analysis, where we vary the width of the prior (i.e., prior scale of a Cauchy distribution) for the standardized effect size of the expected bias and report the corresponding Bayes factors as a function of the prior scale. The results from these sensitivity analyses are reported in Appendix~\ref{app:SensAnal}.

The question of the present work is whether and how data aggregation can bias estimated Bayes factors. In the following sections, we will investigate two issues with data aggregation, namely (issue 1) violations of the sphericity assumption and (issue 2) by-subject aggregation in the presence of item variance.

\hypertarget{issue-1-bayes-factors-for-aggregated-data-are-biased-if-random-effects-variances-are-unequal-violated-sphericity-assumption}{%
\section{Issue 1: Bayes factors for aggregated data are biased if random effects variances are unequal (violated sphericity assumption)}\label{issue-1-bayes-factors-for-aggregated-data-are-biased-if-random-effects-variances-are-unequal-violated-sphericity-assumption}}

LMMs do not require aggregated data. However, it is possible to apply them to aggregated data. For such aggregated data, it is not possible to estimate all random effect variances for the different effects (contrasts). Instead, only some of the random effects are informed by the data (Bergh, Wagenmakers, \& Aust, 2022). E.g., in a one-factor design, only random intercepts are informed, but not random slopes; in a two-factor design, only random slopes for the main effects are informed by the data, but not random slopes for the interaction. The reason for this is that the aggregation collapses the standard deviation associated with random slopes into the residual standard deviation. The important point here is that this collapsing is due to the data aggregation, and does not only apply to repeated measures ANOVA, but also to LMMs fit on aggregated data. Therefore, LMMs fit to aggregated data also assume sphericity, i.e., they assume that at least some of the random slope variances (which cannot be estimated in the aggregated data) are the same.

At the same time, Bayesian omnibus ANOVA-type comparisons can also be performed for the non-aggregated data. Specifically, it is possible to use Bayesian model comparison for LMMs to compare for example (a) a model where a 4-level factor is modeled via three contrasts to (b) a model where this factor is not captured, i.e., the three contrasts are removed. The Bayes factor from this model comparison provides a Bayesian omnibus ANOVA-style test.

Here, we show that if the sphericity assumption is violated, Bayes factors based on LMMs and ANOVA for aggregated data can be biased, whereas LMMs/ANOVA for non-aggregated data are accurate or at least less biased when all random effects are included in the model.

\hypertarget{bayesian-lmms-contrast-based-analyses}{%
\subsection{Bayesian LMMs: contrast-based analyses}\label{bayesian-lmms-contrast-based-analyses}}

\hypertarget{demonstration-via-simulations-an-exemplary-data-set}{%
\subsubsection{Demonstration via simulations: an exemplary data set}\label{demonstration-via-simulations-an-exemplary-data-set}}

First, we illustrate a Bayesian LMM analysis using an exemplary simulated data set. The \texttt{designr} package (Rabe, Kliegl, \& Schad, 2020) allows us to create an artificial experimental design, which specifies a fixed factor \texttt{X} with three levels \texttt{X1} to \texttt{X3}, a random factor with \(20\) subjects, and \(10\) observations for each subject per condition, yielding a total of \(600\) data points.

The \texttt{hypr} package (Rabe, Vasishth, Hohenstein, Kliegl, \& Schad, 2020; Schad, Vasishth, Hohenstein, \& Kliegl, 2020) can then be used to specify a treatment contrast for the 3-level factor \texttt{X}, where experimental conditions \texttt{X2} and \texttt{X3} are each compared to the control condition \texttt{X1}, and where the intercept is the grand mean.\footnote{The corresponding hypr-command is: \texttt{hypr(X1\textasciitilde X2, X1\textasciitilde X3)}.} That the intercept is coded as the grand mean is non-standard for treatment contrasts (by default the intercept assesses the mean response in the baseline condition), but advantageous in LMMs, since the random intercept variance then assesses the variance in mean responses across subjects, which is usually of greater interest than the variance in the baseline condition. The contrasts are coded as numeric covariates in R.

The \texttt{designr} package then allows us to simulate artificial data from an LMM using the \texttt{simLMM} function. For the fixed effects (see Table \ref{table:coefficients}), we assume an intercept (grand mean) of \(200\) and treatment effects of \(20\) for the comparison of each experimental condition with the baseline/control condition. For the subject random effects, we assume a standard deviation of \(20\) for the intercept. Critically, for the variances of random slopes, we assume that the effect (first contrast) comparing the control to the first experimental group (\texttt{X2}, we will refer to this as \texttt{c2vs1} below) has a large standard deviation of \(90\) across subjects, whereas the effect (second contrast) comparing the control group to the second experimental group (\texttt{X3}, we will refer to this as \texttt{c3vs1}) has a small standard deviation of \(10\) across subjects. For simplicity, correlations between random effects are set to \(0\); the standard deviation of the normally distributed residual noise is set to \(50\).

\begin{table}[htbp]
\begin{center}
\begin{tabular}{l c}
\hline
 & Simulation parameters \\
\hline
(Intercept)           & $200$ \\
c2vs1                 & $20$  \\
c3vs1                 & $20$  \\
\hline
Num. obs.             & $600$ \\
Num. groups: subj     & $20$  \\
\hline
SD: subj (Intercept)  & $20$  \\
SD: subj.1 c2vs1      & $90$  \\
SD: subj.2 c3vs1      & $10$  \\
SD: Residual          & $50$  \\
\hline
\end{tabular}
\caption{Parameters used to simulate data from an LMM.}
\label{table:coefficients}
\end{center}
\end{table}

In the exemplary simulation, we ensure that when fitting a (frequentist) LMM to the simulated data, the fixed effects estimated by the model are exactly the true underlying values. This is implemented by setting the argument \texttt{empirical} in the \texttt{simLMM} function (\texttt{designr}) to \texttt{TRUE}. The random effects estimates, however, are not exact.

In Appendix~\ref{app:freqSphericity}, we provide results from frequentist analyses using LMMs, that illustrate the biases in frequentist test statistics that arise in aggregated analyses when the sphericity assumption is violated.

\hypertarget{an-exemplary-analysis-using-bayesian-lmms}{%
\subsubsection{An exemplary analysis using Bayesian LMMs}\label{an-exemplary-analysis-using-bayesian-lmms}}

As discussed above, when the sphericity assumption is violated, fitting LMMs on aggregated data yields biased results.
Here, we perform Bayesian LMM analyses with null hypothesis Bayes factor testing to illustrate such a bias for the exemplary data set.
For the Bayesian modeling, we assume the following priors of the model parameters (equations see below):

\begin{align}
\beta_{(Intercept)}       &\sim Normal(200, 20) \\
\beta_{Contrasts}         &\sim Normal(0, 20) \\
\sigma_{Random \, slopes} &\sim Normal_+(0, 50) \\
\sigma_{Residual}         &\sim Normal_+(0, 20) \\
\rho_{Random \, slopes}   &\sim LKJ(2)
\end{align}

Contrary to frequentist approaches, the priors allow us to estimate random effects correlations in Bayesian models even with limited data, and we therefore include random effects correlations in our Bayesian modeling.

To implement null hypothesis Bayes factor analyses in the present example analysis, one (full) Bayesian LMM implements the alternative hypothesis (H1) by using both contrasts (\texttt{c2vs1} and \texttt{c3vs1}) as fixed effects. Two additional Bayesian LMMs each capture one null hypothesis (H0), where in each Bayesian null model one of the two contrasts is excluded from the set of fixed effects, effectively setting its value to zero.

These Bayes factor analyses are performed for the non-aggregated data and for the aggregated data. For the random effects, we assume random intercepts and slopes as well as their correlations for the non-aggregated data, and for the aggregated data we fit only random intercepts. Random effect specification is unchanged for the Bayesian null models.

The results from these analyses show that when using aggregated data, then the null hypothesis Bayes factors for both contrasts are more or less the same (small differences originate from the sampling process): the null hypothesis Bayes factor for the fixed effect with a large standard deviation of the random slope (\texttt{c2vs1}) has a value of BF10 = 1/1.15, and the null hypothesis Bayes factor for the fixed effect with a small standard deviation of the random slope (\texttt{c3vs1}) has a value of BF10 = 1/1.14 .
Since both Bayes factors are close to 1 they are undecisive.
This result exactly mirrors the frequentist result (see Appendix~\ref{app:freqSphericity}): the evidence for an effect does not depend on its random slope variance, since the obtained Bayesian LMM (based on aggregated data) cannot capture these differences in random slopes.

By contrast, when analyzing non-aggregated data using Bayesian LMMs, the null hypothesis Bayes factors for the two effects are very different from each other (as they should be): the null hypothesis Bayes factor for the contrast with a large standard deviation of the random slope (\texttt{c2vs1}) has a value of BF10 = 0.93 (BF01 = 1.08), and thus does not provide much evidence for either H0 or H1. This is expected since the random slope variance for this effect was very large, such that the small effect size is not supported by the data. By contrast, the null hypothesis Bayes factor for the fixed effect with a small standard deviation of the random slope (\texttt{c3vs1}) has a value of BF10 = 115.80, indicating strong evidence for the alternative hypothesis (H1). Given that the effect varies only very little across subjects (i.e., small random slope variance), even a small simulated effect of \(20\) provides strong evidence for the alternative hypothesis in this limited data set. Thus, when non-aggregated data are used, the differences in random slope variances can be captured and modeled via the random effects, and used to compute correct null hypothesis Bayes factors.

One limitation of this exemplary analysis, is that it does not provide a formal test of whether or to what extend the obtained Bayes factors are accurate or biased. Simulation-based calibration (SBC) can be used to investigate this issue.

\paragraph{Running SBC for models fit to non-/aggregated data}

Applying SBC to our current example of null-hypothesis testing, we define a prior on the model space, e.g., we can define the prior probabilities for a null and an alternative model, specifying how likely each model is a priori. Here, we specify that a priori the alternative and the null hypotheses are both equally likely with a probability of 50\% each.
From these priors, we can randomly draw one hypothesis (model), here, \(n_{sim} = 250\) times. Thus, in each of \(250\) draws we randomly choose one model (either H0 or H1), with the probabilities given by the model priors. For each draw, we first sample model parameters from their prior distributions. For this, we use the same priors as in the exemplary analysis shown above. As the only exception, we set the prior for the standard deviation of the random slopes to 150 (instead of 50; for results with a standard deviation of 50 see Appendix~\ref{app:SBCspher1}, which show that an overly small prior standard deviation of 50 may produce biases; for an explanation of this, see the Appendix). In the simulation process, we always set the standard deviations for the random slopes to values of \(10\) and \(90\) for the two contrasts, where as above, each (treatment) contrast tests the difference between one experimental condition to the baseline condition, and the intercept assesses the grand mean.

Then, we use the sampled model parameters to simulate artificial data. For each simulated artificial data set, we then compute marginal likelihoods and Bayes factors (between the models H1 and H0) using bridge sampling, and we then compute the posterior probabilities for each hypothesis using the true prior model probabilities (i.e., how likely each model is a posteriori). As the last and critical step in SBC, we then compare the posterior model probabilities (averaged across all \(250\) simulations) to the prior model probabilities of 50\%.

A key result in SBC is that if the computation of marginal likelihoods and model posteriors is performed accurately, and does not exhibit bias, then the data-averaged posterior model probabilities should be the same as the prior model probabilities.

\begin{figure}

{\centering \includegraphics{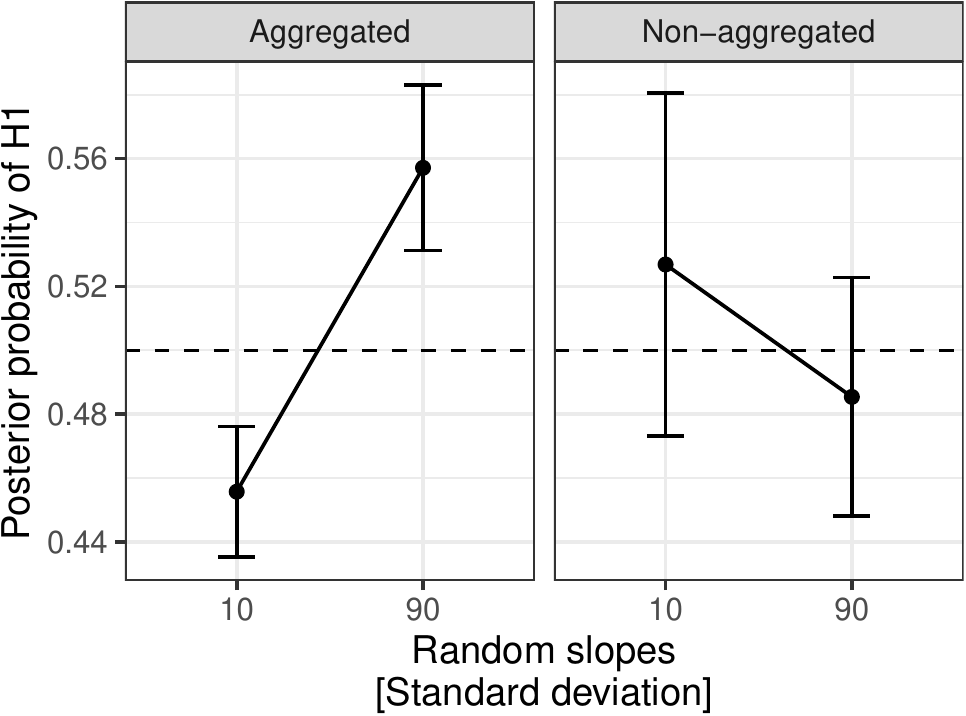} 

}

\caption{Simulation 1.1: Results from SBC for model inference when sphericity is violated. The average posterior model probability together with 95 percent confidence intervals is shown for effects with a small (10) versus a large (90) standard deviation of the random slopes, reflecting small or large variation of the effect across subjects. The horizontal broken line is the prior probability for the H1, and deviations from this line indicate estimation bias. Results are shown from null hypothesis Bayes factor analyses based on aggregated (left panel) versus non-aggregated (right panel) data. They show that aggregating data for null hypothesis tests can lead to biased Bayes factors, which deviate from the true Bayes factor. Bayes factors are more accurate for non-aggregated analyses.}\label{fig:sphericitySBC}
\end{figure}

The results (see Fig.~\ref{fig:sphericitySBC} and Fig.~\ref{fig:sphericitySBCBF}) show that for the aggregated data, the posterior model probabilities differ from the prior probability of \(0.5\). For the contrast with a large standard deviation of the random slopes (i.e., \(90\)), the average posterior model probability is clearly \emph{larger} than the prior value of \(0.5\) (Bayesian t-test using the R-package \texttt{BayesFactor} with default priors: \(BF_{10} =\) 587), suggesting that the posterior model probability may provide too much evidence for H1, and that the corresponding null hypothesis Bayes factor may accordingly be too large (larger than the true Bayes factor). Moreover, for the contrast with a small standard deviation of the random slopes (i.e., \(10\)), the average posterior model probability is clearly \emph{smaller} than the prior value of \(0.5\) (\(BF_{10} =\) 424), demonstrating that the posterior model probability provides too much evidence for H0, and that the corresponding null hypothesis Bayes factor is accordingly too small (smaller than the true Bayes factor). Indeed, the posterior model probabilities strongly differ between the two effects/contrasts: \(BF_{10} =\) 207260087. These results clearly show that when the random slope variances differ between contrasts (i.e., the sphericity assumption is violated), then using aggregated data to perform null hypothesis Bayes factors testing leads to biased Bayes factors, mirroring the results from frequentist analyses of empirical type I (\(\alpha\)) errors.

By contrast, the results for non-aggregated analyses (Fig.~\ref{fig:sphericitySBC}, right panel) with full random effect structure show that the posterior model probability does not seem to differ between the effect with a large (\(90\)) versus a small (\(10\)) standard deviation of the random slopes (\(BF_{01} =\) 5.30), and that both do not seem to differ from the prior value of \(0.5\) (standard deviation of \(10\): \(BF_{01} =\) 9.70; standard deviation of \(90\): \(BF_{01} =\) 8.10). Moreover, the evidence suggests that on average these two posterior model probabilities do not differ from the prior value of \(0.5\) (\(BF_{01} =\) 12.40). Indeed, the difference in posterior model probabilities between large versus small random effect variances seemed to be larger for aggregated data than for non-aggregated data (\(BF_{10} =\) 8640).

Next, we tested whether our results are specific to our first simulated data set, or whether they also hold for other experimental data from a similar experimental design, and with prior distributions informed by a concrete empirical data set. To this end, we used a data set on Pavlovian-instrumental transfer (PIT, Schad et al., 2020) (see below for a description of PIT). We again used a repeated-measures design with three factor levels, but now prior distributions were directly informed by a concrete empirical data set. The SBC results (see Appendix~\ref{app:SBCPIT3}) were the same as in our first set of simulations: when the sphericity assumption was violated, null hypothesis Bayes factors computed from aggregated analyses were biased, whereas these biases could be avoided by running Bayesian LMMs on the non-aggregated data.

\paragraph{SBC simulations for Pavlovian-instrumental transfer with five factor levels}

We next aimed to investigate a situation with more than three within-subject factor levels. For this, we again studied Pavlovian-instrumental transfer (PIT). The basic finding in PIT studies is that subjects perform more button presses (to instrumentally approach a stimulus) when appetitive Pavlovian conditioned stimuli (CSs; i.e., stimuli previously paired with appetitive outcomes such as the win of money) are presented in the background, and that subjects perform less button presses when aversive CSs (e.g., created by pairing with loss of money) are presented in the background. The experimental PIT design we investigated varied CS value parametrically with values including \(-2, -1, 0, +1, +2\) Euros, yielding five levels of a repeated measures factor. We performed SBC for Bayes factors on this parametric experimental design. We coded the five-level factor with a treatment contrast, which compares each factor level to the baseline of \(0\) Euros.

We used effect size estimates based on data from Schad et al. (2020) (first \(30\) subjects) to inform prior distribution. The priors were:\footnote{These were the same priors as in our previous PIT analysis with three factor levels, reported in the Appendix, with the only difference that the prior standard deviation for the residual variance had a smaller standard deviation of $5$ instead of $10$ ($\sigma_{Residual} \sim Normal_+(0, 5)$). This has the advantage that a smaller residual noise term leads to a higher sensitivity of the analysis.}

\begin{align}
\beta_{(Intercept)}       &\sim Normal(5, 5) \\
\beta_{Contrasts}         &\sim Normal(0, 2) \\
\sigma_{Random \, slopes} &\sim Normal_+(0, 10) \\
\sigma_{Residual}         &\sim Normal_+(0, 5) \\
\rho_{Random \, slopes}   &\sim LKJ(2)
\end{align}

For the simulations, to reduce computing time, we used \(10\) simulated subjects, \(3\) trials per condition, and \(200\) SBC simulation runs.
For the four within-subject contrasts, we assumed different variances across subjects: for the contrast comparing \(+2\) to \(0\) Euros, we assumed a large standard deviation of \(10\), for the two contrasts comparing \(+1\) and \(-1\) to \(0\) Euros, we assumed a relatively small standard deviation of \(2\), and for the contrast comparing \(-2\) to \(0\) Euros, we assumed a very small standard deviation of \(0.1\).

\begin{figure}

{\centering \includegraphics{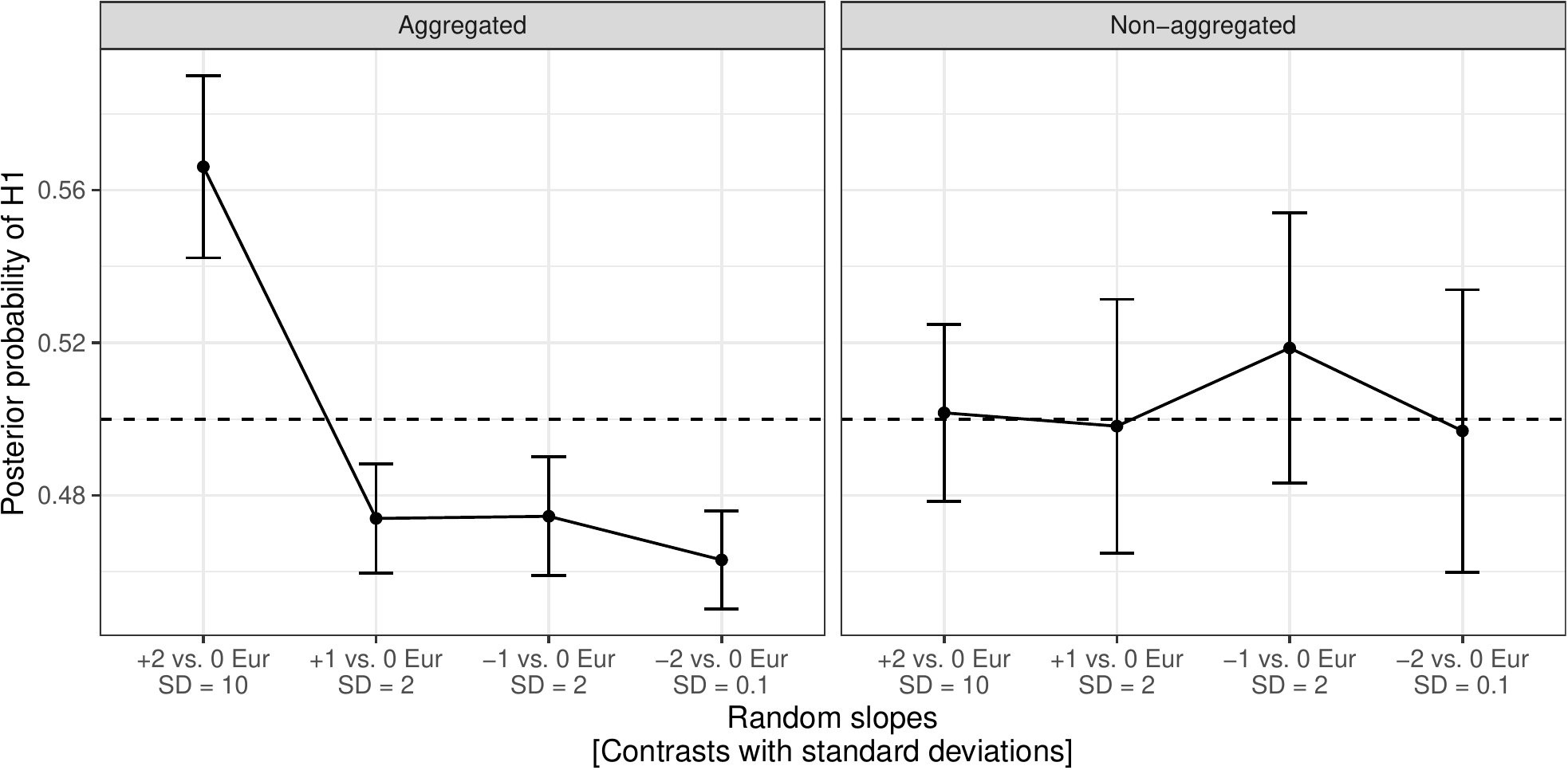} 

}

\caption{Simulation 1.3: A third simulation example, again on Pavlovian-instrumental transfer, now with 5 factor levels: results from SBC for model inference when sphericity is violated. The average posterior model probability together with 95 percent confidence intervals is shown for contrasts comparing +2 to 0 Euros (SD = 10), +1 to 0 Euros (SD = 2), -1 to 0 Euros (SD = 2), and -2 to 0 Euros (SD = 0.1), reflecting differing variations of the effects across subjects. The horizontal broken line is the prior probability for the H1, and deviations from this line indicate estimation bias. Results are shown from null hypothesis Bayes factor analyses based on aggregated data (left panel) and on non-aggregated data (right panel). They show that aggregating data for null hypothesis tests can lead to biased Bayes factors, which deviate from the true Bayes factor. Bayes factors are accurate for non-aggregated analyses.}\label{fig:sphericitySBCpit5}
\end{figure}

The results (see Fig.~\ref{fig:sphericitySBCpit5} and Fig.~\ref{fig:sphericitySBCpit5BF}) show that Bayes factors are biased in the aggregate analyses for all contrasts. For the contrast with a large standard deviation of the random slopes (SD \(=10\); \(+2\) vs.~\(0\) Euros) the average posterior model probability was clearly larger than the prior model probability of \(0.5\), indicating that the Bayes factor was estimated as being too large. For the contrasts with a small standard deviation (SD =\(2\); comparing \(+1\) and \(-1\) vs.~\(0\) Euros) or a very small standard deviation (SD =\(0.1\); comparing \(-2\) vs.~\(0\) Euros) of the random slopes, the average posterior model probability was clearly smaller than the prior value of \(0.5\), indicating that the Bayes factor was estimated as being too small for these contrasts. Again, no biases are visible in the non-aggregated analyses.

\paragraph{$2 \times 2$ design: analyses of the two-step decision-making task}

Next, we wanted to test the influence of violations of the sphericity assumption in a factorial design involving two factors, and we specifically chose a \(2 \times 2\) design for analysis. The simulations are based on the two-step decision task (Daw, Gershman, Seymour, Dayan, \& Dolan, 2011), and the parameters used are inspired by our own data on the task (Schad et al., 2014). Briefly, in the task, subjects perform decisions in two stages. The key theoretical question is whether subjects repeat first-stage choices from one trial to the next (1 = repeated, 0 = not repeated), and whether this is influenced by whether the last trial was rewarded (rewarded = 0.5, not rewarded = -0.5), whether there was a common (0.5) or a rare (-0.5) transition, and whether there is an interaction between reward and transition frequency. A main effect of reward is thought to reflect model-free (i.e., habitual) decisions, an interaction of reward \(\times\) transition is thought to reflect model-based (i.e., goal-directed) decisions, and no main effect of transition is expected. In the present simulations, we assumed that the data consisted of the proportion of repeated first-stage choices across five blocks for each of 10 subjects, recorded separately for rewarded/unrewarded trials, as well as for common/rare transitions. A better analysis of the original two-step task data would involve running a generalized linear mixed model with a logistic link function on the non-aggregated repetition data, but for simplicity, here we stick to proportions, assuming they follow a normal distribution (which in reality they don't). However, the aim of our simulations was not to provide a realistic analysis of the two-step data, but simply to use that design to develop prior distributions in order to use them in a simulation-based calibration analysis.

We assumed the following priors:

\begin{align}
\beta_{(Intercept)}       &\sim Normal(0.7, 0.1) \\
\beta_{Contrasts}         &\sim Normal(0, 0.2) \\
\sigma_{Random \, slopes} &\sim Normal_+(0, 0.2) \\
\sigma_{Residual}         &\sim Normal_+(0, 0.5) \\
\rho_{Random \, slopes}   &\sim LKJ(2)
\end{align}

Prior probabilities for the models were set to 50\% each.

When analyzing a 2 \(\times\) 2 design with aggregated data using LMMs, there are different possible approaches to data analysis. First, we can simply ignore random slopes as we did in the examples with only one factor level - this is indeed the default setting in Bayesian repeated measures ANOVA in the R-package \texttt{BayesFactor}, and it used to be the default approach in JASP (Bergh et al., 2022). However, in a 2 \(\times\) 2 design, there are two data points available for each subject and each factor level. Therefore, theoretically, it is possible to estimate random slopes for the two main effects. (These estimates based on two data points may not be very stable.) The interaction term, however, is confounded with the residual noise again, and random slopes are not fully informed by the data (and cannot be estimated in frequentist LMMs) (Bergh et al., 2022). We expected that although some between-subject variance can be captured in this analysis, two data points are not enough to capture this source of variance fully (i.e., of all random slopes), and we therefore expected that the biases should still exist when the interaction term has large variance. Therefore, we assumed true standard deviations of the random slopes for SD \(= 0\) for the main effects, and of SD \(= 0.4\) for the interaction. We used 200 runs of SBC.

\begin{figure}

{\centering \includegraphics{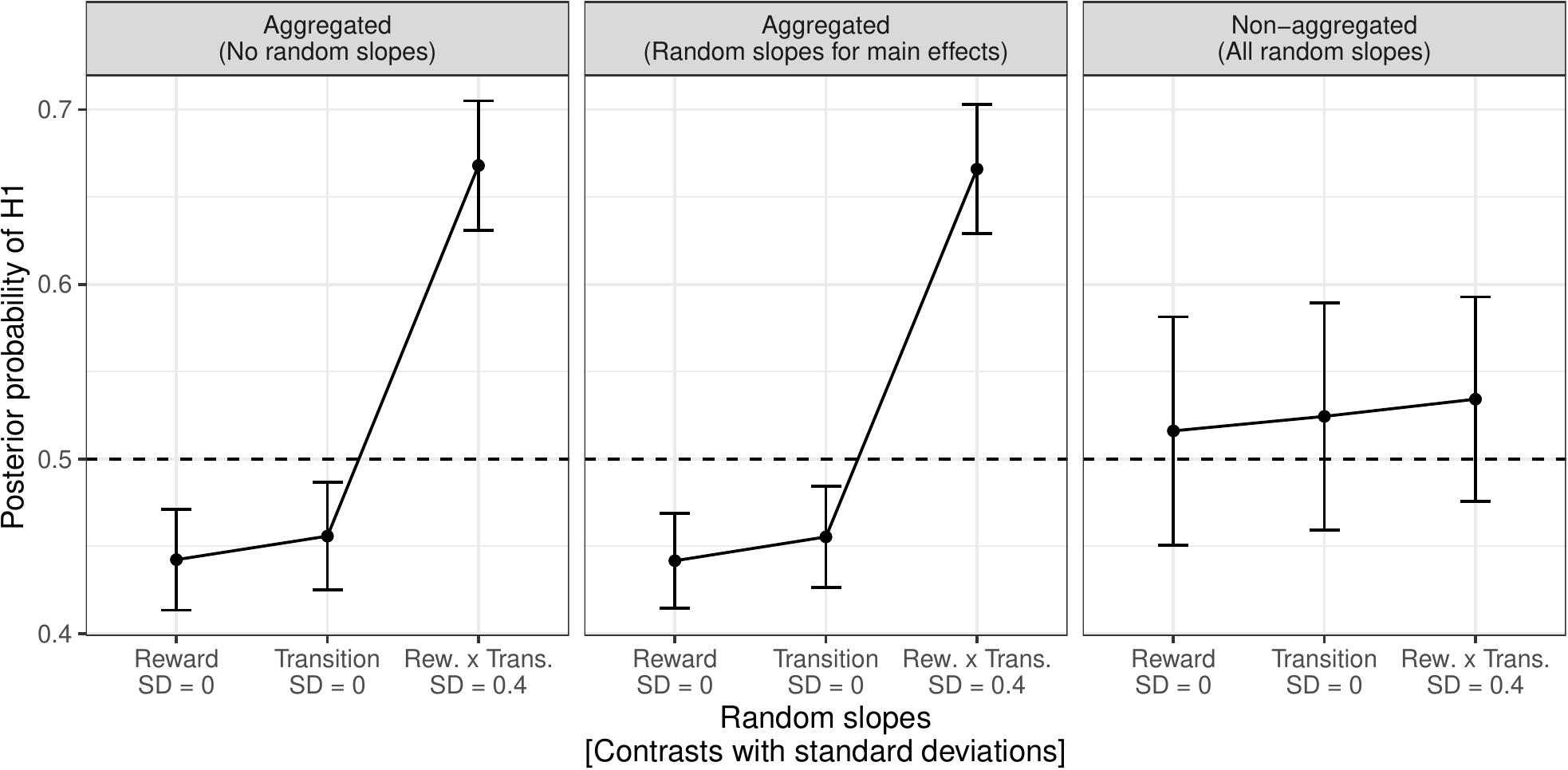} 

}

\caption{Simulation 1.4: A fourth simulation example, employing a 2 x 2 design based on the two-step decision task: results from SBC for model inference when sphericity is violated. The average posterior model probability together with 95 percent confidence intervals is shown for the main effects of reward (last trial was rewarded/not rewarded, reflecting model-free choices; SD = 0) and transition frequency (last trial induced a common/rare transition; SD = 0), and their interaction (reflecting model-based choices; SD = 0.4), reflecting differing variations of the effects across subjects. The horizontal broken line is the prior probability for the H1, and deviations from this line indicate estimation bias. Results are shown from null hypothesis Bayes factor analyses based on aggregated data (left panel), on aggregated data with random slopes for the main effects but not the interaction (middle panel), and for non-aggregated data (right panel). They show that aggregating data for null hypothesis tests can lead to biased Bayes factors, which deviate from the true Bayes factor. This effect is present when no random slopes are estimated and when random slopes for the main effects are estimated. Bayes factors are accurate for non-aggregated analyses.}\label{fig:sphericitySBC2step}
\end{figure}

The results (see Fig.~\ref{fig:sphericitySBC2step} and Fig.~\ref{fig:sphericitySBC2stepBF}) showed that, as before, when the sphericity assumption was violated, Bayes factors computed using aggregated data were biased. Specifically, the two effects with a small true random slope variance (main effects reward and transition; SD \(= 0\)) exhibited a conservative bias, and the effect with a large true random slope variance (the interaction reward \(\times\) transition; SD \(= 0.4\)) exhibited an anti-conservative bias in the Bayes factor. Interestingly, the same biases were present also when random slopes for the main effects were included. Similar biases were absent when estimating Bayes factors using non-aggregated data and estimating all random slope terms. Further simulations (not shown) revealed that when the true random slope variance was mixed for the main effects (SD \(= 0\) and SD \(= 0.4\)) and small for the interaction (SD \(= 0\)), then using aggregated data and estimating random slopes for the main effects was sufficient to avoid biased estimates.

These results further extend and replicate our previous findings to a new experimental design with more than one factor, supporting our main claim that Bayes factors from aggregated analyses can be biased when the sphericity assumption is violated.

\paragraph{Interim summary}

Taken together, these analyses suggest that if random slope variances differ between contrasts, then null hypothesis Bayes factor analyses based on aggregated data may yield biased Bayes factors: for contrasts with a large random slope variance, the Bayes factor may be too large (i.e., liberal) compared to its true value, whereas for contrasts with a small random slope variance, the Bayes factor may be too small (i.e., conservative). This problem can be avoided or at least reduced by analyzing the non-aggregated data and using an appropriate random effects structure (Schad et al., 2022).

Next, we turn to Bayesian ANOVA, i.e., omnibus tests of factors, here with more than two levels. Again, we investigate the situation that the sphericity assumption is violated, and we study whether and how omnibus ANOVA Bayes factors are biased when analyzing aggregated versus non-aggregated data.

\hypertarget{bayesian-anova-testing-the-influence-of-a-factor-with-more-than-two-levels-involving-several-contrasts-when-the-sphericity-assumption-is-violated}{%
\subsection{Bayesian ANOVA: testing the influence of a factor with more than two levels involving several contrasts when the sphericity assumption is violated}\label{bayesian-anova-testing-the-influence-of-a-factor-with-more-than-two-levels-involving-several-contrasts-when-the-sphericity-assumption-is-violated}}

In the analyses above, we tested Bayes factors for individual a priori contrasts. An alternative approach is to perform a test for the influence of a factor, by comparing a model where all contrasts representing that factor are included to a null model, where all contrasts are simultaneously removed. This is the Bayesian equivalent to an omnibus ANOVA F-test. We and others have previously argued (Hays, 1978; Schad et al., 2020) that in most situations, researchers have very specific hypotheses about the comparisons between different experimental conditions. In this situation, a priori contrasts should be tested \emph{instead} of an omnibus ANOVA test. Omnibus ANOVA tests should only be done when no specific hypotheses are available concerning which condition means differ from which other condition means. Thus, generally speaking, we think that ANOVA omnibus tests are often not needed in cognitive science. However, given that many researchers still use ANOVA omnibus tests, here we investigate their properties in the Bayesian context.

Here we investigate, whether the Bayesian ANOVA - when performed on aggregated data - exhibits biases, when the sphericity assumption is violated (as is happening in frequenist ANOVA).
More specifically, we investigated two potential reasons for why violations of the sphericity assumption can lead to biased Bayes factors: (i) data aggregation; (ii) model assumptions (see Appendix~\ref{app:BayesFactor}).

\hypertarget{biased-bayes-factors-due-to-aggregated-data}{%
\subsubsection{Biased Bayes factors due to aggregated data}\label{biased-bayes-factors-due-to-aggregated-data}}

To investigate the first question, i.e., the role of aggregated data, we used the \texttt{brms} package with the same simulations reported above, i.e., those testing a PIT effect. We looked at the simulations testing a three-level factor for PIT (see Appendix~\ref{app:SBCPIT3}), and next, we looked at the simulations testing a five-level factor for PIT. In each analysis, we used the same parameters as above, with the only difference that there was just a single null model in each analysis, which had all the fixed-effects contrasts for the factor removed, providing a Bayes factor test of the null hypothesis that the condition means don't differ from each other.

\begin{figure}

{\centering \includegraphics{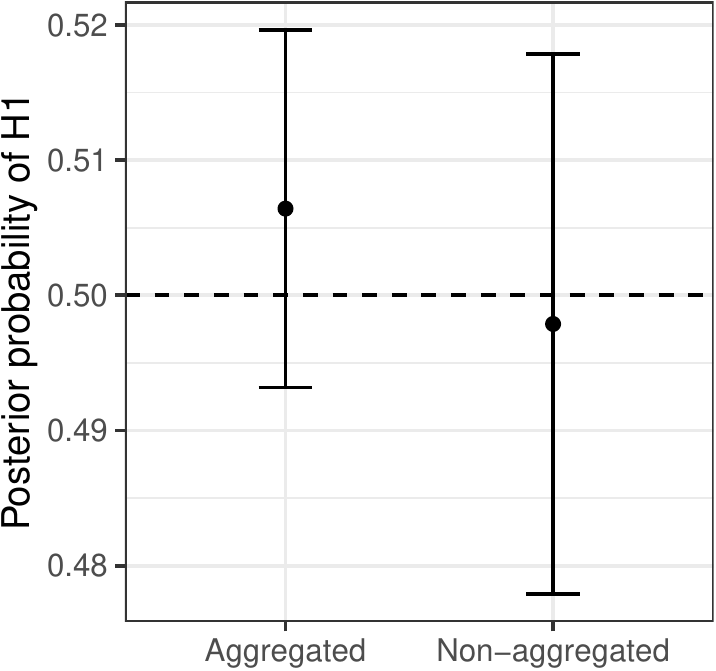} 

}

\caption{Simulation 1.5: Testing evidence for a factor overall, i.e., testing all contrasts at once, by comparing the full model against a null model, where all fixed-effects contrats are removed. Data are from the simulation reported above, i.e., on Pavlovian-instrumental transfer, with three factor leves. Results show that aggregating data for null hypothesis tests does not lead to biased Bayes factors in this example simulation, possibly suggesting that conservative and anti-conservative biases cancel each other out.}\label{fig:sphericitySBCpit3F}
\end{figure}

\begin{figure}

{\centering \includegraphics{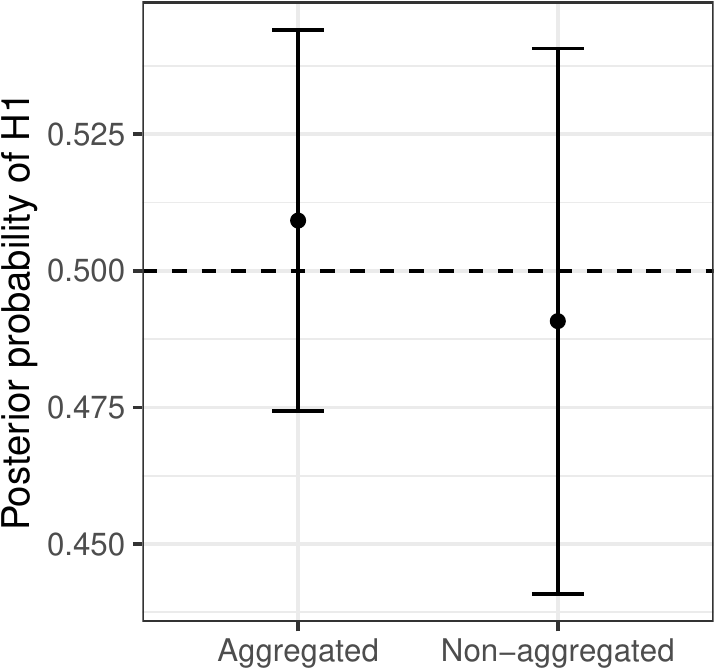} 

}

\caption{Simulation 1.6: Data are from the simulation reported above, i.e., on Pavlovian-instrumental transfer, with five factor levels. Testing evidence for a factor overall, i.e., testing all contrasts at once, by comparing the full model against a null model, where all fixed-effects contrats are removed. Results show that aggregating data for null hypothesis tests does not lead to biased Bayes factors in this example simulation, suggesting that conservative and anti-conservative biases cancel each other out.}\label{fig:sphericitySBCpit5F}
\end{figure}

The results showed that the Bayes factor based on aggregated data was not biased, neither for the three-level PIT factor (Fig.~\ref{fig:sphericitySBCpit3F} and Fig.~\ref{fig:sphericitySBCpit3FBF}), nor for the five-level PIT factor (Fig.~\ref{fig:sphericitySBCpit5F} and Fig.~\ref{fig:sphericitySBCpit5FBF}). This may suggest that the conservative and anti-conservative biases that we observed for a priori contrasts cancel each other out when all contrasts are tested together. This result contrasts with frequentist tests, where a violation of the sphericity assumption is known to lead to an anti-conservative bias. It thus seems that priors and Bayes factors can - at least in the two example analyses shown here - protect against liberal biases.

\hypertarget{biased-bayes-factors-due-to-model-assumptions-bayesian-anova-using-the-bayesfactor-package}{%
\subsubsection{\texorpdfstring{Biased Bayes factors due to model assumptions: Bayesian ANOVA using the \texttt{BayesFactor} package}{Biased Bayes factors due to model assumptions: Bayesian ANOVA using the BayesFactor package}}\label{biased-bayes-factors-due-to-model-assumptions-bayesian-anova-using-the-bayesfactor-package}}

Next, we aimed to further study the consequences of a violated sphericity assumption in the context of Bayesian ANOVA. For this, we made use of the R-package \texttt{BayesFactor} (Morey \& Rouder, 2022; Rouder, Morey, Speckman, \& Province, 2012). Again, we performed SBC by simulating data based on the model underlying the \texttt{BayesFactor} package, which effectively implements a specific type of Bayesian linear mixed-effects model (for details see Appendix~\ref{app:BayesFactor}). An important difference of the \texttt{BayesFactor} package to the \texttt{brms} models discussed above is that it assumes sphericity for all contrasts coding one factor, i.e., it assumes that all these contrasts share the same (prior) standard deviation. Priors were taken as the default priors from the \texttt{BayesFactor} package.

We simulated data from two repeated-measures experiments; for details on simulations and results see Appendix~\ref{app:BayesFactor}. For the first experiment, we again assumed one within-subject factor with three levels and again assumed a violation of sphericity in the simulation. The SBC results showed that aggregating data for null hypothesis tests lead to biased Bayes factors that were too small, reflecting a conservative bias. Bias was reduced when fitting the model to non-aggregated data, but it was still present. This may be the case because - as stated above - Bayesian ANOVA in the \texttt{BayesFactor} package assumes that for all contrasts coding the same factor, random slope variances are the same. Bias was absent when the sphericity assumption was not violated in the simulated data.

In a second simulation experiment (Appendix~\ref{app:BayesFactor}), we analyzed a 2 \(\times\) 2 repeated measures design. As our previous analysis of the two-step task, we fitted models (a) to aggregated data without using random slopes, (b) to aggregated data with random slopes for the main effects, and (c) to non-aggregated data using the full random effects structure. As before, we assumed that the true random slope variances were small for the two main effects (prior scales of \(0.001\)), and were large for the interaction term (prior scale of \(1\)). The results were similar to those from the \texttt{brms} package reported above (see analysis of the two-step task): for the aggregated data, Bayes factor estimates for the main effects were too small, reflecting conservative tests - and this was true irrespective of whether random slopes for the main effects were included in the model. At the same time, the Bayes factor estimates for the interaction term were too large, reflecting anti-conservative tests. The biases for the main effects were absent when fitting the model on non-aggregated data. However, Bayes factors for the interaction term were too small, reflecting a conservative bias. This bias disappeared in simulated data where the sphericity assumption was not violated.

\paragraph{Summary of Issue 1}

Taken together, these analyses suggest that if random slope variances differ between contrasts, then null hypothesis Bayes factor analyses based on aggregated data can yield biased Bayes factors: for contrasts with a large random slope variance, the Bayes factor may be too large (i.e., liberal) compared to its true value, whereas for contrasts with a small random slope variance, the Bayes factor may be too small (i.e., conservative). Moreover, when performing an omnibus test of the influence of a factor comprising more than two levels (i.e., more than one contrast), where random slope variances differ from each other, we observed conservative biases for the \texttt{BayesFactor} package, presumably (at least partly) because the underlying model assumes sphericity. That biases were absent for the \texttt{brms} models may reflect that the model does not assume sphericity in this test. Moreover, it may reflect a balance between conservative and anti-conservative biases, and should not be taken for granted in general. The problem with biased Bayes factor estimates can be avoided (or at least reduced, see \texttt{BayesFactor} package) by running Bayesian LMMs on unaggregated data and using an appropriate random effects structure (Schad et al., 2022). However, while for the \texttt{BayesFactor} package, fitting the model to non-aggregated data reduced the bias of Bayes factors, the bias was still present and only disappeared if it happened that sphericity was present in the data (i.e., no violation).

\hypertarget{issue-2-bayesian-hypothesis-tests-for-aggregated-data-are-biased-if-item-variance-is-present}{%
\section{Issue 2: Bayesian hypothesis tests for aggregated data are biased if item variance is present}\label{issue-2-bayesian-hypothesis-tests-for-aggregated-data-are-biased-if-item-variance-is-present}}

A second factor that can lead to biased Bayes factors when aggregating data is the presence of a second random factor, e.g., when item variance is present in addition to subject variance. It is well known for frequentist analyses that data aggregation (per subject and condition) in the presence of item variance leads to an inflation of alpha error (Clark, 1973; Forster \& Dickinson, 1976). Importantly, this problem occurs not only for repeated measures ANOVA, but also when fitting LMMs (random intercept-only) to aggregated data.

The traditional (frequentist) procedure for dealing with this situation is to run repeated measures ANOVA not only with subjects as random factor (\(F1\)), but to additionally run a second ANOVA with items as random factor (\(F2\)). The min-\(F'\) value then provides a traditional frequentist test statistic, which combines results from both (\(F1\) and \(F2\)) analyses. LMMs approach this issue by fitting the data on the individual-trial level, and by estimating random effects for subjects and for items simultaneously, providing a single test statistic (i.e., a single t-/p-value) for a given effect.

In Appendix~\ref{app:FreqItem}, we illustrate these issues observed in frequentist analyses of aggregated data using artificial data simulation together with LMM analyses. These results for frequentist models have been reported before (e.g., Clark, 1973; Forster \& Dickinson, 1976). In the next step, we demonstrate that this liberal bias in analysis results -- when data aggregation is used despite the presence of item variance -- is not constrained to frequentist models, but equally applies to Bayes factor null hypothesis tests.

\hypertarget{exemplary-data-simulation-and-analysis}{%
\subsection{Exemplary data simulation and analysis}\label{exemplary-data-simulation-and-analysis}}

Aggregating data to the by-subject level, and ignoring item variance even when it is present, leads to a liberal bias in the analysis.
To illustrate this effect of by-subject aggregation in the presence of item variance, we simulate data from a simple latin-square design based on Gibson and Wu (2013), which includes one two-level within-subject and within-item factor (X; sum-contrast coded, Schad et al., 2020). As in the original data set we assume that the data have \(42\) subjects and \(16\) items, yielding a total of \(672\) data points. We simulate log-normally distributed response times (\texttt{designr} package) in a reading task from a generalized linear mixed-effects model with crossed random effects for subjects and items with the following model parameters (which are defined in log-space; see Table \ref{table:simulation2}). The simulation uses the argument \texttt{empirical=TRUE} such that the simulated fixed effects correspond exactly to the indicated values.

\begin{table}[htbp]
\begin{center}
\begin{tabular}{l c}
\hline
& Simulation parameters \\
\hline
(Intercept)           & $6$ \\
X                     & $-0.10$ \\
\hline
Num. obs.             & $672$  \\
Num. groups: subj     & $42$   \\
Num. groups: item     & $16$   \\
\hline
SD: subj (Intercept)  & $0.24$   \\
SD: subj.1 X          & $0.11$   \\
SD: item (Intercept)  & $0.18$   \\
SD: item.1 X          & $0.25$   \\
SD: Residual          & $0.51$   \\
\hline
\end{tabular}
\caption{Parameters used to simulate data from an LMM with a log-normally distributed dependent variable.}
\label{table:simulation2}
\end{center}
\end{table}

\hypertarget{bayes-factor-null-hypothesis-test}{%
\subsubsection{Bayes factor null hypothesis test}\label{bayes-factor-null-hypothesis-test}}

Here we demonstrate that the same liberal bias observed in frequentist analyses of aggregated data is also present in Bayes factor null hypothesis tests based on aggregated data.

In a first step, we analyze the exemplary artificial data set by computing Bayes factors. As the critical second step, we then perform SBC for model inference to investigate bias in Bayes factor estimates.

For the Bayesian modeling, we assume priors that we had previously derived using prior predictive checks (Schad et al., 2021):

\begin{align}
\beta_{(Intercept)} &\sim Normal(6, 0.6) \\
\beta_{X} &\sim Normal(0, 0.1) \\
\sigma_{Random\,slopes} &\sim Normal_+(0, 0.1) \\
\sigma_{Residual} &\sim Normal_+(0, 0.5) \\
\rho_{Random\,slopes} &\sim LKJ(2)
\end{align}

We use \texttt{brms} to fit Bayesian LMMs with a lognormal likelihood to the simulated data -- once based on by-subject aggregated data and once on non-aggregated data. For the non-aggregated analysis, we use a full random effects structure, including correlated random intercepts and slopes for subjects as well as for items. For the by-subject aggregated analysis, we estimate only random intercepts for subjects.

The results from this analysis show that the null hypothesis Bayes factor for the non-aggregated analysis is BF10 = 1.92, yielding no evidence for the H0 or the H1. By contrast, the null hypothesis Bayes factor for the aggregated analysis is BF10 = 32.54, and thus clearly larger, seemingly providing strong evidence for the alternative hypothesis. This shows that in an aggregated analysis ignoring item variance, the evidence against the null hypothesis can be stronger than in a non-aggregated analysis. Again, the question arises as to which of these two Bayes factors is correct / unbiased. To investigate this issue, we turn to SBC.

\hypertarget{sbc}{%
\subsubsection{SBC}\label{sbc}}

\begin{figure}

{\centering \includegraphics{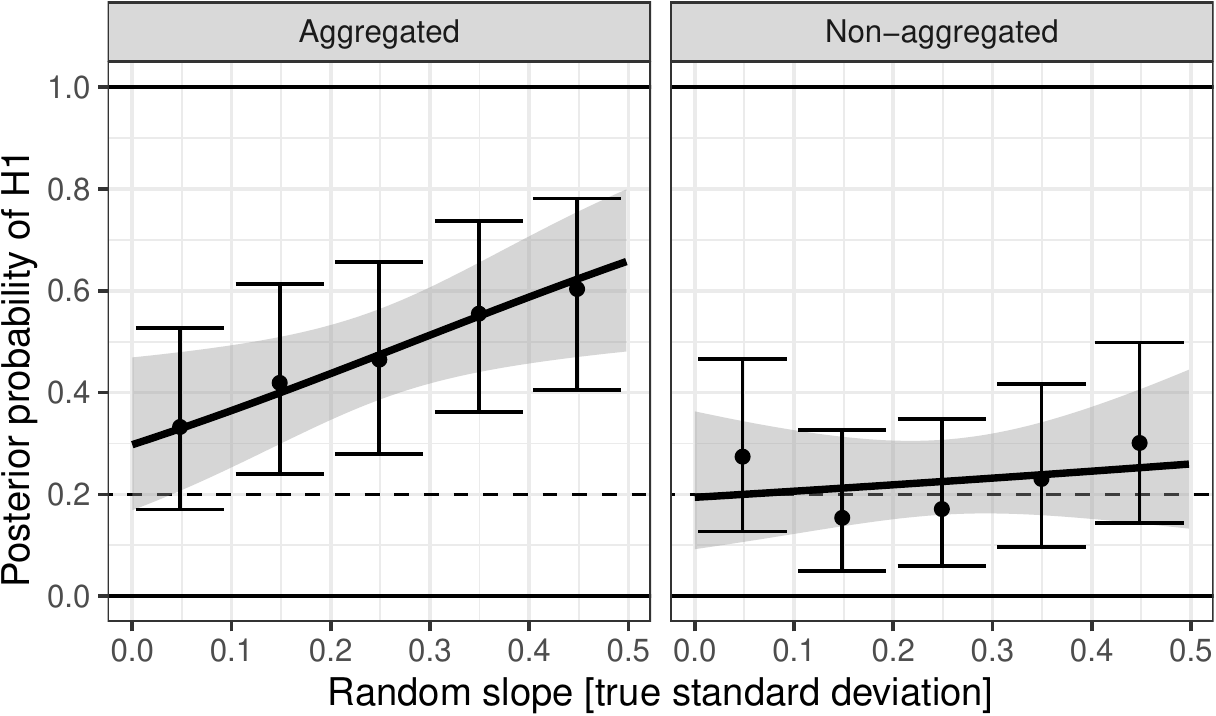} 

}

\caption{Results from SBC (simulation 2.1): the posterior probability for the alternative hypothesis (H1) is shown as a function of the true standard deviation of the item random slope. Results are shown for analysis based on by-subject aggregated data (left panel) and for analysis using non-aggregated data (right panel). Error bars are confidence intervals, regression lines are from a logistic regression. The prior probability for the null is set to 0.2 (this is chosen - lower than before, where it was 0.5 - for pragmatic reasons to demonstrate an expected increase in the average posterior probabilities compared to the prior of 0.2 and to avoid potential ceiling effects). The results show that the posterior probability for H1 increases with larger random slope variances, leading to a deviation from the prior, demonstrating that Bayes factor estimates are too large when item variance is ignored in aggregated analyses.}\label{fig:itemSBC}
\end{figure}

We performed SBC by varying the true standard deviation of the random item slopes from \(0\) to \(0.5\) in small steps of \(0.004\) across \(125\) simulations. The prior probability for the alternative hypothesis was set to a value of \(0.2\). We chose a smaller prior probability here (previously \(0.5\)) because we expected a liberal bias only (i.e., too large Bayes factors; which might be easiest to see when the prior probability of the H1 is low), whereas in Issue 1, we expected liberal as well as conservative biases (i.e., too small and too large Bayes factors; which should be easiest to see with a prior probability of H1 at \(0.5\)). Importantly, the results from SBC should not depend on the choice of the prior probability of the model. The same design and priors as in the exemplary data analyses were used. The results show (see Fig.~\ref{fig:itemSBC} and Fig.~\ref{fig:dpri-item-so}) that for aggregated analyses, when the random item slope variance is close to zero (i.e., true values ranging from \(0\) to \(0.10\)), the null hypothesis Bayes factor is close to the prior of \(0.2\), suggesting unbiased Bayes factor estimation. However, when the random item slope variance is set to be larger than zero, then the posterior probability for H1 is increasingly larger than the prior value of \(0.2\) (Bayesian linear regression using the R-package \texttt{BayesFactor} with default priors: \(BF_{10} =\) 8.30), demonstrating substantial bias in the estimation of Bayes factors, such that estimated Bayes factors are too large, i.e., larger than the true Bayes factor, and are thus incorrect. By contrast, when the same simulated data is analyzed using non-aggregated analyses, then the posterior probability for H1 is on average not different from the prior probability of \(0.2\) (Bayesian t-test with default priors: \(BF_{01} =\) 5.70), and does not increase with increasing random item slope variance (\(BF_{01} =\) 3.90), suggesting that the null hypothesis Bayes factor for non-aggregated analyses using Bayesian LMMs is unbiased.

\paragraph{Item aggregation in the BayesFactor package}

We next used the \texttt{BayesFactor} package to implement corresponding experimental designs (for details on simulations and results, see Appendix ~\ref{app:ItemBayesFactor}). We first used the same design as in the previous section, testing a two-level factor, and then we repeated these analyses using a 4-level factor. In both cases, the results showed that Bayes factors were accurate in non-aggregated analyses. However, in aggregated analyses the posterior probability for H1 increased with larger random slope variances, leading to a deviation from the prior, demonstrating that Bayes factor estimates were too large when item variance was ignored.

\paragraph{Item aggregation when analyzing a $2 \times 2$ design using brms}

As a next step, we used a \(2 \times 2\) design with random subject and random item effects (using a latin-square design). To speed up simulations, we used 16 subjects and 8 items, yielding a total of 128 data points. For the priors, we used the values from the two-step analyses reported above.

\begin{figure}

{\centering \includegraphics{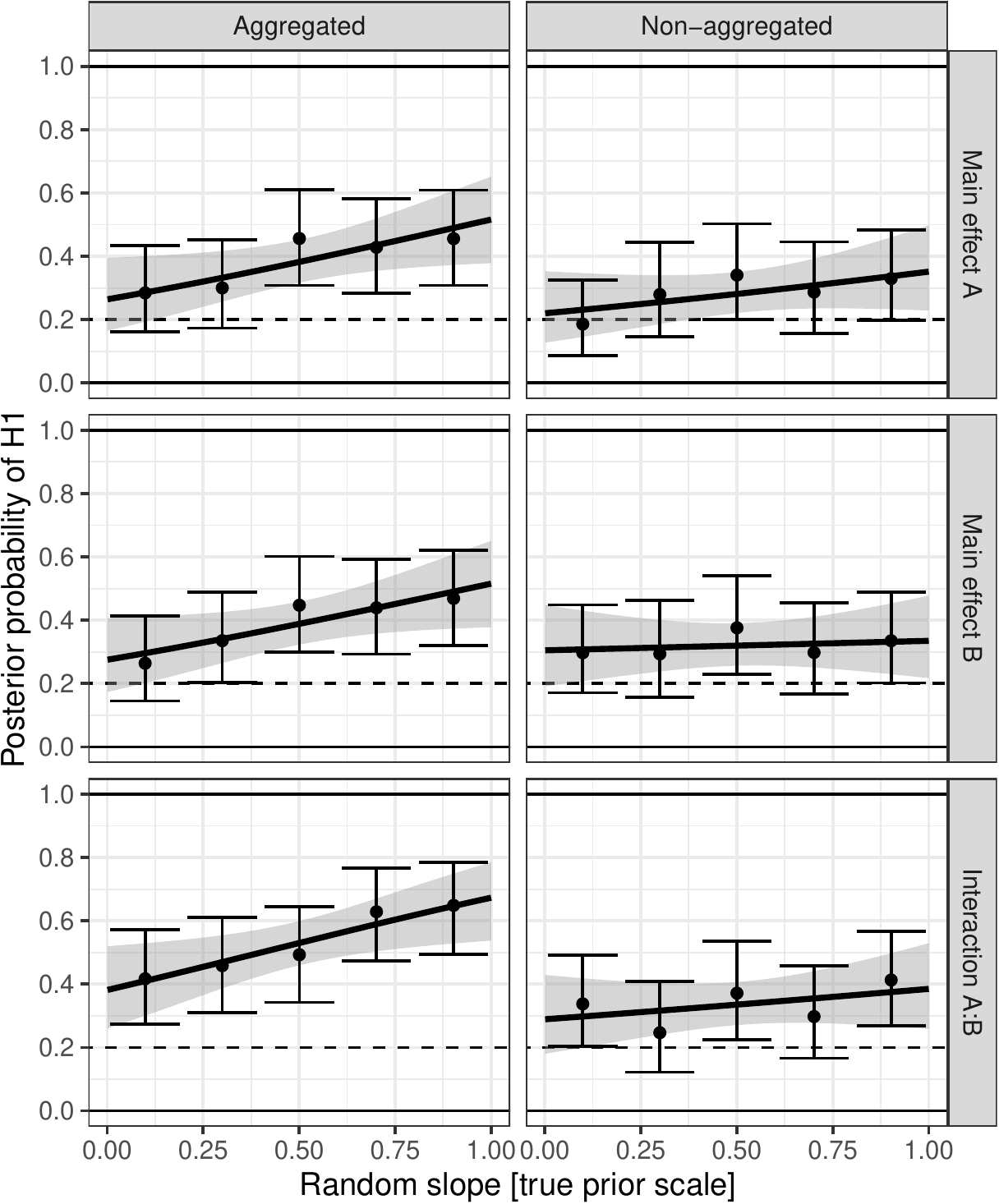} 

}

\caption{Simulation 2.4: Analyses of a 2 x 2 repeated measures design. Results from SBC: the posterior probability for the alternative hypothesis (H1) is shown as a function of the true standard deviation of the item random slope for main effects A (upper panels) and B (middle panels) as well as their interaction (lower panels). Results are shown for analysis based on by-subject aggregated data (left panels) and for analysis using non-aggregated data (right panels). Error bars are confidence intervals, regression lines are from a logistic regression. The prior probability for the null is set to 0.2. The results show that when aggregating across items, the posterior probability for H1 increases with larger item random slope variances, leading to a deviation from the prior, demonstrating that Bayes factor estimates are too large when item variance is ignored in aggregated analyses. The results are not unbiased for non-aggregated analyses, but average bias is reduced at least for some analyses (main effect A and interaction A:B).}\label{fig:itemSBC-BF-2x2}
\end{figure}

The results (see Fig.~\ref{fig:itemSBC-BF-2x2} and Fig.~\ref{fig:dpri-item-F4-2x2}) using 200 simulations showed that when aggregating over items in the analysis despite the presence of random item slopes, then estimated Bayes factors were too large, i.e., the average posterior was larger than the prior of \(0.2\), suggesting a liberal bias. This was present for main effects as well as for the interaction. Moreover, when the random item slope variance increased, this liberal bias also increased, suggesting the bias when testing the fixed effect was particularly large when large random slope variances are present.

By contrast, when non-aggregated analyses were used, and item variance was explicitly modeled, then some of these biases were reduced. Importantly, there was still (liberal) bias, since on average the posterior probabilities differed from the prior values of \(0.2\) (for both main effects and the interaction). However, when the item variance increased, there was no clear increase in posterior model probabilities (Bayes factor sensitivity analyses, see Fig.~\ref{fig:dpri-item-F4-2x2}, did not clearly support the H1 of an increase, but also did not provide clear evidence for H0 of no increase). However, the average bias was reduced in the non-aggregated analyses (for main effect A and the interaction; no clear evidence for main effect B). The slopes (capturing the increase with larger item variances) were not clearly reduced in the non-aggregated analyses compared to the aggregated analysis.

We conclude that running non-aggregated analyses reduces the average bias in posterior model probabilities at least for some of the tested effects. However, the results also demonstrate that \texttt{brms} together with \texttt{bridgesampling} exhibits biases in Bayes factor estimation even in the non-aggregated analyses. We (Schad et al., 2022) have pointed out before, that bridge sampling is not guaranteed to yield an unbiased Bayes factor estimate, and that therefore SBC is needed to determine for an individual analysis, whether Bayes factor estimates are unbiased. It seems that in the current 2 x 2 design with crossed random effects for subjects and items, bridge sampling results are biased even for non-aggregated analyses.

\paragraph{Interim summary for Issue 2}

Our analyses showed that when an effect varies across items, but items are ignored in the analysis by averaging across them, then Bayes factors are too large, reflecting a liberal bias in the estimation. We demonstrated this for a simple 2-level factor design using the R-packages \texttt{brms} and \texttt{BayesFactor}, for Bayesian ANOVA using a 4-level-factor (\texttt{BayesFactor} package), and for a \(2 \times 2\) design (\texttt{brms} package).

\hypertarget{discussion}{%
\section{Discussion}\label{discussion}}

We investigated null hypothesis Bayes factor tests using Bayesian hierarchical / linear mixed-effects models and Bayesian ANOVA using aggregated versus non-aggregated data. We used simulation-based calibration (SBC) to demonstrate that analyses based on aggregate data can lead to biases in the resulting null hypothesis tests, and should therefore be avoided. We demonstrated that biases in inferences do not only occur in frequentist approaches, where these are well-known and routinely corrected using frequentist tools, but that they also apply in Bayesian data analysis using Bayes factors, where they are often neglected (Bergh et al., 2022).

We showed that biases in Bayes factors based on aggregated data occur when the sphericity assumption is violated and random effects variances differ between different contrasts / effects. Indeed, in our experience, empirically, violations of the sphericity assumption occur quite frequently in the cognitive sciences. In these cases, using aggregated data without modeling the full random effects structure will yield a biased result for the null hypothesis Bayes factors, and can make the evidence for the H1 look stronger than it actually is (for effects/contrasts with a large random slope variance), or can make the evidence look weaker than it is (for effects/contrasts with a small random slope variance; and when testing a whole factor at once), reducing sensitivity for detecting a real experimental effect or falsely suggesting evidence for the null model when the alternative model may in fact be true.

The best way to proceed is to run a Bayesian LMM / ANOVA on non-aggregated data and to explicitly estimate random slopes for each effect (or to carefully select the random effect structure based on the data, Bates, Kliegl, Vasishth, \& Baayen, 2015; Matuschek, Kliegl, Vasishth, Baayen, \& Bates, 2017; Oberauer, 2022). This approach was successful in most of our studied approaches - with the exception of two analyses using the \texttt{BayesFactor} package (testing the influence of a factor in Bayesian omnibus ANOVA and testing the interaction in a \(2 \times 2\) design), where bias remained even when using non-aggregated data, and the model only yielded unbiased Bayes factors if sphericity was also a property of the data.

We also studied a second factor leading to biased null hypothesis Bayes factors in by-subject aggregated analyses, namely variance associated with items. We showed that when item variance is present but ignored by aggregating the data per subject and condition, then the resulting null hypothesis Bayes factors will be too large. That is, the evidence for an effect will look larger than it actually is. Again, this problem was reduced or disappeared when explicitly modeling random effects associated with subjects and with items simultaneously by fitting the model to the non-aggregated data. However, in a brms analysis of a 2 \(\times\) 2 design, we found that some bias remained even when explicitly modeling random item slopes.

Our results are in line with previous findings (Barr, Levy, Scheepers, \& Tily, 2013; Schad et al., 2022) that statistical results can be biased when important random effects variances are ignored, i.e., not explicitly modeled. When random slope variances are not modeled, but left out of the model, this can lead to biases in null hypothesis Bayes factors, resulting in a Bayes factor that is too large (or too small), and that makes the evidence for an effect or against an effect look stronger (or weaker) than it actually is. This is the same issue that Barr et al. (2013) discuss for frequentist LMMs. If a researcher wants to omit random slope variances from a Bayesian LMM, then careful model selection is needed, and only random slope terms that are not supported by the data should be dropped from the model (Bates et al., 2015; Matuschek et al., 2017; Oberauer, 2022).

In summary, our work demonstrates that aggregating data to the by-subject level during Bayesian modeling can lead to biased null hypothesis tests based on Bayes factors. There may be some situations (e.g., very large data sets and models), where non-aggregated analyses take an extremely long time, and obtaining stable Bayes factor estimates may be very difficult. In such situations, aggregated analyses may provide a way forward. If an aggregated analysis is used, researchers should provide evidence that the sphericity assumption is justified, i.e., they should make sure that random slope variances do not strongly differ between contrasts/effects. Testing violations of the sphericity assumption in small data sets may run the risk of false negative results, which can again introduce biases in Bayes factor estimation. Therefore, as the most conservative strategy, such biases can be reduced by running non-aggregated analyses. If item variance can in principle be present in a data set, it should be explicitly modeled using non-aggregated Bayesian hierarchical / linear-mixed effects models, software packages for which are easily accessible today (Bürkner, 2017; Carpenter et al., 2017; Morey \& Rouder, 2018).
Bayesian modeling, and (null hypothesis) Bayes factor analyses have many advantages over frequentist p-values (Rouder et al., 2018; Tendeiro \& Kiers, 2019; van Ravenzwaaij \& Wagenmakers, 2021; Wagenmakers et al., 2010). However, some of the statistical caveats learned from frequentist tools need to be considered even when running Bayesian models.

A preprint of this manuscript has been published on arXiv (Schad, Nicenboim, \& Vasishth, 2022a). Reproducible code is available from OSF (Schad, Nicenboim, \& Vasishth, 2022b).

\hypertarget{acknowledgements}{%
\section{Acknowledgements}\label{acknowledgements}}

This work was partly funded by the Deutsche Forschungsgemeinschaft (DFG, German Research Foundation), Sonderforschungsbereich 1287, project number 317633480 (Limits of Variability in Language). We thank Paul-Christian Bürkner and Michael Betancourt for discussion on the formulation of SBC for Bayes factors.

\newpage

\newpage

\hypertarget{references}{%
\section{References}\label{references}}

\begingroup
\setlength{\parindent}{-0.5in}
\setlength{\leftskip}{0.5in}

\hypertarget{refs}{}
\begin{CSLReferences}{1}{0}
\leavevmode\vadjust pre{\hypertarget{ref-barr2013random}{}}%
Barr, D. J., Levy, R., Scheepers, C., \& Tily, H. J. (2013). Random effects structure for confirmatory hypothesis testing: Keep it maximal. \emph{Journal of Memory and Language}, \emph{68}(3), 255--278.

\leavevmode\vadjust pre{\hypertarget{ref-BatesEtAlParsimonious}{}}%
Bates, D., Kliegl, R., Vasishth, S., \& Baayen, H. (2015). \emph{Parsimonious mixed models}. Retrieved from \url{http://arxiv.org/abs/1506.04967}

\leavevmode\vadjust pre{\hypertarget{ref-bates2015fitting}{}}%
Bates, D., Mächler, M., Bolker, B. M., \& Walker, S. C. (2015). Fitting linear mixed-effects models using lme4. \emph{Journal of Statistical Software}, \emph{67}(1), 1--48.

\leavevmode\vadjust pre{\hypertarget{ref-bennettEfficientEstimationFree1976}{}}%
Bennett, C. H. (1976). Efficient estimation of free energy differences from {Monte Carlo} data. \emph{Journal of Computational Physics}, \emph{22}(2), 245--268.

\leavevmode\vadjust pre{\hypertarget{ref-van2022bayesian}{}}%
Bergh, D. van den, Wagenmakers, E.-J., \& Aust, F. (2022). \emph{Bayesian repeated-measures ANOVA: An updated methodology implemented in JASP}.

\leavevmode\vadjust pre{\hypertarget{ref-betancourt2018calibrating}{}}%
Betancourt, M. (2018). Calibrating model-based inferences and decisions. \emph{arXiv Preprint arXiv:1803.08393}.

\leavevmode\vadjust pre{\hypertarget{ref-box1954some}{}}%
Box, G. E. (1954). Some theorems on quadratic forms applied in the study of analysis of variance problems, i. Effect of inequality of variance in the one-way classification. \emph{The Annals of Mathematical Statistics}, \emph{25}(2), 290--302.

\leavevmode\vadjust pre{\hypertarget{ref-Buerkner2017brms}{}}%
Bürkner, P.-C. (2017). {brms}: An {R} package for {Bayesian} multilevel models using {Stan}. \emph{Journal of Statistical Software}, \emph{80}(1), 1--28. \url{https://doi.org/10.18637/jss.v080.i01}

\leavevmode\vadjust pre{\hypertarget{ref-Buerkner2018brms}{}}%
Bürkner, P.-C. (2018). Advanced {Bayesian} multilevel modeling with the {R} package {brms}. \emph{The R Journal}, \emph{10}(1), 395--411. \url{https://doi.org/10.32614/RJ-2018-017}

\leavevmode\vadjust pre{\hypertarget{ref-carpenter2017stan}{}}%
Carpenter, B., Gelman, A., Hoffman, M. D., Lee, D., Goodrich, B., Betancourt, M., \ldots{} Riddell, A. (2017). Stan: A probabilistic programming language. \emph{Journal of Statistical Software}, \emph{76}(1), 1-\/-32.

\leavevmode\vadjust pre{\hypertarget{ref-chow2017bayesian}{}}%
Chow, S.-M., \& Hoijtink, H. (2017). Bayesian estimation and modeling: Editorial to the second special issue on {B}ayesian data analysis. \emph{Psychological Methods}, \emph{22}(4), 609--615.

\leavevmode\vadjust pre{\hypertarget{ref-clark1973language}{}}%
Clark, H. H. (1973). The language-as-fixed-effect fallacy: A critique of language statistics in psychological research. \emph{Journal of Verbal Learning and Verbal Behavior}, \emph{12}(4), 335--359.

\leavevmode\vadjust pre{\hypertarget{ref-cook2006validation}{}}%
Cook, S. R., Gelman, A., \& Rubin, D. B. (2006). Validation of software for {B}ayesian models using posterior quantiles. \emph{Journal of Computational and Graphical Statistics}, \emph{15}(3), 675--692.

\leavevmode\vadjust pre{\hypertarget{ref-daw2011model}{}}%
Daw, N. D., Gershman, S. J., Seymour, B., Dayan, P., \& Dolan, R. J. (2011). Model-based influences on humans' choices and striatal prediction errors. \emph{Neuron}, \emph{69}(6), 1204--1215.

\leavevmode\vadjust pre{\hypertarget{ref-etz2018how}{}}%
Etz, A., Gronau, Q. F., Dablander, F., Edelsbrunner, P. A., \& Baribault, B. (2018). How to become a {B}ayesian in eight easy steps: An annotated reading list. \emph{Psychonomic Bulletin \& Review}, \emph{25}(1), 219--234.

\leavevmode\vadjust pre{\hypertarget{ref-etz2018introduction}{}}%
Etz, A., \& Vandekerckhove, J. (2018). Introduction to {B}ayesian inference for psychology. \emph{Psychonomic Bulletin \& Review}, \emph{25}(1), 5--34.

\leavevmode\vadjust pre{\hypertarget{ref-forster1976more}{}}%
Forster, K. I., \& Dickinson, R. G. (1976). More on the language-as-fixed-effect fallacy: Monte carlo estimates of error rates for {F1}, {F2}, {F'}, and min {F'}. \emph{Journal of Verbal Learning and Verbal Behavior}, \emph{15}(2), 135--142.

\leavevmode\vadjust pre{\hypertarget{ref-ge2018turing}{}}%
Ge, H., Xu, K., \& Ghahramani, Z. (2018). Turing: A language for flexible probabilistic inference. In A. Storkey \& F. Perez-Cruz (Eds.), \emph{Proceedings of the twenty-first international conference on artificial intelligence and statistics} (pp. 1682--1690). PMLR. Retrieved from \url{https://proceedings.mlr.press/v84/ge18b.html}

\leavevmode\vadjust pre{\hypertarget{ref-gibson2013processing}{}}%
Gibson, E., \& Wu, H.-H. I. (2013). Processing {C}hinese relative clauses in context. \emph{Language and Cognitive Processes}, \emph{28}(1-2), 125--155.

\leavevmode\vadjust pre{\hypertarget{ref-goodrich2020rstanarm}{}}%
Goodrich, B., Gabry, J., Ali, I., \& Brilleman, S. (2020). Rstanarm: Bayesian applied regression modeling via stan. \emph{R Package Version}, \emph{2}(1).

\leavevmode\vadjust pre{\hypertarget{ref-greenhouse1959methods}{}}%
Greenhouse, S. W., \& Geisser, S. (1959). On methods in the analysis of profile data. \emph{Psychometrika}, \emph{24}(2), 95--112.

\leavevmode\vadjust pre{\hypertarget{ref-Gronau2020bridgesampling}{}}%
Gronau, Q. F., Singmann, H., \& Wagenmakers, E.-J. (2020). {bridgesampling}: An {R} package for estimating normalizing constants. \emph{Journal of Statistical Software}, \emph{92}(10), 1--29. \url{https://doi.org/10.18637/jss.v092.i10}

\leavevmode\vadjust pre{\hypertarget{ref-hays1978statistics}{}}%
Hays, W. L. (1978). \emph{Statistics for the social sciences} (Vol. 410). New York: Holt, Rinehart; Winston.

\leavevmode\vadjust pre{\hypertarget{ref-heck2020review}{}}%
Heck, D. W., Boehm, U., Böing-Messing, F., Bürkner, P.-C., Derks, K., Dienes, Z., \ldots{} others. (2022). A review of applications of the {B}ayes factor in psychological research. \emph{Psychological Methods}.

\leavevmode\vadjust pre{\hypertarget{ref-hoijtink2017bayesian}{}}%
Hoijtink, H., \& Chow, S.-M. (2017). Bayesian hypothesis testing: Editorial to the special issue on {B}ayesian data analysis. \emph{Psychological Methods}, \emph{22}(2), 211--216.

\leavevmode\vadjust pre{\hypertarget{ref-huynh1976estimation}{}}%
Huynh, H., \& Feldt, L. S. (1976). Estimation of the {B}ox correction for degrees of freedom from sample data in randomized block and split-plot designs. \emph{Journal of Educational Statistics}, \emph{1}(1), 69--82.

\leavevmode\vadjust pre{\hypertarget{ref-JASP2022}{}}%
JASP Team. (2022). \emph{{JASP (Version 0.16.1){[}Computer software{]}}}. Retrieved from \url{https://jasp-stats.org/}

\leavevmode\vadjust pre{\hypertarget{ref-jeffreys1939theory}{}}%
Jeffreys, H. (1939). \emph{Theory of probability}. Oxford: Clarendon Press.

\leavevmode\vadjust pre{\hypertarget{ref-kass1995bayes}{}}%
Kass, R. E., \& Raftery, A. E. (1995). Bayes factors. \emph{Journal of the American Statistical Association}, \emph{90}(430), 773--795.

\leavevmode\vadjust pre{\hypertarget{ref-R-lmerTest}{}}%
Kuznetsova, A., Brockhoff, P. B., \& Christensen, R. H. B. (2017). {lmerTest} package: Tests in linear mixed effects models. \emph{Journal of Statistical Software}, \emph{82}(13), 1--26. \url{https://doi.org/10.18637/jss.v082.i13}

\leavevmode\vadjust pre{\hypertarget{ref-lee2011cognitive}{}}%
Lee, M. D. (2011). How cognitive modeling can benefit from hierarchical {B}ayesian models. \emph{Journal of Mathematical Psychology}, \emph{55}(1), 1--7.

\leavevmode\vadjust pre{\hypertarget{ref-lunn2000winbugs}{}}%
Lunn, D. J., Thomas, A., Best, N., \& Spiegelhalter, D. (2000). {WinBUGS}-{A} {B}ayesian modelling framework: {C}oncepts, structure, and extensibility. \emph{Statistics and Computing}, \emph{10}(4), 325--337.

\leavevmode\vadjust pre{\hypertarget{ref-ly2016harold}{}}%
Ly, A., Verhagen, J., \& Wagenmakers, E.-J. (2016). Harold {J}effreys's default {B}ayes factor hypothesis tests: Explanation, extension, and application in psychology. \emph{Journal of Mathematical Psychology}, \emph{72}, 19--32.

\leavevmode\vadjust pre{\hypertarget{ref-magezi2015linear}{}}%
Magezi, D. A. (2015). Linear mixed-effects models for within-participant psychology experiments: An introductory tutorial and free, graphical user interface ({LMMgui}). \emph{Frontiers in Psychology}, \emph{6}, 2.

\leavevmode\vadjust pre{\hypertarget{ref-matuschek2017balancing}{}}%
Matuschek, H., Kliegl, R., Vasishth, S., Baayen, H., \& Bates, D. (2017). Balancing {T}ype {I} error and power in linear mixed models. \emph{Journal of Memory and Language}, \emph{94}, 305--315.

\leavevmode\vadjust pre{\hypertarget{ref-mauchly1940significance}{}}%
Mauchly, J. W. (1940). Significance test for sphericity of a normal n-variate distribution. \emph{The Annals of Mathematical Statistics}, \emph{11}(2), 204--209.

\leavevmode\vadjust pre{\hypertarget{ref-mengSimulatingRatiosNormalizing1996}{}}%
Meng, X., \& Wong, W. H. (1996). Simulating ratios of normalizing constants via a simple identity: {A} theoretical exploration. \emph{Statistica Sinica}, \emph{6}(4), 831--860.

\leavevmode\vadjust pre{\hypertarget{ref-MoreyBayesFactor}{}}%
Morey, R. D., \& Rouder, J. N. (2018). \emph{BayesFactor: Computation of {B}ayes factors for common designs}. Retrieved from \url{https://CRAN.R-project.org/package=BayesFactor}

\leavevmode\vadjust pre{\hypertarget{ref-BFpackage2022}{}}%
Morey, R. D., \& Rouder, J. N. (2022). \emph{BayesFactor: Computation of bayes factors for common designs}. Retrieved from \url{https://CRAN.R-project.org/package=BayesFactor}

\leavevmode\vadjust pre{\hypertarget{ref-mulder2016editors}{}}%
Mulder, J., \& Wagenmakers, E.-J. (2016). Editors' introduction to the special issue {``{B}ayes factors for testing hypotheses in psychological research: Practical relevance and new developments.''} \emph{Journal of Mathematical Psychology}, \emph{72}, 1--5.

\leavevmode\vadjust pre{\hypertarget{ref-NicenboimVasishth2016}{}}%
Nicenboim, B., \& Vasishth, S. (2016). {Statistical methods for linguistic research: {Foundational} Ideas - {Part} {II}}. \emph{Language and Linguistics Compass}, \emph{10}(11), 591--613. \url{https://doi.org/10.1111/lnc3.12207}

\leavevmode\vadjust pre{\hypertarget{ref-Oberauer2022BayesFactor}{}}%
Oberauer, K. (2022). The importance of random slopes in mixed models for bayesian hypothesis testing. \emph{Psychological Science}, \emph{33}(4), 648--665.

\leavevmode\vadjust pre{\hypertarget{ref-plummer2003jags}{}}%
Plummer, M. (2003). JAGS: A program for analysis of {B}ayesian graphical models using {G}ibbs sampling. \emph{Proceedings of the 3rd International Workshop on Distributed Statistical Computing}, \emph{124}, 1--10. Vienna, Austria.

\leavevmode\vadjust pre{\hypertarget{ref-raaijmakers1999deal}{}}%
Raaijmakers, J. G., Schrijnemakers, J. M., \& Gremmen, F. (1999). How to deal with {``the language-as-fixed-effect fallacy''}: Common misconceptions and alternative solutions. \emph{Journal of Memory and Language}, \emph{41}(3), 416--426.

\leavevmode\vadjust pre{\hypertarget{ref-R-designr}{}}%
Rabe, M. M., Kliegl, R., \& Schad, D. J. (2020). \emph{{d}esignr: Balanced factorial designs}. Retrieved from \url{https://CRAN.R-project.org/package=designr}

\leavevmode\vadjust pre{\hypertarget{ref-R-hypr_b}{}}%
Rabe, M. M., Vasishth, S., Hohenstein, S., Kliegl, R., \& Schad, D. J. (2020). {h}ypr: An {R} package for hypothesis-driven contrast coding. \emph{The Journal of Open Source Software}, \emph{5}(48), 2134. \url{https://doi.org/10.21105/joss.02134}

\leavevmode\vadjust pre{\hypertarget{ref-rouder2018bayesian}{}}%
Rouder, J. N., Haaf, J. M., \& Vandekerckhove, J. (2018). Bayesian inference for psychology, part {IV}: Parameter estimation and {B}ayes factors. \emph{Psychonomic Bulletin \& Review}, \emph{25}(1), 102--113.

\leavevmode\vadjust pre{\hypertarget{ref-rouder2012default}{}}%
Rouder, J. N., Morey, R. D., Speckman, P. L., \& Province, J. M. (2012). Default bayes factors for ANOVA designs. \emph{Journal of Mathematical Psychology}, \emph{56}(5), 356--374.

\leavevmode\vadjust pre{\hypertarget{ref-salvatier2016probabilistic}{}}%
Salvatier, J., Wiecki, T. V., \& Fonnesbeck, C. (2016). Probabilistic programming in python using PyMC3. \emph{PeerJ Computer Science}, \emph{2}, e55.

\leavevmode\vadjust pre{\hypertarget{ref-schad2020toward}{}}%
Schad, D. J., Betancourt, M., \& Vasishth, S. (2021). Toward a principled {B}ayesian workflow in cognitive science. \emph{Psychological Methods}, \emph{26}(1), 103--126. \url{https://doi.org/10.1037/met0000275}

\leavevmode\vadjust pre{\hypertarget{ref-schad2014processing}{}}%
Schad, D. J., Jünger, E., Sebold, M., Garbusow, M., Bernhardt, N., Javadi, A.-H., \ldots{} others. (2014). Processing speed enhances model-based over model-free reinforcement learning in the presence of high working memory functioning. \emph{Frontiers in Psychology}, 1450. \url{https://doi.org/10.3389/fpsyg.2014.01450}

\leavevmode\vadjust pre{\hypertarget{ref-Schad2022BayesFactor}{}}%
Schad, D. J., Nicenboim, B., Bürkner, P.-C., Betancourt, M., \& Vasishth, S. (2022). Workflow techniques for the robust use of {B}ayes factors. \emph{Psychological Methods}. \url{https://doi.org/10.1037/met0000472}

\leavevmode\vadjust pre{\hypertarget{ref-schad2022dataOSF}{}}%
Schad, D. J., Nicenboim, B., \& Vasishth, S. (2022b). Data aggregation can lead to biased inferences in {B}ayesian linear mixed models. \emph{OSF Repository}. Retrieved from \url{https://osf.io/mjf47/}

\leavevmode\vadjust pre{\hypertarget{ref-schad2022dataArXiv}{}}%
Schad, D. J., Nicenboim, B., \& Vasishth, S. (2022a). Data aggregation can lead to biased inferences in {B}ayesian linear mixed models. \emph{arXiv Preprint arXiv:2203.02361}.

\leavevmode\vadjust pre{\hypertarget{ref-schad2020dissociating}{}}%
Schad, D. J., Rapp, M. A., Garbusow, M., Nebe, S., Sebold, M., Obst, E., \ldots{} others. (2020). Dissociating neural learning signals in human sign-and goal-trackers. \emph{Nature Human Behaviour}, \emph{4}(2), 201--214. \url{https://doi.org/10.1038/s41562-019-0765-5}

\leavevmode\vadjust pre{\hypertarget{ref-schad2020capitalize}{}}%
Schad, D. J., Vasishth, S., Hohenstein, S., \& Kliegl, R. (2020). How to capitalize on a priori contrasts in linear (mixed) models: A tutorial. \emph{Journal of Memory and Language}, \emph{110}, 104038. \url{https://doi.org/10.1016/j.jml.2019.104038}

\leavevmode\vadjust pre{\hypertarget{ref-Talts:2018aa}{}}%
Talts, S., Betancourt, M., Simpson, D., Vehtari, A., \& Gelman, A. (2018). Validating {B}ayesian inference algorithms with simulation-based calibration. \emph{arXiv Preprint arXiv:1804.06788}.

\leavevmode\vadjust pre{\hypertarget{ref-tendeiro2019review}{}}%
Tendeiro, J. N., \& Kiers, H. A. (2019). A review of issues about null hypothesis {B}ayesian testing. \emph{Psychological Methods}, \emph{24}(6), 774-\/-795.

\leavevmode\vadjust pre{\hypertarget{ref-tendeiro2021white}{}}%
Tendeiro, J. N., \& Kiers, H. A. (2021). On the white, the black, and the many shades of gray in between: Our reply to van {R}avenzwaaij and {W}agenmakers (2021). \emph{PsyArXiv Preprint PsyArXiv:tjxvz}.

\leavevmode\vadjust pre{\hypertarget{ref-van2021bayes}{}}%
van Doorn, J., Aust, F., Haaf, J. M., Stefan, A. M., \& Wagenmakers, E.-J. (2021). Bayes factors for mixed models. \emph{Computational Brain \& Behavior}, 1--13.

\leavevmode\vadjust pre{\hypertarget{ref-van2019advantages}{}}%
van Ravenzwaaij, D., \& Wagenmakers, E.-J. (2021). Advantages masquerading as {`issues'} in {B}ayesian hypothesis testing: A commentary on {T}endeiro and {K}iers (2019). \emph{Psychological Methods}. \url{https://doi.org/10.1037/met0000415}

\leavevmode\vadjust pre{\hypertarget{ref-vandekerckhove2018bayesian}{}}%
Vandekerckhove, J., Rouder, J. N., \& Kruschke, J. K. (2018). Bayesian methods for advancing psychological science. \emph{Psychonomic Bulletin \& Review}, \emph{25}(1), 1--4.

\leavevmode\vadjust pre{\hypertarget{ref-VasishthEtAl2017EDAPS}{}}%
Vasishth, S., Nicenboim, B., Beckman, M. E., Li, F., \& Kong, E. (2018). Bayesian data analysis in the phonetic sciences: {A} tutorial introduction. \emph{Journal of Phonetics}, \emph{71}, 147--161. \url{https://doi.org/10.1016/j.wocn.2018.07.008}

\leavevmode\vadjust pre{\hypertarget{ref-wagenmakersBayesianHypothesisTesting2010}{}}%
Wagenmakers, E.-J., Lodewyckx, T., Kuriyal, H., \& Grasman, R. (2010). Bayesian hypothesis testing for psychologists: {A} tutorial on the {Savage}-{D}ickey method. \emph{Cognitive Psychology}, \emph{60}(3), 158--189. \url{https://doi.org/10.1016/j.cogpsych.2009.12.001}

\end{CSLReferences}

\endgroup

\newpage

\hypertarget{appendix-appendix}{%
\appendix}

\hypertarget{app:freqSphericity}{%
\section{Frequentist simulations: Sphericity Assumption}\label{app:freqSphericity}}

\hypertarget{an-exemplary-simulation-using-frequentist-lmms}{%
\subsection{An exemplary simulation using frequentist LMMs}\label{an-exemplary-simulation-using-frequentist-lmms}}

A first analysis of one exemplary simulated data set illustrates how LMMs fit to non-aggregated versus aggregated data fare when there is a violation of sphericity. We show that a non-aggregated LMM analysis can estimate the underlying differences in variances, such that the estimated standard errors for the fixed effects can capture this difference in variances, and thus inform the resulting t- and p-values. By contrast, an LMM based on aggregated data cannot capture this difference in random effects variances, such that the difference in random effect variances is neither reflected in the standard errors of the fixed effects, nor in the resulting t- and p-values. This leads to biases in their estimation.

In the exemplary simulation, we ensure that when fitting an LMM to the simulated data, the fixed effects estimated by the model are exactly the true underlying values. This is implemented by setting the argument \texttt{empirical} in the \texttt{simLMM} function (\texttt{designr}) to \texttt{TRUE}. The random effects estimates, however, are not exact.

\begin{table}[htbp]
\begin{center}
\begin{tabular}{l c c l c l}
\hline
 & Simulation parameters & \multicolumn{2}{c}{Non-aggregated} & \multicolumn{2}{c}{Aggregated} \\
\hline
(Intercept)           & $200$ & $200.00^{***}$ & $(5.07)$  & $200.00^{***}$ & $(6.55)$ \\
c2vs1                 & $20$  & $20.00$        & $(\boldsymbol{21.37})$ & $20.00$        & $(\boldsymbol{16.05})$        \\
c3vs1                 & $20$  & $20.00\boldsymbol{^{***}}$  & $(\boldsymbol{5.01})$  & $20.00$        & $(\boldsymbol{16.05})$        \\
\hline
Num. obs.             & $600$ & \multicolumn{2}{c}{$600$} & \multicolumn{2}{c}{$60$}   \\
Num. groups: subj     & $20$  & \multicolumn{2}{c}{ $20$} & \multicolumn{2}{c}{$20$}   \\
\hline
SD: subj (Intercept)  & $20$  & \multicolumn{2}{c}{$20.7$} & \multicolumn{2}{c}{$0.0$}  \\
SD: subj.1 c2vs1      & $90$  & \multicolumn{2}{c}{$92.9$} & \multicolumn{2}{c}{$$}     \\
SD: subj.2 c3vs1      & $10$  & \multicolumn{2}{c}{$0.0$} & \multicolumn{2}{c}{$$}      \\
SD: Residual          & $50$  & \multicolumn{2}{c}{$50.1$} & \multicolumn{2}{c}{$50.7$} \\
\hline
\multicolumn{6}{l}{\scriptsize{$^{***}p<0.001$; $^{**}p<0.01$; $^{*}p<0.05$}}
\end{tabular}
\caption{Parameters used to simulate data from an LMM as well as results from frequentist LMM analyses of the exemplary simulated data set using aggregated and non-aggregated data. Brakets show standard errors.}
\label{table:coefficientsAppendix}
\end{center}
\end{table}

The exemplary simulated data set can be analyzed using LMMs on non-aggregated and on aggregated data. The non-aggregated LMM analysis, using the packages \texttt{lme4} (Bates, Mächler, Bolker, \& Walker, 2015) and \texttt{lmerTest} (Kuznetsova, Brockhoff, \& Christensen, 2017), estimates fixed effects -- for the intercept and for the comparison of each experimental condition to the baseline condition (\texttt{c2vs1} and \texttt{c3vs1}) -- and (uncorrelated) random effects (for the intercept and each comparison) varying across subjects. In frequentist analyses, we specify random effects as uncorrelated, because estimating random effects correlations from limited data often leads to problems with model fitting, and because estimating random effects correlations is not expected to influence test statistics of fixed effects.

The results of the non-aggregated LMM (see Table \ref{table:coefficientsAppendix}, middle column) show that the estimated fixed effects are exactly as specified (i.e., they have a value of exactly 20), due to the argument \texttt{empirical=TRUE}.
Moreover, the random effects variances are estimated as 92.90 and as 0, close to the true values of \(90\) and \(10\). This reflects the differences assumed in the data simulation, i.e., a violation of sphericity, which is not a problem in the current non-aggregated analysis. Importantly, this difference in random slope variances translates into a difference in the standard errors of the fixed effects, which are estimated as
21.37 and as 5.01. This difference is important for statistical inference: the first contrast varies strongly across subjects; therefore, its standard error is large, as it should be. The second contrast varies very little across subjects; therefore, its standard error is small, as it should be. This difference in standard errors accordingly translates into differences in t- and p-values. Given that both effect estimates have exactly the same size (of 20), if an effect varies strongly across subjects (i.e., \texttt{c2vs1}), then the p-value should be relatively large, which is the case here. By contrast, if an effect varies very little across subjects (i.e., \texttt{c3vs1}), then the p-value should be relatively small, which is the case here.

These differences disappear when aggregating the data by subject and condition, and running an LMM on the aggregated data.

Table \ref{table:coefficientsAppendix} (right column) shows that only the intercept variance can be estimated from the aggregated data, but differences in random slope variances cannot be captured. Therefore, the standard errors of the fixed effects are exactly the same for both effects (i.e., they have a value of 16). Differences in the random effect variances between these effects cannot be separated from the residual noise any more. Accordingly, the t- and p-values are the same for both effects -- which is incorrect given the large difference in random slope variances.

\hypertarget{estimation-of-type-i-and-type-ii-errors}{%
\subsubsection{Estimation of type I and type II errors}\label{estimation-of-type-i-and-type-ii-errors}}

One key question from these analyses is whether the p-values obtained based on the LMM analyses of (non-)aggregated data are accurate or biased. To investigate this in frequentist analyses, we need to estimate empirical alpha error rates based on simulating data from the null hypothesis. Specifically, this can be tested by simulating data based on the null hypothesis (i.e., that the estimate is zero for both contrasts) many times (here \(1000\) times), analyzing the data using LMMs (using aggregated and non-aggregated data), and testing whether the fixed effect estimates are significantly different from zero (at an \(\alpha = 0.05\) level) in \(5\)\% of cases, as is theoretically expected.

\begin{figure}

{\centering \includegraphics{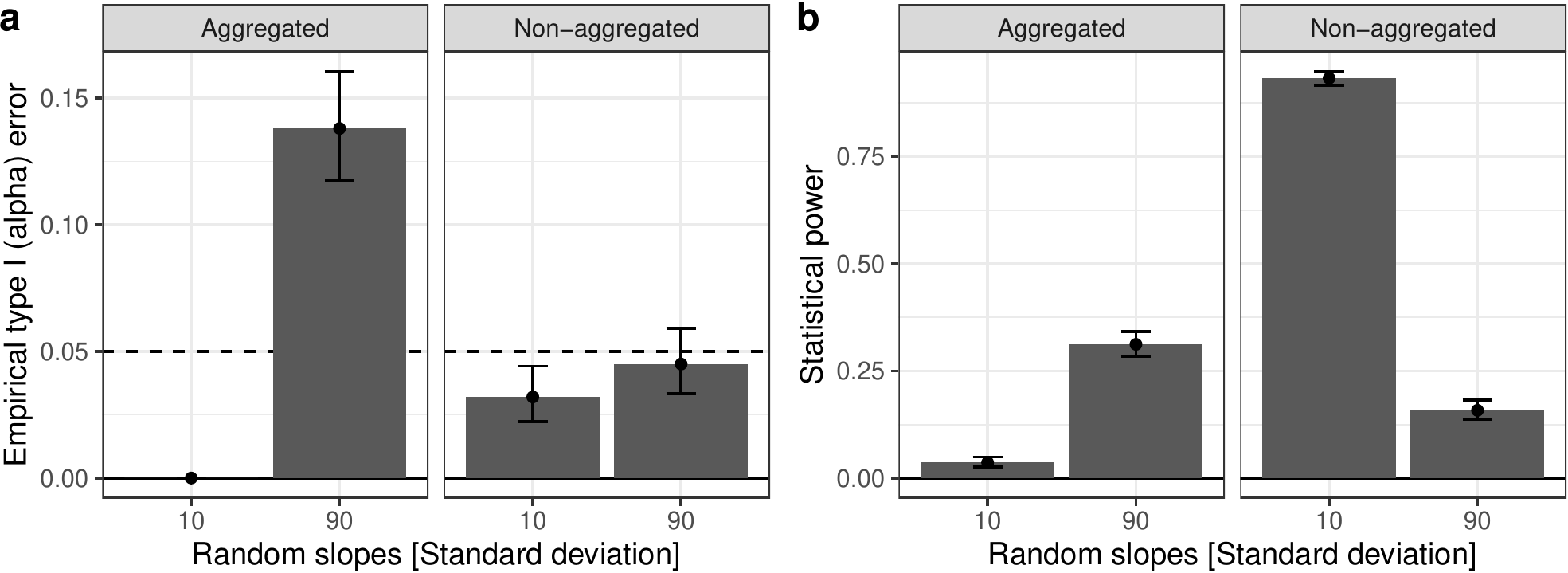} 

}

\caption{Empirical type I (alpha) error and statistical power when the standard deviation of the random slopes is small (10) versus large (90). Results are shown for LMMs fit to aggregated data versus to non-aggregated data. (a) The empirical type I (alpha) error is accurate and roughly at 5 percent for non-aggregated data, but biases are present for the aggregated data, leading to biased p-values, which are either too small (large random slope variance, 90) or too large (small random slope variance, 10). (b) For the effect with a small random slope (standard deviation of 10), statistical power is much reduced for aggregated data, but is good for the non-aggregated analysis. For the effect with a large random slope (standard deviation of 90), there seems to be an artificial increase in power in the aggregated analyses; however, this is increase is invalid since the empirical alpha error is inflated. (a+b) Error bars show 95 percent confidence intervals.}\label{fig:sphericityPower}
\end{figure}

The results show (see Fig.~\ref{fig:sphericityPower}a, right panel) that for the LMM analyses based on non-aggregated data, the empirical type I (\(\alpha\)) error is roughly at 5\% for both effects: i.e., the empirical type I error is at roughly 5\% for the effect where the true standard deviation of the random slopes is small (with a value of 10; empirical alpha error = 0.03) and also for the effect where the true standard deviation of the random slopes is large (with a value of 90; empirical alpha error = 0.04). This analysis demonstrates that, as expected, the non-aggregated analysis yields unbiased p-values.

By contrast, p-values based on the aggregated analysis are biased (see Fig.~\ref{fig:sphericityPower}a, left panel): here, the empirical type I (\(\alpha\)) error clearly deviates from its normative 5\% level for both effects. For the effect where the true standard deviation of the random slopes is small (with a value of 10), the empirical type I (alpha) error has a value of 0 and is thus much too small, leading to an overly and strongly conservative bias in the p-value. For the effect where the true standard deviation of the random slopes is large (with a value of 90), the empirical type I (alpha) error has a value of 0.14, and is thus clearly larger than the normative 5\% level, expressing a liberal bias in the p-value, such that effects turn out significant far too often. This shows that the p-values based on the aggregated analysis are biased and should not be used. The reason for this is that the aggregated analysis cannot separate random slope variances from residual noise. For the fixed effect with the larger random slope variance, the standard error is estimated to be too small, and for the fixed effect with the smaller random slope variance, the standard error is estimated to be too large. In both cases, the random slope variance cannot be estimated. By contrast, the non-aggregated analysis estimates the different random slope variances of both effects and can adjust the standard errors accordingly to obtain empirical type I (alpha) errors of roughly 5\% for both effects despite their differing random slope variances.

One might argue that for the aggregated analysis, the effect with a small random slope variance has a conservative p-value, which can still be used. However, statistical power (i.e., 1 - type II error) for this effect is much reduced compared to the non-aggregated analysis (see Fig.~\ref{fig:sphericityPower}b; the true effect for both contrasts is set to a value of 20 in the simulations). Thus, when sphericity is violated, aggregated analyses can yield either too many false positive results (for the effect with a large random slope variance) or can have low power (for the effect with a small random slope variance).

These analyses demonstrate that when using data aggregation in frequentist LMMs, the empirical type I (alpha) error is not at its nominal value when the sphericity assumption is violated, i.e., the p-values are biased. The safest thing to do is to not aggregate.

\hypertarget{app:SBCspher1}{%
\section{Additional analyses for the first SBC example in issue 1 (simulation 1.1)}\label{app:SBCspher1}}

For the first SBC example in issue 1 (simulation 1.1), we performed additional SBC analyses, with slightly changed priors. Here, the standard deviation of the prior normal distribution characterizing the random slopes had a value of 50 (as the prior in the exemplary data analysis) instead of 150 (which was used in the SBC analysis in simulation 1.1).

The results are similar to the analyses with a standard deviation of 150 reported in the main text. There appears to be one difference in the analyses of the non-aggregated data:
There seems to be a small difference in the same direction as in the aggregated analysis; specifically, the frequentist 95\% confidence interval excludes the prior value of \(0.5\), suggesting that there is indeed bias present. This result was supported by a significant difference from the prior value of \(0.5\) in a frequentist t-test (\(t(\) 213 \() =\) 2.45, \(p =\) 0.02), was not significant in a rank-based Wilcoxon test (\(V =\) 11418, \(p =\) 0.93), and a default Bayes factor showed that the result was inconclusive (\(BF_{10}=\) 1.43).
We speculate that if this effect should be real, it may derive from the common prior on the standard deviation of random slopes, which in this simulation was assumed to be a truncated normal distribution with mean \(0\) and standard deviation \(50\), which may lead to a reduced estimate for the standard deviation of the effect with a large standard deviation (of \(90\)), thus, underestimating the variance and leading to a liberal bias in the Bayes factor analysis. Despite this potential mechanism, the difference seen visually is small. This interpretation is supported by the analyses in the main text, where the prior standard deviation is larger, and the effect was not present.

These analyses suggest that other causes (such as assuming a too small prior standard deviation for the random slopes) might potentially also introduce bias in Bayes factor analyses, even when running Bayesian LMMs on unaggregated data.

\begin{figure}

{\centering \includegraphics{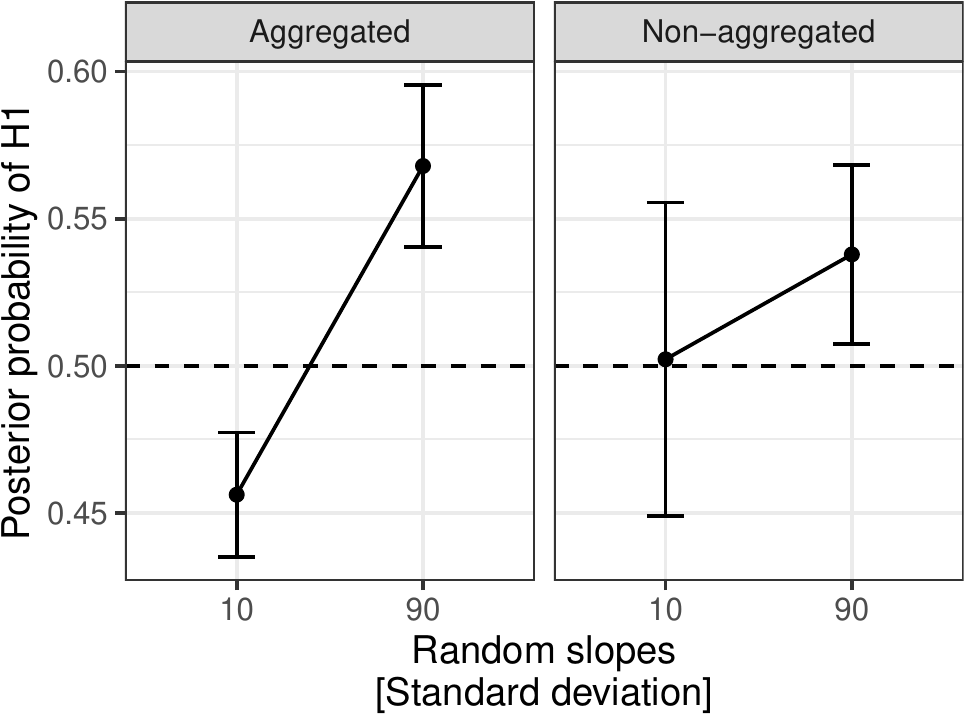} 

}

\caption{Issue 1: Results from SBC for model inference when sphericity is violated, and a smaller prior standard deviation for the random slopes (of 50 instead of 150) is used. The average posterior model probability together with 95 percent confidence intervals is shown for effects with a small (10) versus a large (90) standard deviation of the random slopes, reflecting small or large variation of the effect across subjects. The horizontal broken line is the prior probability for the H1, and deviations from this line indicate estimation bias. Results are shown from null hypothesis Bayes factor analyses based on aggregated (left panel) versus non-aggregated (right panel) data.}\label{fig:sphericitySBCapp}
\end{figure}

\hypertarget{app:SBCPIT3}{%
\section{SBC simulations using a second experimental design with empirically informed priors: Pavlovian-instrumental transfer}\label{app:SBCPIT3}}

We next aimed to test whether our results are specific to our first simulated data set, or whether they also hold for other experimental data from a similar experimental design, but with prior distributions informed by a concrete empirical data set. To this end, we used a data set on Pavlovian-instrumental transfer (PIT, Schad et al., 2020).
The basic finding in PIT studies is that subjects perform more button presses (to instrumentally approach a stimulus) when appetitive Pavlovian conditioned stimuli (CSs; i.e., stimuli previously paired with appetitive outcomes such as the win of money) are presented in the background, and that subjects perform less button presses when aversive CSs (e.g., created by pairing with loss of money) are presented in the background.

We used effect size estimates based on data from Schad et al. (2020) (first \(30\) subjects) to inform prior distribution. The priors were:

\begin{align}
\beta_{(Intercept)}       &\sim Normal(5, 5) \\
\beta_{Contrasts}         &\sim Normal(0, 2) \\
\sigma_{Random \, slopes} &\sim Normal_+(0, 10) \\
\sigma_{Residual}         &\sim Normal_+(0, 10) \\
\rho_{Random \, slopes}   &\sim LKJ(2)
\end{align}

We again assumed that the standard deviations of the random slopes differed from each other, i.e., a violation of the sphericity assumption: for the contrast comparing appetitive versus neutral CSs, we assumed a standard deviation of \(10\), and for the contrast comparing aversive versus neutral CSs, we assumed a standard deviation of \(0\). For the simulations, we used a small number of subjects (\(n_{subj} = 15\)) and a small number of trials per subject and condition (\(n_{rep} = 3\); i.e., a total of \(135\) data points) to speed up computations, and performed \(n_{sim} = 500\) SBC simulations.

\begin{figure}

{\centering \includegraphics{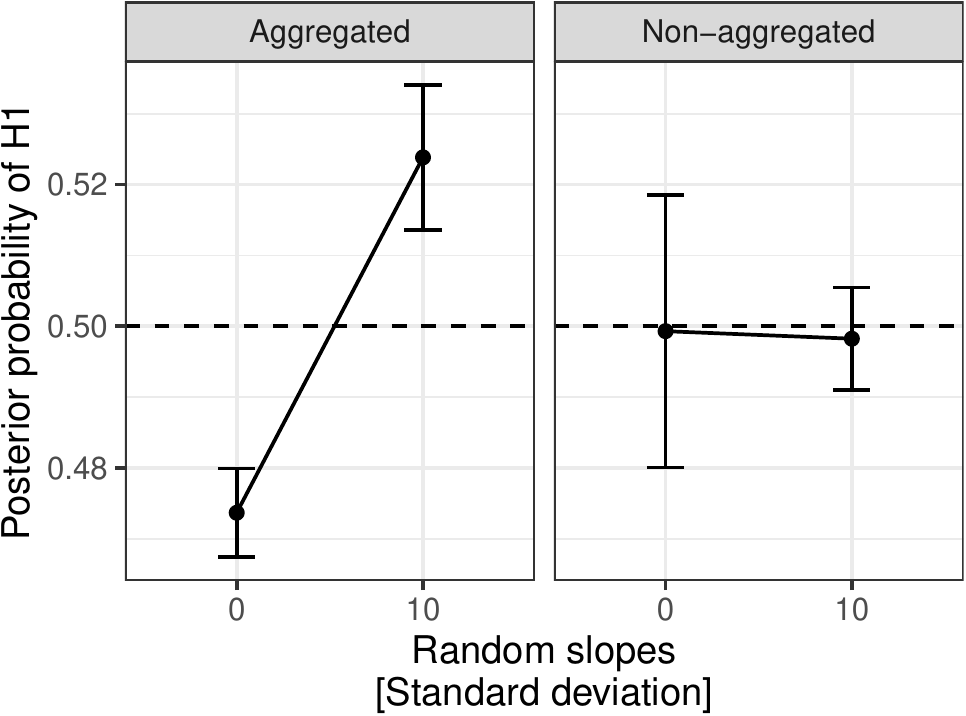} 

}

\caption{Simulation 1.2: A second simulation example on a Pavlovian-instrumental transfer paradigm: results from SBC for model inference when sphericity is violated. The average posterior model probability together with 95 percent confidence intervals is shown for effects with a small (0) versus a large (10) standard deviation of the random slopes, reflecting small or large variation of the effect across subjects. The horizontal broken line is the prior probability for the H1, and deviations from this line indicate estimation bias. Results are shown from null hypothesis Bayes factor analyses based on aggregated data (left panel) versus on non-aggregated data (with random slopes; right panel). They show that aggregating data for null hypothesis tests can lead to biased Bayes factors, which deviate from the true Bayes factor. Bayes factors are more accurate when analyzing non-aggregated data.}\label{fig:sphericitySBCpit}
\end{figure}

\begin{figure}

{\centering \includegraphics{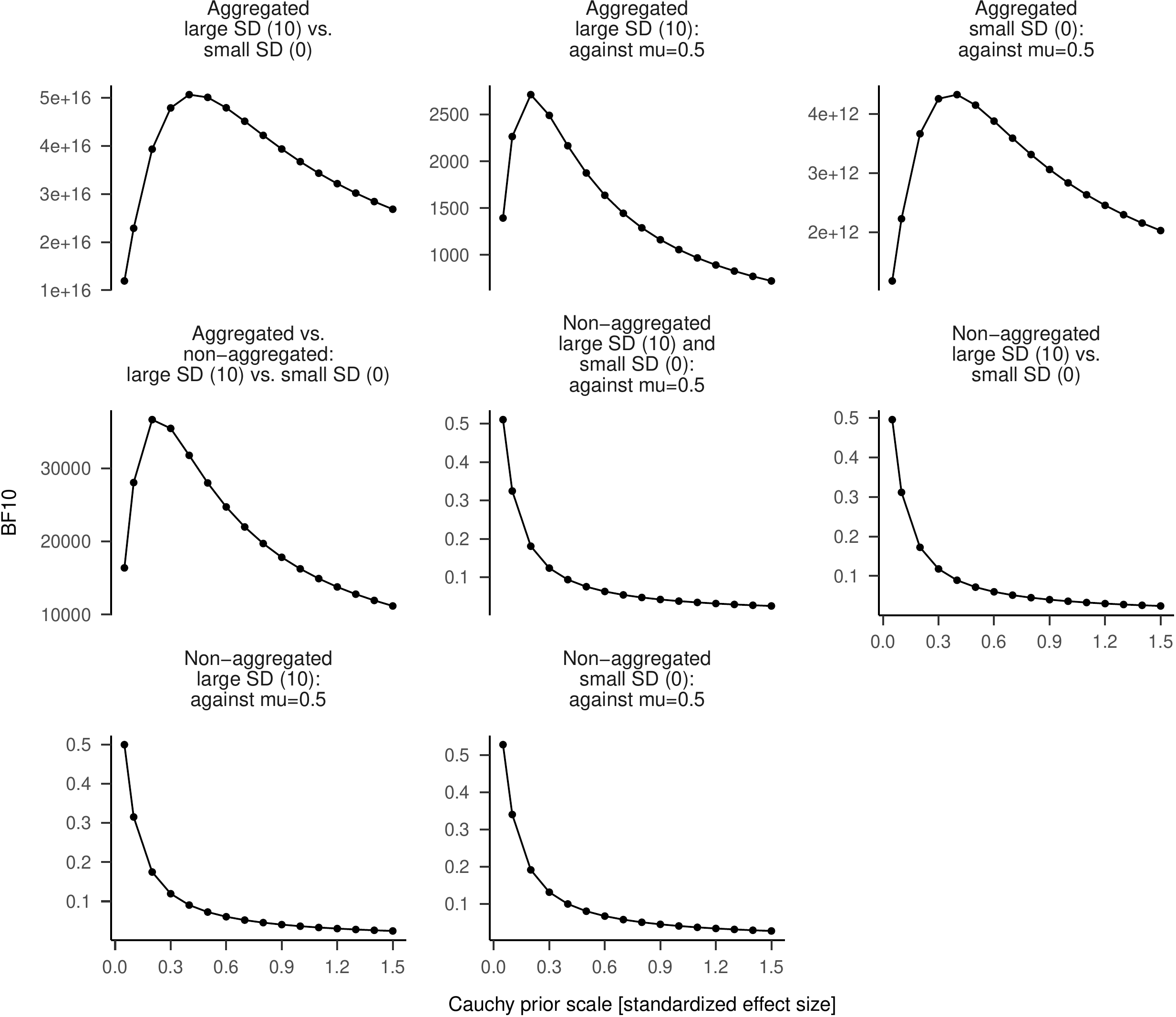} 

}

\caption{Simulation 1.2: Second simulation example on Pavlovian-instrumental transfer: results from SBC for model inference when the sphericity assumption is violated. Shown are results from a sensitivity analysis on the posterior model probabilities, where the prior width is varied across simulations. Results support biased Bayes factors for aggregated analyses, but accurate Bayes factors for non-aggregated analyses. SD = standard deviation; against mu=0.5: comparing the alternative hypothesis against the null that the posterior mean is 0.5.}\label{fig:sphericitySBCpitBF}
\end{figure}

The results (see Fig.~\ref{fig:sphericitySBCpit} and ~\ref{fig:sphericitySBCpitBF}) show that again, Bayes factors are biased in the aggregate analyses (Fig.~\ref{fig:sphericitySBCpit}, left panel), but are unbiased in the non-aggregated analyses, when all random slopes can be estimated (Fig.~\ref{fig:sphericitySBCpit}, right panel).

\hypertarget{app:BayesFactor}{%
\section{Bayesian ANOVA using the BayesFactor package: Sphericity assumption}\label{app:BayesFactor}}

In the \texttt{BayesFactor} package, we used the \texttt{lmBF} function for all analyses. The alternative \texttt{anovaBF} function simply provides a wrapper to the \texttt{lmBF} function.
The \texttt{BayesFactor} package focuses on providing ANOVA-type comparisons, and we therefore use Bayes factors to provide evidence for an alternative model where a factor is included, against a null model, where the full factor is excluded, i.e., all condition means are assumed to be equal.

The Bayesian linear mixed-effects model underlying the Bayesian ANOVA in the \texttt{BayesFactor} package assumes the following linear model:
\(\mathbf{y} = \mu + \sigma_{\epsilon} \mathbf{X} \mathbf{\theta} + \mathbf{\epsilon}\), where \(\mu\) is an intercept parameter capturing the grand mean, \(\mathbf{X}\) is the design matrix, \(\mathbf{\theta}\) is a vector of standardized effect parameters, and \(\mathbf{\epsilon} \sim N(0, \sigma_{\epsilon})\) reflects normally distributed residuals with residual standard deviation \(\sigma_{\epsilon}\).
Thus, the model's coefficients (\(\theta\)) are scaled by the residual standard deviation \(\sigma_{\epsilon}\), thus providing scaled effect sizes.

What is the hierarchical structure and what are the priors for this model's parameters? The model implements a Jeffreys-Zellner-Siow (JZS) Bayes factor. In the JZS Bayes factor, the prior on the intercept \(\mu\) is completely flat, and the prior on the residual noise term \(\sigma_{\epsilon}\) is the noninformative Jeffreys prior \(p(\sigma_{\epsilon}^2) \propto 1/\sigma_{\epsilon}^2\).

For the scaled effect parameters \(\theta\), they are assumed to come from a normal distribution with mean zero: \(\theta \sim Normal(0, \sigma_{\theta})\). The standard deviation of this normal distribution, \(\sigma_{\theta}\), has as a prior distribution a scaled inverse chi-square distribution with one degree of freedom \(\sigma_{\theta} \sim InvChiSquare(df=1, scale=s)\). Combining the normal prior for \(\theta\) with the scaled inverse chi-square distribution for \(\sigma_{\theta}\) implies a scaled Cauchy prior distribution for \(\theta\).

In the design matrix \(\mathbf{X}\), for random effects (e.g., for subjects), each subject has one column, which is coded as \(1\) for the data points of the given subject, and coded \(0\) otherwise.
For fixed effects with \(\alpha\) conditions, each condition is coded with one column (1/0 coding; \(\mathbf{X}_{\alpha}\)). Then, a sum-to-zero constraint is imposed. The covariance matrix is defined as: \(\Sigma_{\alpha} = \mathbf{I}_{\alpha} - \mathbf{J}_{\alpha} / \alpha\); \(\mathbf{I}_{\alpha}\) is the identity matrix, \(\mathbf{J}_{\alpha}\) is a square matrix with \(\alpha\) rows and columns, and all entries set to one. An eigenvalue decomposition yields \(\Sigma_{\alpha} = Q_{\alpha} \mathbf{I}_{\alpha-1} Q_{\alpha}'\), where \(Q_{\alpha}\) is a matrix of eigenvectors. A new parameter vector of \(\alpha-1\) effects is defined by \(\alpha^* = Q_{\alpha}' \alpha\). The new design matrix is \(\mathbf{X}_{\alpha}^* = \mathbf{X}_{\alpha} Q_{\alpha}\). The resulting linear model equation is:
\(\mathbf{y} = \mu + \sigma_{\epsilon} \mathbf{X}_{\alpha}^*\mathbf{\alpha}^* + \mathbf{\epsilon}\).
This parametrization effectively implements a scaled version of Helmert contrasts, where the vector length for each contrast equals \(1\).
For details see Rouder et al. (2012).
For simplicity, here we denote all effect parameters for random effects \(\theta\) and for fixed effects \(\mathbf{\alpha}^*\) simply as \(\theta\), and the design matrix simply as \(\mathbf{X}\).

In the \texttt{BayesFactor} package, by default the prior scale for fixed effects is assumed to be \(0.5\), and the prior scale for random effects is assumed to be \(1\).

Importantly, in the definition of priors (\(\theta \sim Normal(0, \sigma_{\theta})\) or equivalently \(\alpha* \sim Normal(0, \sigma_{\theta})\)), the prior standard deviation \(\sigma_{\theta}\) is assumed to be identical for different effects coding the same factor (i.e., drawn from its prior only once), which effectively implements a multivariate Cauchy prior. This means, that the \texttt{BayesFactor} package assumes that the standard deviations of all random slope terms capturing one factor are identical, i.e., it assumes sphericity for different effects coding the same factor. At this level, it cannot capture differences in variance between contrasts coding the same factor.

For effect size parameters coding the effects of different factors, the prior standard deviation \(\sigma_{\theta}\) is assumed to be independent from each other, effectively coding an independent Cauchy distribution. Thus, the standard deviations of contrasts coding different factors are assumed to be independent. They can thus differ from each other, and differences in random slope variances can be explicitly estimated and modeled.

\paragraph{BayesFactor package: one three-level repeated-measures factor}

We simulated data from a repeated-measures experiment. We again assumed one within-subject factor with three levels, used a design with 30 subjects, and with three repetitions (trials) per subject and per factor level, leading to a total of nine trials per subject. The priors were taken as the default priors from the \texttt{BayesFactor} package.

The true prior scales for the random slopes were set to values of \(0.1\) and \(10\) for the two contrasts. The results were the same when using values of \(0.001\) and \(1\) instead.

\begin{figure}

{\centering \includegraphics{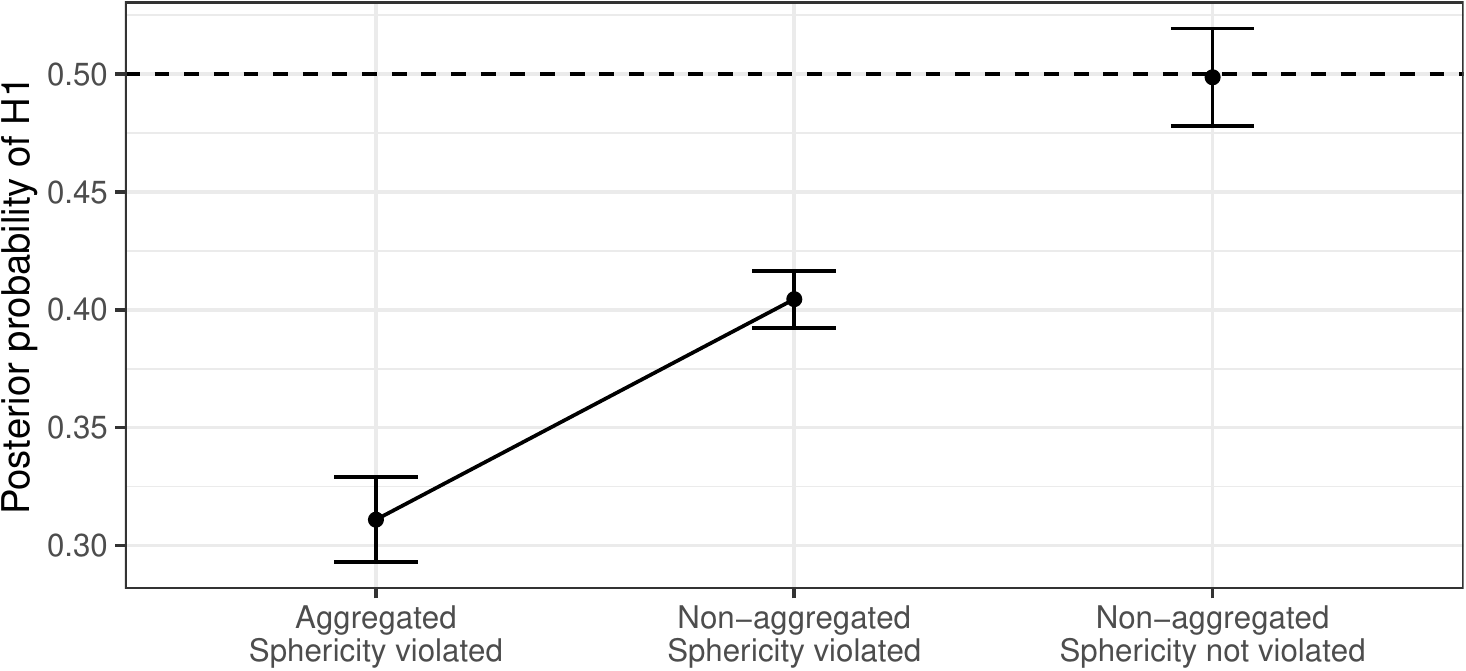} 

}

\caption{Simulation 1.7: Simulation-based calibration for the Bayes factor package, using a simple design with one repeated measures factor with 3 levels. Testing evidence for the factor overall, i.e., testing all contrasts at once, by comparing the full model against a null model, where all fixed-effects contrats are removed. Results show that aggregating data for null hypothesis tests leads to biased Bayes factors that are too small, reflecting a conservative bias. Bias is reduced when fitting the model to non-aggregated data, but it is still present. Bias is absent when the sphericity assumption is not violated - this is shown in the right-most data point. Important: this data point uses a different simulated data set - with no violation of sphericity - whereas the left and middle points use data where the sphericity assumption is violated.}\label{fig:sphericitySBC-BayesFactor-F3}
\end{figure}

\begin{figure}

{\centering \includegraphics{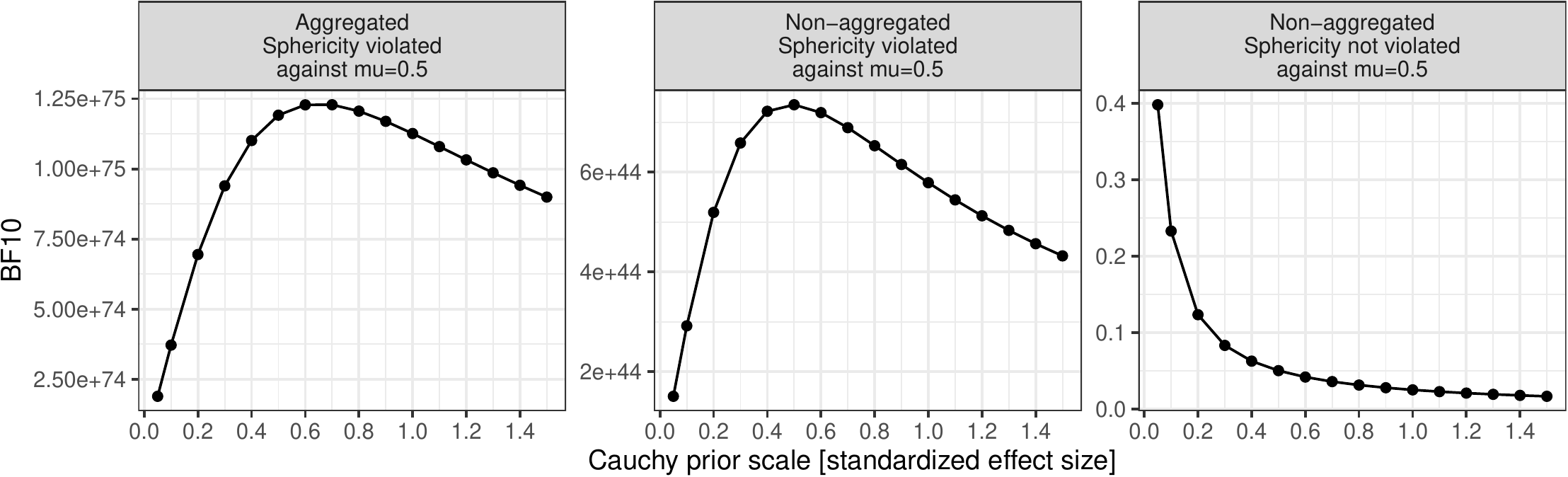} 

}

\caption{Simulation 1.7: Data are on a repeated measures factor with 3 levels. Testing evidence for a factor overall, i.e., testing all contrasts at once, by comparing the full model against a null model, where all fixed-effects contrats are removed. Results support accurate Bayes factors when the sphericity assumption is not violated, but conservative bias otherwise.}\label{fig:sphericitySBC-BayesFactor-F3-BF}
\end{figure}

The results showed that when the sphericity assumption was violated, Bayes factors estimated from aggregated data were too small, i.e., they had a conservative bias (see Fig.~\ref{fig:sphericitySBC-BayesFactor-F3} and Fig.~\ref{fig:sphericitySBC-BayesFactor-F3-BF}). This bias was reduced when the model was fit to non-aggregated data. However, conservative bias was still present and Bayes factors were not accurate. This may be the case because - as stated above - Bayesian ANOVA in the \texttt{BayesFactor} package assumes that for all contrasts coding the same factor, random slope variances are the same. I.e., the model does not allow us to estimate differences in random slopes between contrasts of the same factor, and thus even the non-aggregated analysis assumes sphericity. Note that the bias is still reduced compared to non-aggregated analyses. One possible reason may be the following: our non-aggregated analyses involve estimation of random slopes. In this case, random effects variance components (characterizing the estimated variance in the population) are assumed to be identical between contrasts, reflecting the sphericity assumption. However, the model also estimates individual random effects for each subject. These individual random effects estimates are not constrained to follow the group-level variance. Instead, they may capture some differences in variances. This may reduce the bias of this analysis compared to our aggregated analyses, where no random slopes were estimated.

Last, in simulations with a different data set built so that sphericity is not violated, we found no bias in Bayes factor estimation, validating our SBC simulations as well as the computation of Bayes factors in the \texttt{BayesFactor} package.

\paragraph{BayesFactor package: a 2 $\times$ 2 repeated-measures design}

Next, we used SBC for the \texttt{BayesFactor} package for analyzing a 2 \(\times\) 2 repeated measures design. For the simulations, we assumed four repetitions for each of the four design cells, yielding 16 data points for each of 30 subjects. As our previous analysis of the 2 \(\times\) 2 design (see the two-step task), we fitted models (a) to aggregated data without using random slopes, (b) to aggregated data with random slopes for the main effects, and (c) to non-aggregated data using the full random effects structure. As before, we assumed that the true random slope variances were small for the two main effects (prior scales of \(0.001\)), and were large for the interaction term (prior scale of \(1\)).

\begin{figure}

{\centering \includegraphics{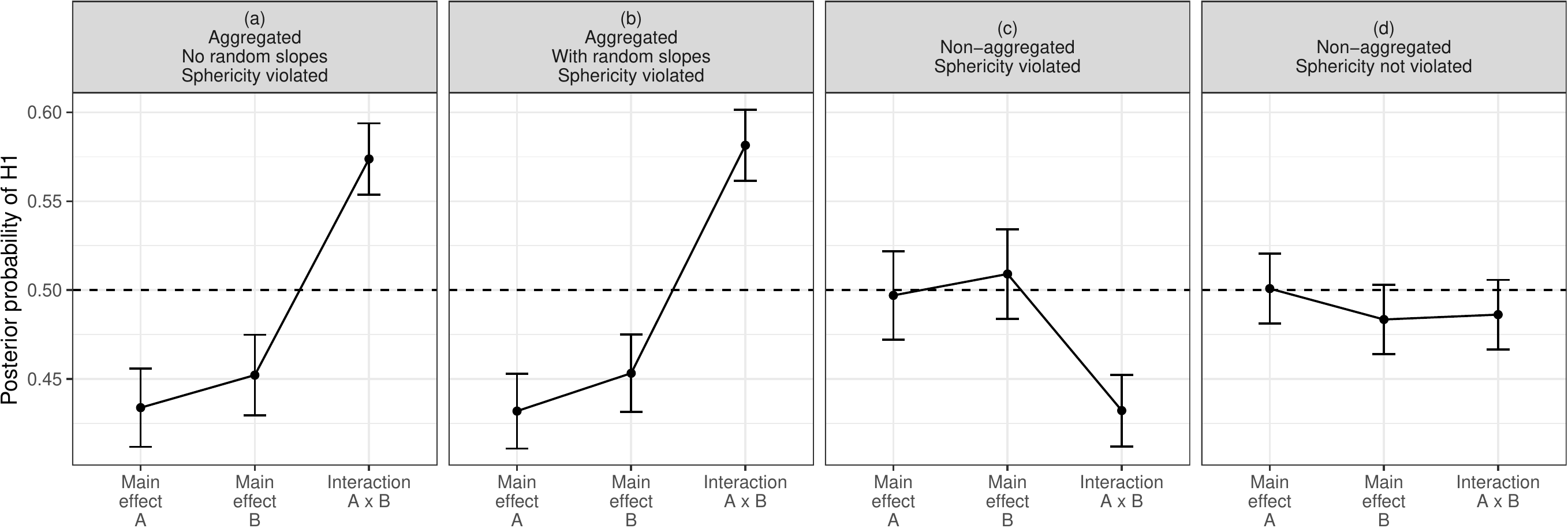} 

}

\caption{Simulation 1.8: Simulation-based calibration for the Bayes factor package, using a 2 x 2 repeated measures design. True random slope variances are simluated to be small for the main effects and large for the interaction. Results show that aggregating data for null hypothesis tests leads to biased Bayes factors that are too small for the main effects, reflecting a conservative bias. For the interaction, the Bayes factors are too large, reflecting a liberal bias. This holds true when fitting the model on aggregated data without random slopes and when including random slopes for the main effects. When fitting the model to non-aggregated data and including all random slopes, there is no bias for main effects; for the interaction, Bayes factors are too small. No bias is present when the sphericitiy assumption is not violated.}\label{fig:sphericitySBC-BayesFactor-2x2}
\end{figure}

\begin{figure}

{\centering \includegraphics{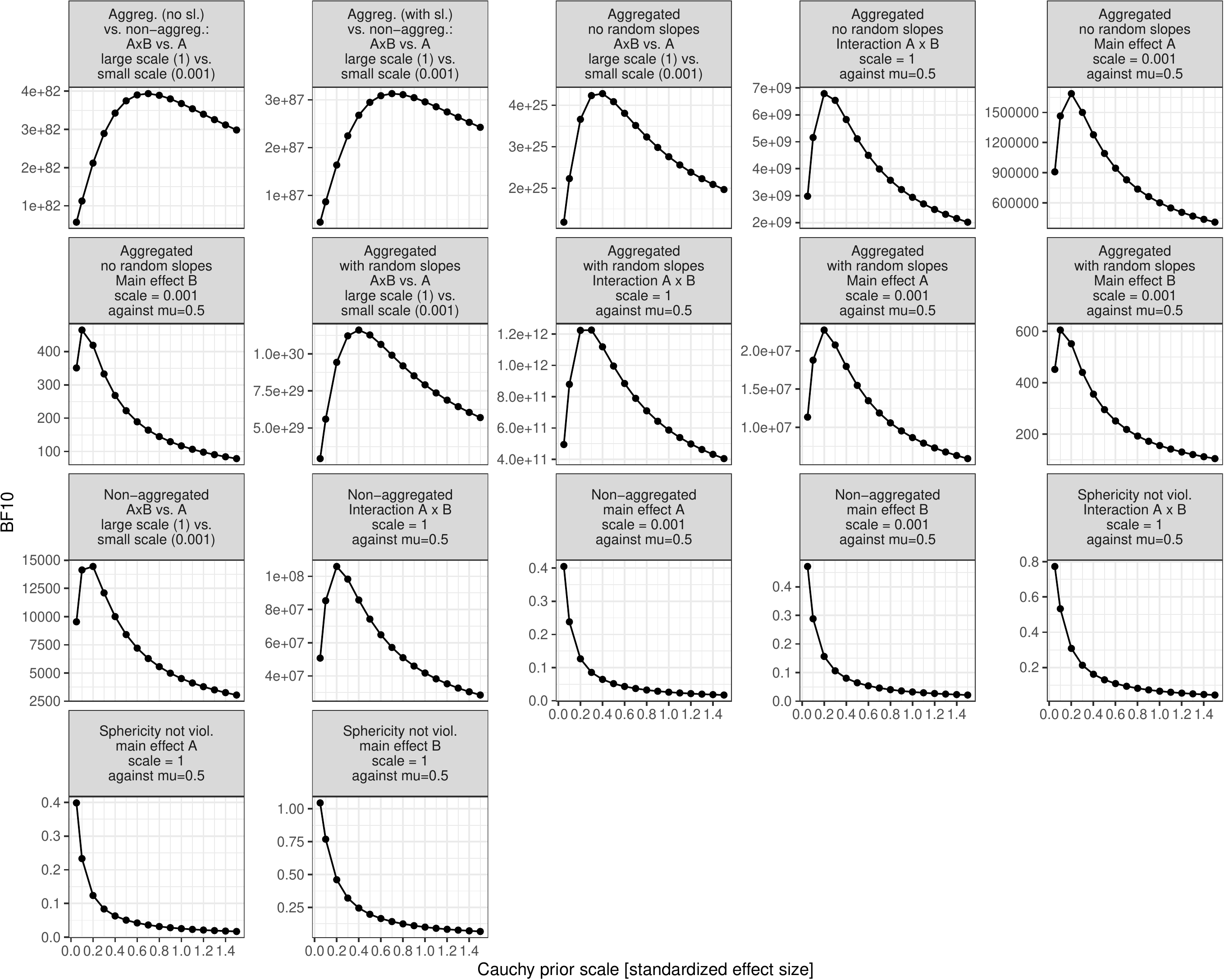} 

}

\caption{Simulation 1.8: SBC using a 2 x 2 repeated measures design. Shown are results from sensitivity analyses on the posterior model probabilities, where the prior width is varied across simulations. Results support biased Bayes factors for aggregated analyses, but accurate Bayes factors for non-aggregated analyses. Aggr. = aggregated; non-aggr. = non-aggregated; AxB = interaction between factors A and B; A = main effect of factor A; scale = prior scale; no sl. = no random slopes; with sl. = with random slopes; not viol. = not violated; ME = main effects.}\label{fig:sphericitySBC-BF-2x2}
\end{figure}

The results (see Fig.~\ref{fig:sphericitySBC-BayesFactor-2x2} and Fig.~\ref{fig:sphericitySBC-BF-2x2}) using \(1000\) simulations were similar to those from the \texttt{brms} package reported above (see analysis of the two-step task): when estimating Bayes factors from the aggregated data (Fig.~\ref{fig:sphericitySBC-BayesFactor-2x2}, panels a + b), the Bayes factor estimates for the main effects were too small, reflecting conservative tests - and this was true irrespective of whether random slopes for the main effects were included in the model. At the same time, the Bayes factor estimates for the interaction term were too large, reflecting anti-conservative tests.

The biases for the main effects were absent when fitting the model on non-aggregated data and estimating the full random effects structure (Fig.~\ref{fig:sphericitySBC-BayesFactor-2x2}, panel c). This may reflect the fact that in the \texttt{BayesFactor} package, random slope variances are estimated separately for main effects stemming from different factors, such that violations of sphericity can be accounted for by the model. However, Bayes factors for the interaction term were too small, reflecting a conservative bias.

To validate our procedure and the \texttt{BayesFactor} package, we also ran analyses, where sphericity was assumed in the simulation (i.e., no deviation from sphericity). These analyses showed (Fig.~\ref{fig:sphericitySBC-BayesFactor-2x2}, panel d) that no bias was present when the sphericity assumption was not violated in the simulated data, and non-aggregated analyses were performed.

\hypertarget{app:FreqItem}{%
\section{Item aggregation: Frequentist simulations}\label{app:FreqItem}}

It is well known for frequentist repeated measures ANOVA that data aggregation (per subject and condition) in the presence of item variance leads to an inflation of type I (alpha) error (Clark, 1973; Forster \& Dickinson, 1976). Here, we demonstrate that this problem occurs not only for repeated measures ANOVA, but also when fitting LMMs to aggregated data. The traditional procedure for dealing with this situation is to run repeated measures ANOVA not only with subjects as random factor (\(F1\)), but to additionally run a second ANOVA with items as random factor (\(F2\)). A min-\(F'\) value then provides the traditional frequentist test statistic, which combines results from both (\(F1\) and \(F2\)) analyses. Here, we demonstrate that in LMMs, this problem can be solved in a more elegant way by fitting the data on the individual-trial level, and by estimating random effects for subjects and for items simultaneously, providing a single test statistic (i.e., a single t-/chi-square-value) for a given effect.

One approach to this data is to ignore item variability, and to perform an LMM analysis using aggregated data: this approach relies on computing the mean reading time for each subject and condition and fitting an LMM (R-packages \texttt{lme4} / \texttt{lmerTest}) with a fixed effect for factor \texttt{X} to the log-transformed averaged reading times. As discussed above, in the aggregated data it is not possible to estimate random slopes, leaving only random intercepts for subjects in the estimation procedure. The results from this analysis show a highly significant effect of factor \texttt{X} (\(b=\) -0.10, \(SE=\) 0.03, \(t=\) -3.70, \(p < .001\)). An alternative is to model item variance explicitly in an LMM analysis of the non-aggregated data, which includes random intercepts and slopes for subjects and also for items. With such consideration of item variance, the fixed effect of factor \texttt{X} is no longer significant (\(b=\) -0.10, \(SE=\) 0.05, \(t=\) -1.90, \(p=\) 0.08).

This exemplary data set leaves unclear which of the two p-values is correct. This can be tested via empirical type I (\(\alpha\)) error simulations, where data are simulated based on the null hypothesis of no (fixed) effect of factor \texttt{X}. Then, the effect of factor \texttt{X} should be significant in \(5\)\% of the simulated data sets at a significance level of \(\alpha = 0.05\). We performed \(1000\) such simulations, while varying the true standard deviation of the random item slopes from \(0\) to \(0.5\).

\begin{figure}

{\centering \includegraphics{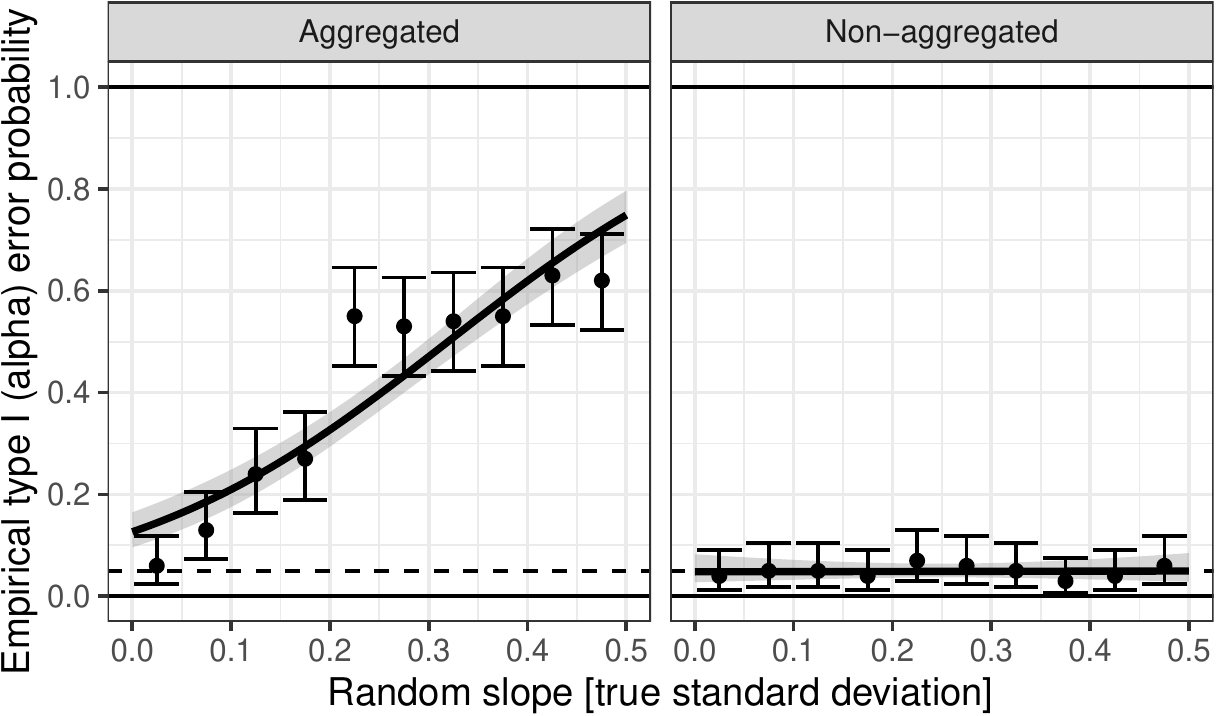} 

}

\caption{The empirical type I (alpha) error is shown as a function of the true standard deviation of random item slopes for linear mixed-effects models fitted to data aggregated to the by-subject level (left panel), effectively ignoring item variance, versus to non-aggregated data (right panel), where item random effects are explicitly modeled. The x-axis represents the range of variation observed in data sets we have analyzed in the past. Regression lines are from a logistic regression; error bars show 95 percent confidence intervals.}\label{fig:itemAlpha}
\end{figure}

The results from this analysis (see Fig.~\ref{fig:itemAlpha}, left panel) show that for aggregated analyses, the empirical type I (\(\alpha\)) error steadily increases with increasing random slope variance (\(b=\) 6.04, \(SE=\) 0.52, \(z=\) 11.60, \(p < .001\)). Thus, if the random slope variance is high, the data analysis based on aggregated data will likely yield a significant result even when no effect is present. By contrast, if non-aggregated analysis is used, and item variance is explicitly modeled via item random effects, then the empirical type I (\(\alpha\)) error does not increase with increasing random item slope variance (\(b=\) 0.07, \(SE=\) 1.01, \(z=\) 0.10, \(p=\) 0.95), and on average does not differ from the nominal \(5\)\%-level (\(\chi^2(1)=\) 0.02, \(p=\) 0.88), demonstrating that the p-value for non-aggregated data yields an unbiased result.

\hypertarget{app:ItemBayesFactor}{%
\section{Item aggregation: BayesFactor package}\label{app:ItemBayesFactor}}

\paragraph{Item aggregation in the BayesFactor package, using a two-level repeated measures factor with crossed random effects for subjects and items}

\begin{figure}

{\centering \includegraphics{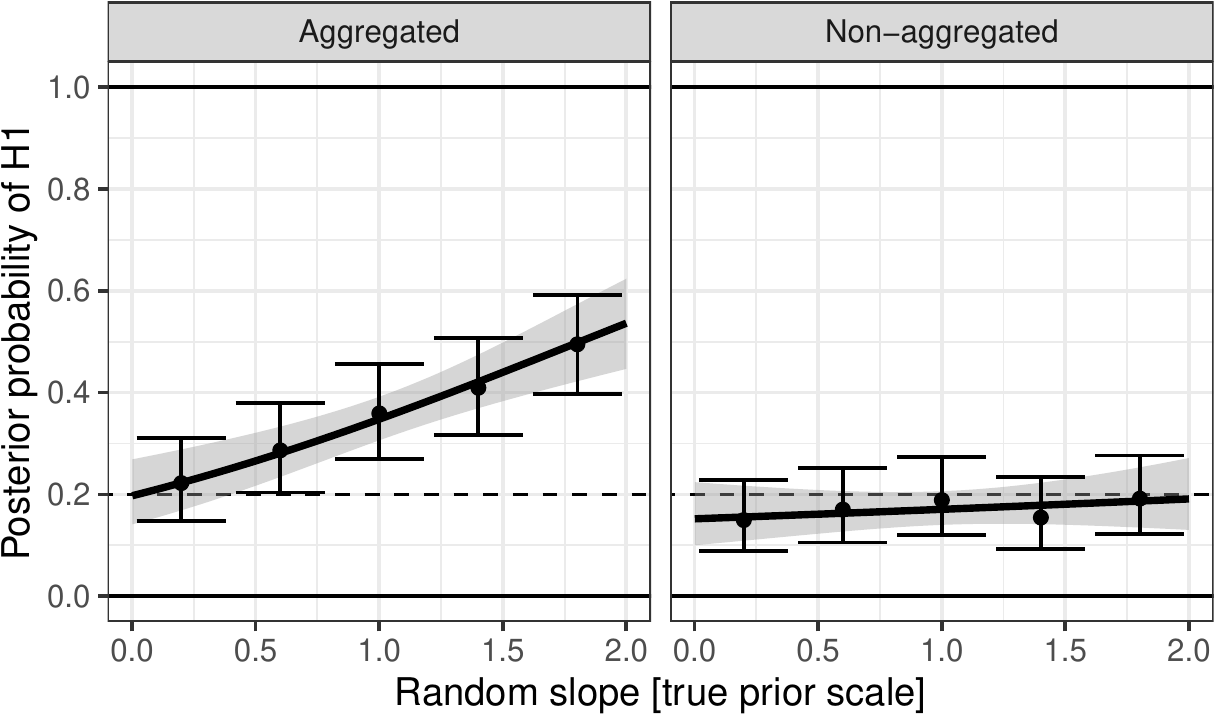} 

}

\caption{Simulation 2.2: Repeating the previous analyses using the BayesFactor package. Results from SBC: the posterior probability for the alternative hypothesis (H1) is shown as a function of the true standard deviation of the item random slope. Results are shown for analysis based on by-subject aggregated data (left panel) and for analysis using non-aggregated data (right panel). Error bars are confidence intervals, regression lines are from a logistic regression. The prior probability for the null is set to 0.2. The results show that in aggregated analyses the posterior probability for H1 increases with larger random slope variances, leading to a deviation from the prior, demonstrating that Bayes factor estimates are too large when item variance is ignored in aggregated analyses. Bayes factors are accurate in non-aggregated analyses.}\label{fig:itemSBC-BF-so}
\end{figure}

\begin{figure}

{\centering \includegraphics{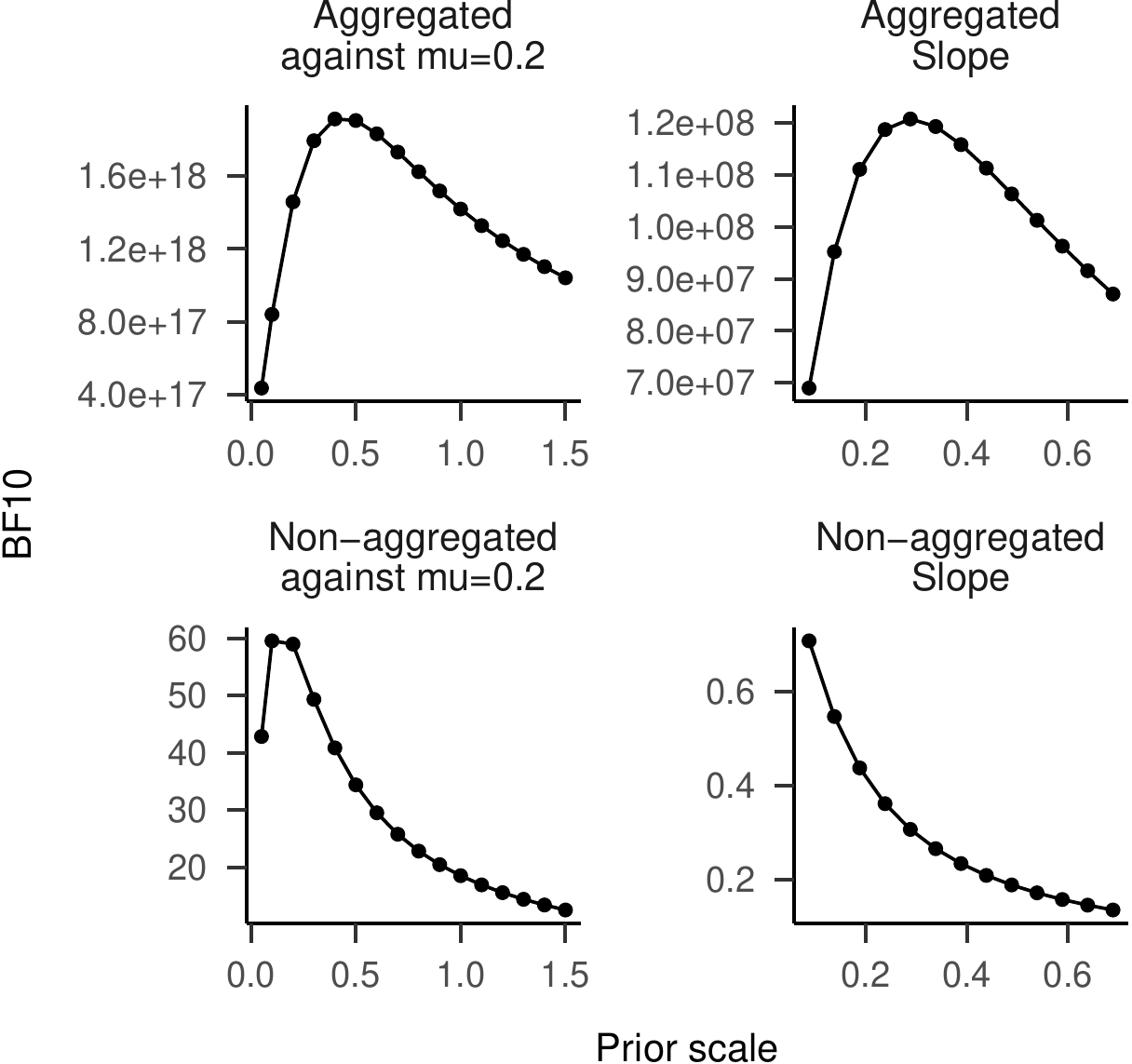} 

}

\caption{Simulation 2.2: BayesFactor package: Results from SBC for model inference when item variance is present. Shown are results from a sensitivity analysis on the posterior model probabilities, where the prior scale of the random item slopes is varied across simulations. Results suggest biased Bayes factors for aggregated analyses, but relatively more accurate Bayes factors for non-aggregated analyses.}\label{fig:dpri-item-so-BF}
\end{figure}

We next used the same experimental design as above (Gibson \& Wu, 2013) to investigate the effects of aggregating across items when using the \texttt{BayesFactor} package to estimate Bayes factors. We found similar results as with the \texttt{brms} package: specifically, we found (see Fig.~\ref{fig:itemSBC-BF-so} and Fig.~\ref{fig:dpri-item-so-BF}) using \(500\) simulations that when the true prior scale for the random item slopes was close to zero, then Bayes factors estimated from aggregated and from non-aggregated analyses were close to the nominal value of \(0.2\). However, when the true prior scale for the random item slopes increased to values larger than zero, then Bayes factors estimated from aggregated analyses also increased beyond the prior value of \(0.2\), to much higher values, reflecting a liberal / anti-conservative bias in the Bayes factor estimates. By contrast, when estimating Bayes factors from non-aggregated data, and estimating the full random effects structure, then estimated Bayes factors did not increase with increasing true prior scales for the item random slopes, indicating unbiased estimates. (However, Bayes factor analyses of the SBC results provided evidence that the Bayes factors exhibited a small but reliable conservative bias, see Fig.~\ref{fig:dpri-item-so-BF}.)

Next, we were interested in looking at more complex designs with random slopes for subjects and items. Accordingly, we extended the analyses to situations of a four-level factor and to a \(2 \times 2\) design.

\paragraph{Item aggregation in the BayesFactor package using a four-level factor}

As a next step, we used the same design (now using 40 subjects and 20 items) as in the previous simulations, but instead of having a repeated-measures factor with two levels, we expanded this to have a factor with four levels. We used the \texttt{BayesFactor} package to test the effect of the whole factor by comparing an alternative model with the factor included to a null model where the fixed-effect of the factor was removed. While the previous analyses focussed on individual contrast coefficients (i.e., capturing the comparison between two means), this example of a four-level factor implements a Bayesian omnibus ANOVA-type test. For the model parameters, we used the default priors of the \texttt{BayesFactor} package.

\begin{figure}

{\centering \includegraphics{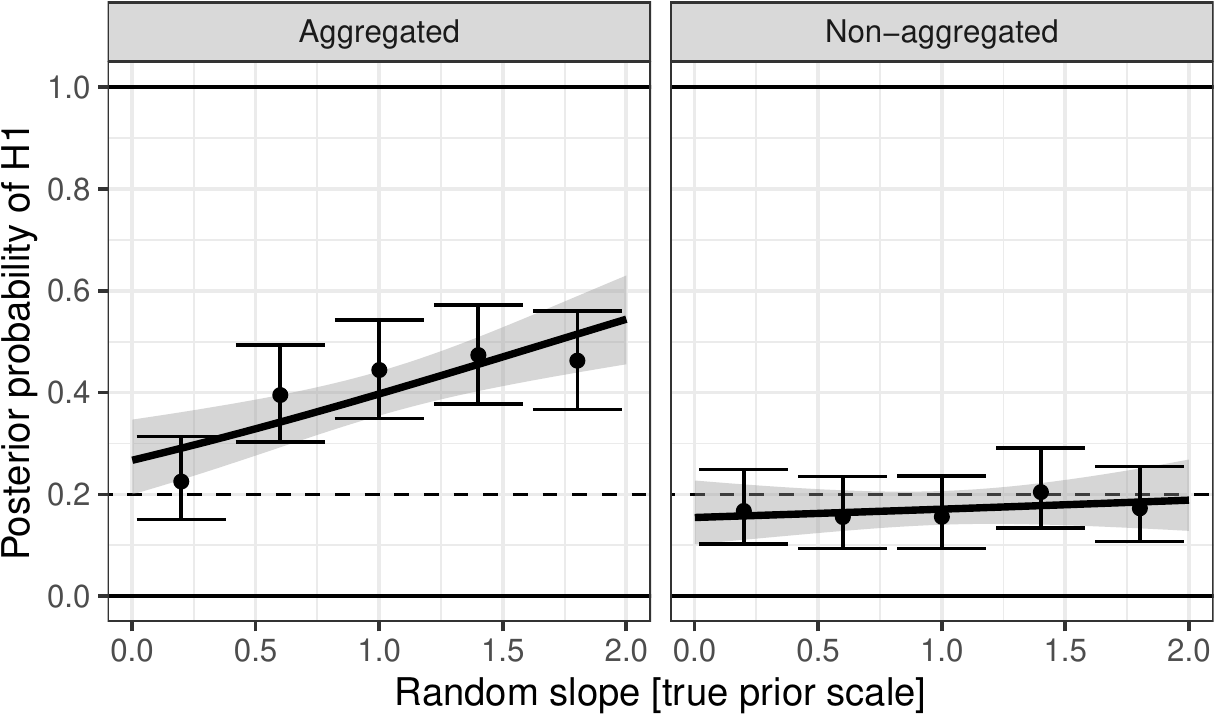} 

}

\caption{Simulation 2.3: Repeating the previous analyses using a 4-level factor, using an omnibus ANOVA-type Bayesian test. Results from SBC: the posterior probability for the alternative hypothesis (H1) is shown as a function of the true standard deviation of the item random slope. Results are shown for analysis based on by-subject aggregated data (left panel) and for analysis using non-aggregated data (right panel). Error bars are confidence intervals, regression lines are from a logistic regression. The prior probability for the null is set to 0.2. The results show that the posterior probability for H1 increases with larger random slope variances, leading to a deviation from the prior, demonstrating that Bayes factor estimates are too large when item variance is ignored in aggregated analyses.}\label{fig:itemSBC-BF-F4}
\end{figure}

\begin{figure}

{\centering \includegraphics{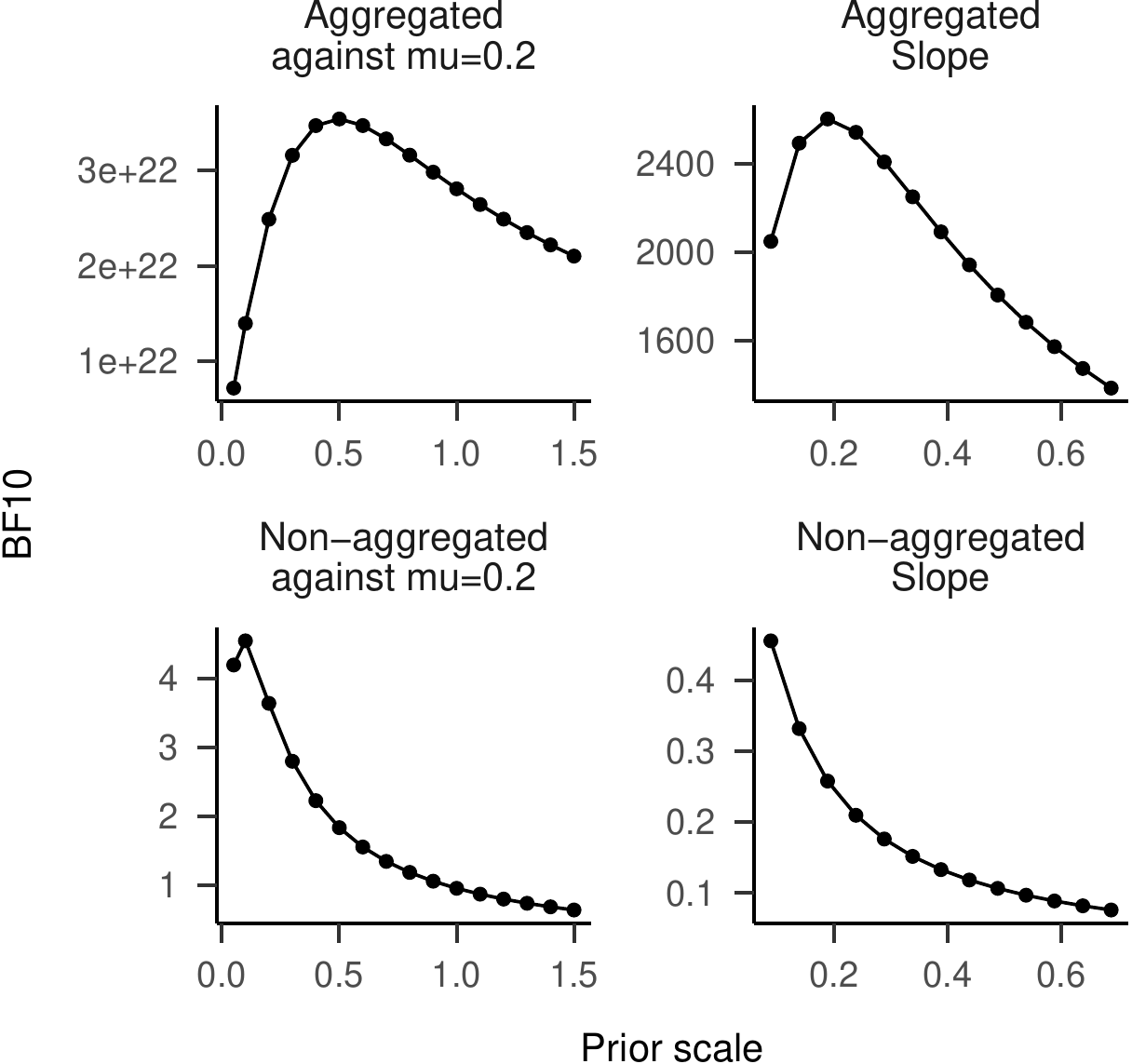} 

}

\caption{Simulation 2.3: BayesFactor package: Results from SBC for model inference when item variance is present. Shown are results from a sensitivity analysis on the posterior model probabilities, where the prior scale of the random item slopes is varied across simulations. Results suggest biased Bayes factors for aggregated analyses, but relatively more accurate Bayes factors for non-aggregated analyses.}\label{fig:dpri-item-F4-BF}
\end{figure}

The results (see Fig.~\ref{fig:itemSBC-BF-F4} and Fig.~\ref{fig:dpri-item-F4-BF}) using 500 simulations showed that when aggregating over items in the analysis despite the presence of random item slopes, then estimated Bayes factors were too large, i.e., the average posterior was larger than the prior of \(0.2\), suggesting a liberal bias. Moreover, when random item slope variance increased, this liberal bias also increased, suggesting the bias when testing the fixed effect was particularly large when large random slope variances are present.

By contrast, when non-aggregated analyses were used and item variance was explicitly modeled, then Bayes factor estimates were unbiased, i.e., the posterior mean did not differ from the prior of \(0.2\). (In fact, Fig.~\ref{fig:dpri-item-F4-BF} shows that Bayes factors were slightly too small.) Importantly, in non-aggregated analyses, increasing random item slope variance did not induce a liberal bias (i.e., no increased average posterior model probabilities), in contrast to the aggregated analyses.

\hypertarget{app:SensAnal}{%
\section{Sensitivity analyses}\label{app:SensAnal}}

\begin{figure}

{\centering \includegraphics{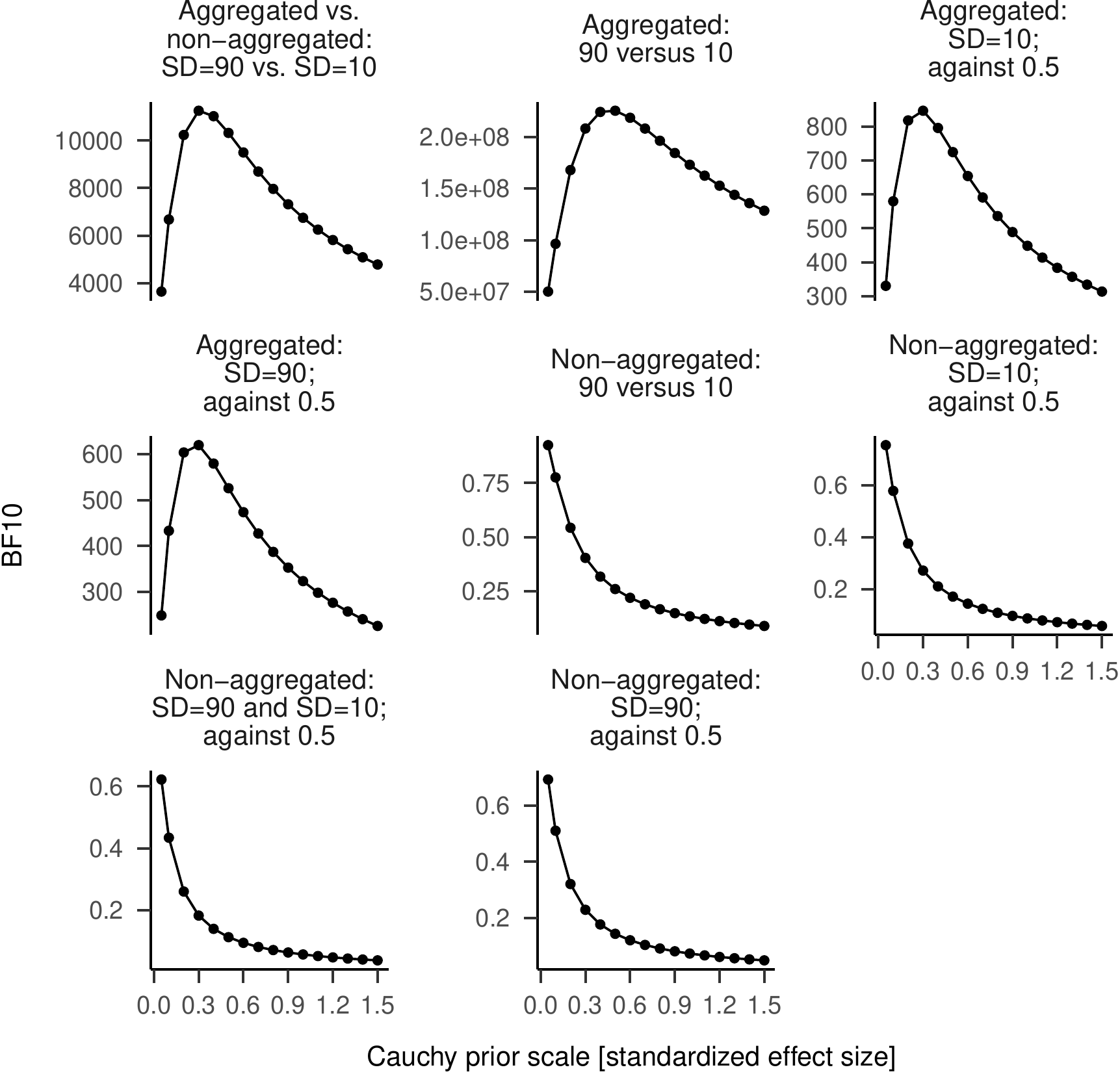} 

}

\caption{Simulation 1.1: Results from SBC for model inference when the sphericity assumption is violated. Shown are results from a sensitivity analysis on the posterior model probabilities, where the prior width is varied across simulations. Results support biased Bayes factors for aggregated analyses, but accurate Bayes factors for non-aggregated analyses. SD = standard deviation, against 0.5 = comparing an alternative hypothesis against the null hypothesis that the posterior mean is 0.5.}\label{fig:sphericitySBCBF}
\end{figure}

\begin{figure}

{\centering \includegraphics{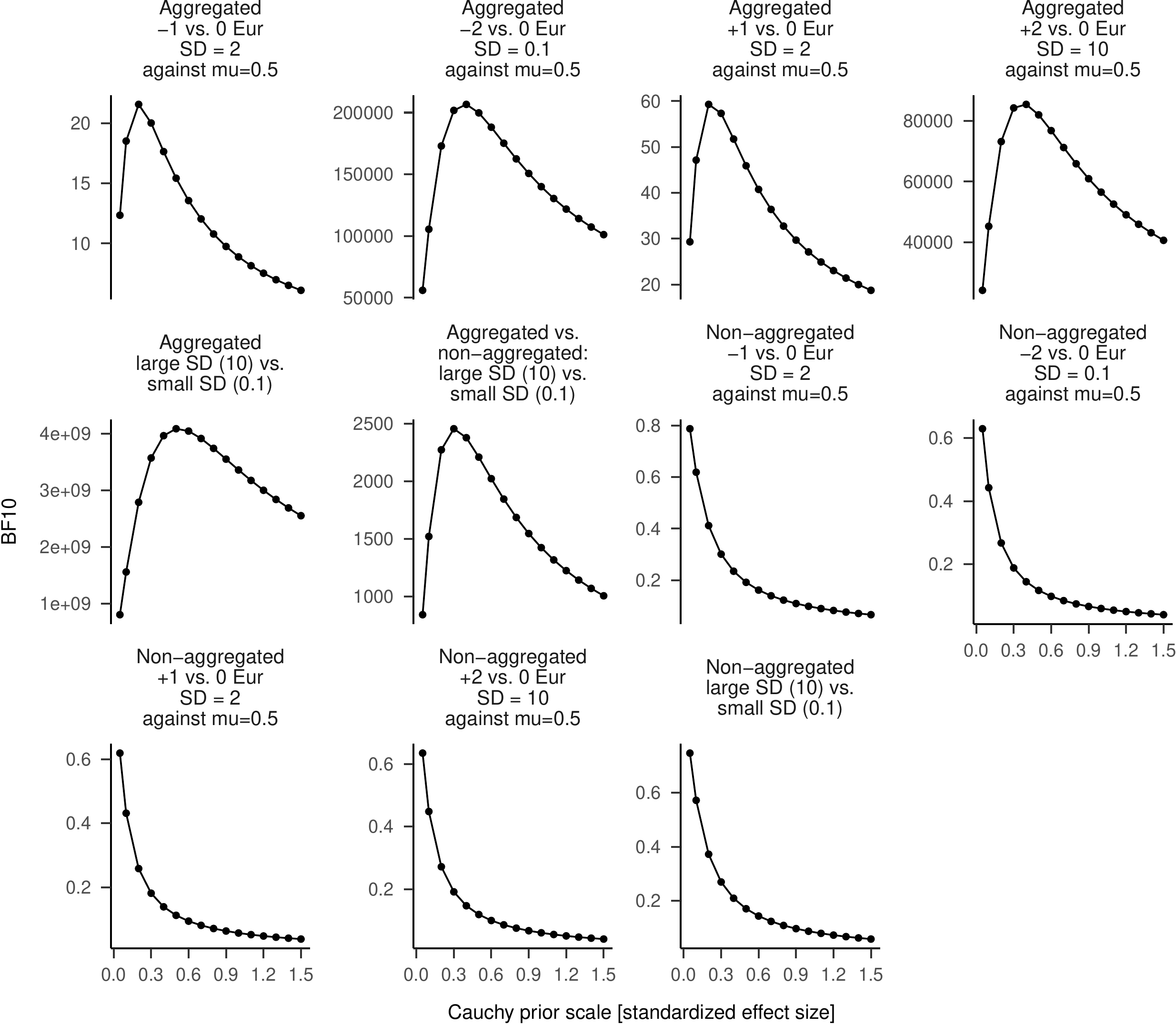} 

}

\caption{Simulation 1.3: Second simulation example on Pavlovian-instrumental transfer, now with 5 factor levels: results from SBC for model inference when the sphericity assumption is violated. Shown are results from a sensitivity analysis on the posterior model probabilities, where the prior width is varied across simulations. Results support biased Bayes factors for aggregated analyses, but accurate Bayes factors for non-aggregated analyses. SD = standard deviation; against mu=0.5: comparing the alternative hypothesis against the null that the posterior mean is 0.5.}\label{fig:sphericitySBCpit5BF}
\end{figure}

\begin{figure}

{\centering \includegraphics{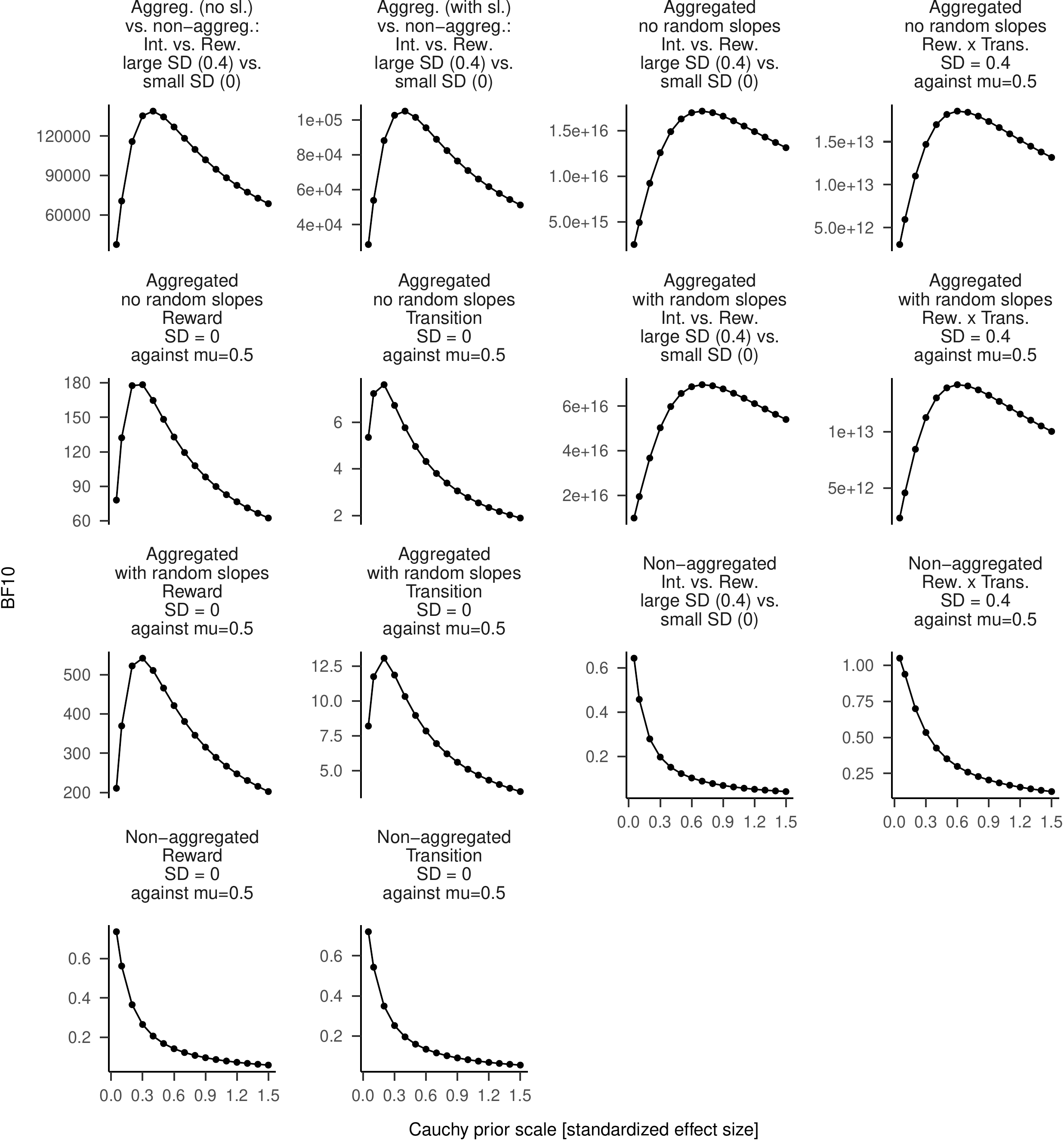} 

}

\caption{Simulation 1.4: Simulations of the two-step task, using a 2 x 2 design: results from SBC for model inference when the sphericity assumption is violated. Shown are results from a sensitivity analysis on the posterior model probabilities, where the prior width is varied across simulations. Results support biased Bayes factors for aggregated analyses, but accurate Bayes factors for non-aggregated analyses. Aggreg. = aggregated; non-aggreg. = non-aggregated; no sl. = no random slopes; with sl. = with random slopes for main effects; Int. = Interaction reward x transition; Rew. = reward; Trans. = transition; SD = standard deviation.}\label{fig:sphericitySBC2stepBF}
\end{figure}

\begin{figure}

{\centering \includegraphics{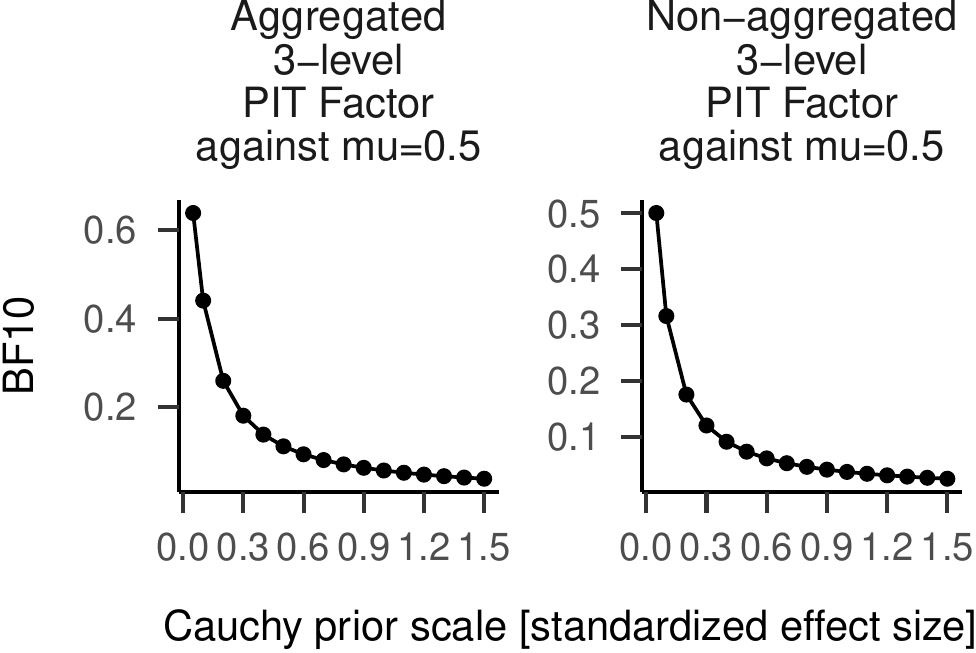} 

}

\caption{Simulation 1.5: Testing evidence for a factor overall, i.e., testing all contrasts at once, by comparing the full model against a null model, where all fixed-effects contrasts are removed. Data are from the simulation reported above, i.e., on Pavlovian-instrumental transfer, with three factor levels. Results support accurate Bayes factors overall.}\label{fig:sphericitySBCpit3FBF}
\end{figure}

\begin{figure}

{\centering \includegraphics{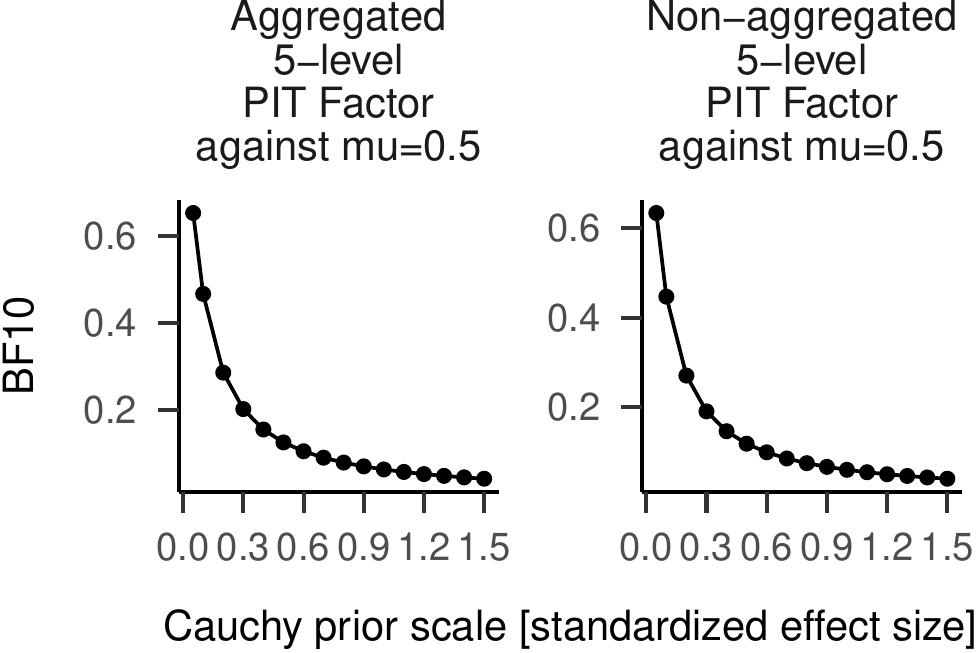} 

}

\caption{Simulation 1.6: Data are on Pavlovian-instrumental transfer with five factor levels. Testing evidence for a factor overall, i.e., testing all contrasts at once, by comparing the full model against a null model, where all fixed-effects contrats are removed. Results support accurate Bayes factors overall.}\label{fig:sphericitySBCpit5FBF}
\end{figure}

\begin{figure}

{\centering \includegraphics{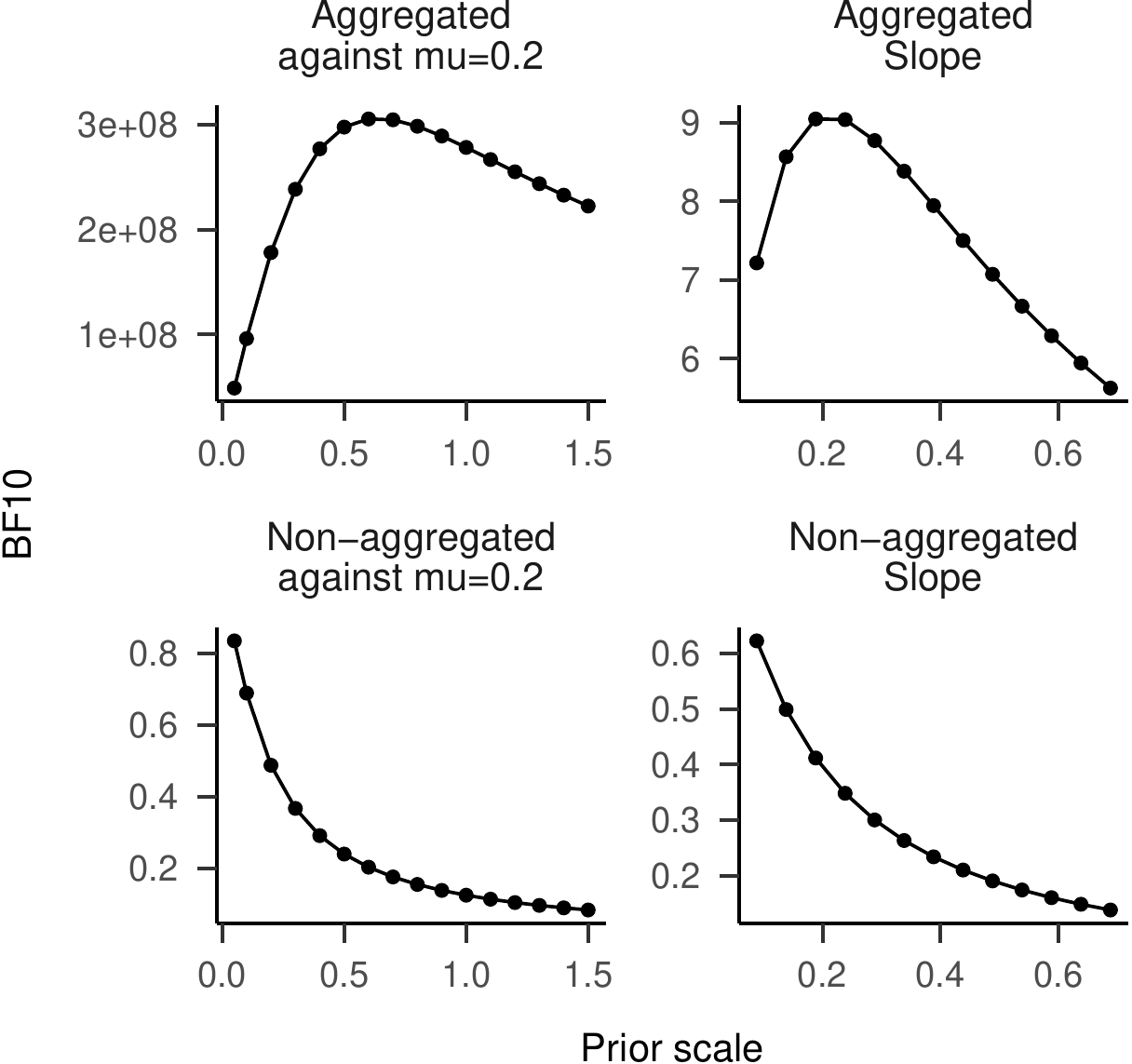} 

}

\caption{Simulation 2.1: Results from SBC for model inference when item variance is present. Shown are results from a sensitivity analysis on the posterior model probabilities, where the prior scale of the random item slopes is varied across simulations. Results suggest biased Bayes factors for aggregated analyses, but relatively more accurate Bayes factors for non-aggregated analyses.}\label{fig:dpri-item-so}
\end{figure}

\begin{figure}

{\centering \includegraphics{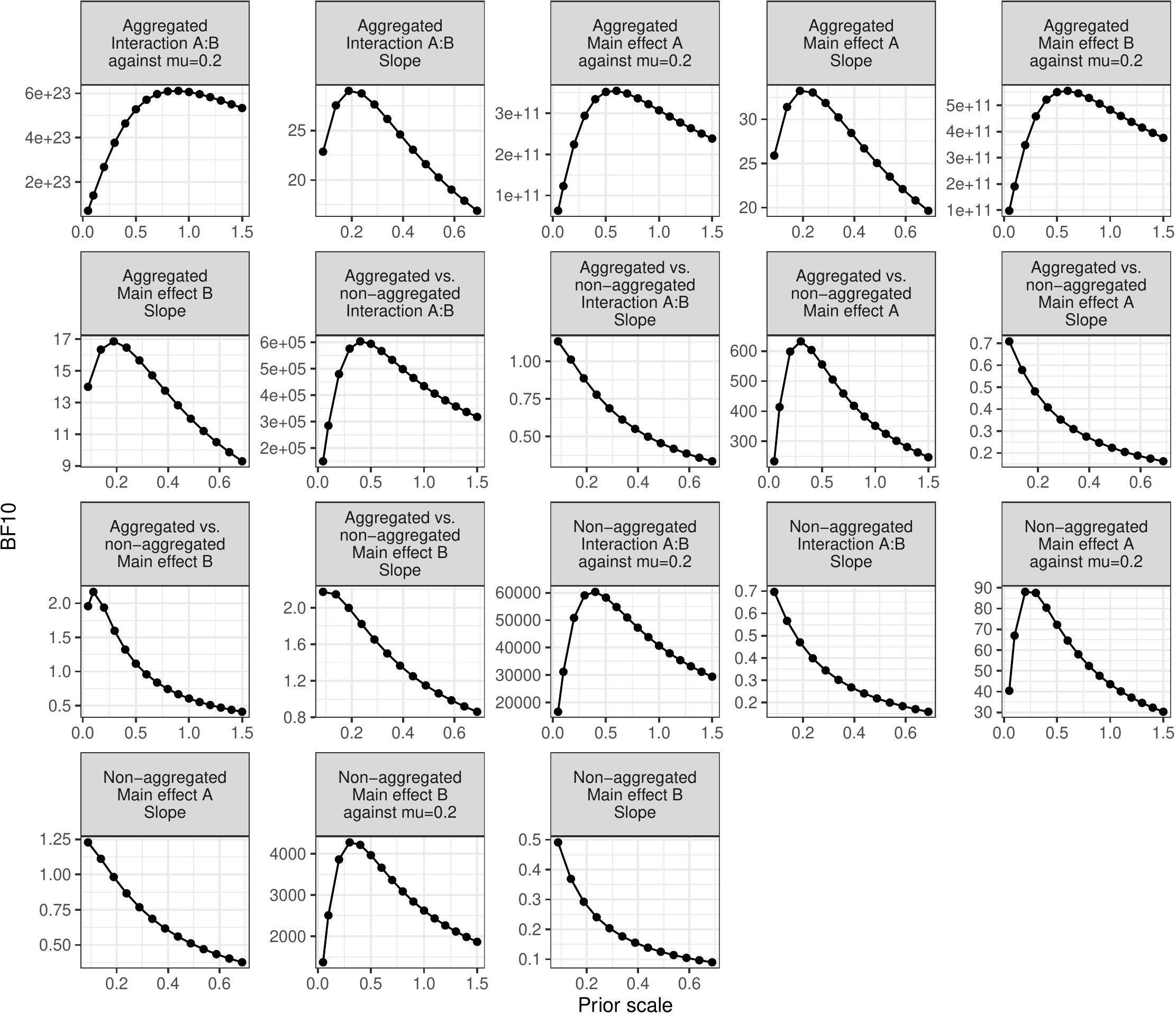} 

}

\caption{Simulation 2.4: brms package: Results from SBC for model inference when item variance is present. Shown are results from a sensitivity analysis on the posterior model probabilities, where the prior scale of the random item slopes is varied across simulations. Results suggest biased Bayes factors for aggregated analyses, that bias is (partly) reduced for non-aggregated analyses, but that also non-aggregated analyses still exhibit bias.}\label{fig:dpri-item-F4-2x2}
\end{figure}

\end{document}